 \documentclass[prc,amsmath,amsfonts,showpacs,multicol,eqsecnum,nofootinbib]{revtex4}
 \usepackage{graphicx}  
 \usepackage{pstcol}    
 \usepackage{axodraw}   
 \usepackage{bm}

\begin{document}
\hyphenation{Rijken}
\hyphenation{Nijmegen}
 
 \title{ Constituent Quark Model and Nucleon-Nucleon Potentials}                                                
\author{Th.A.\ Rijken}
\affiliation{Institute for Theoretical Physics Nijmegen, University of Nijmegen,
         Nijmegen, The Netherlands } 
 
\date{version of: \today}

\begin{abstract}                                       
In these notes, while focusing on the meson-nucleon vertices,
 we give a derivation of the nucleon-nucleon (NN) potentials 
from meson-exchange between the quarks. 
To establish such a relation the quark-quark-meson (QQM) interactions are
properly defined. Hitherto, the coefficients 
in the Pauli-spinor expansion of the meson-nucleon-nucleon (NNM)
vertices are equated with those of the QQM-vertices. 
In these notes we employ the description of the nucleon with Dirac-spinors
in the SU(6) semi-relativistic "constituent" quark-model (CQM) as formulated by
LeYouanc, {\it et al}.
It appears that the "constituent" quark model, {\it i.e.} $m_Q=M_N/3$,
 is able to produce the same ratio's for the central-, spin-spin-, tensor-, 
spin-orbit-, and quadratic-spin-orbit Pauli-invariants as in the 
phenomenological NNM-vertices.
In order to achieve this, the scalar-, magnetic-vector, and axial-vector
interactions require, besides the standard ones, an extra coupling
to the quarks without the introduction of new parameters. In the case of the axial-vector mesons   
an extra coupling to the quarks is necessary, which is related to  
the quark orbital-angular momentum contribution to the nucleon spin.
Furthermore, a momentum correlation between the quark interacting
with the meson and the remaining quark pair, and a gaussian QQM form factor,
are necessary, to avoid "spurious" terms.

From these results we have a formulation of the QQ-interactions which
are directly related to the nucleon-nucleon extended-soft-core (ESC) 
interactions. This can be utilized in for example a study of 
	mixed quark and nuclear matter in neutron stars.
\end{abstract}
\maketitle
 
\twocolumngrid
\section{Introduction}                                     
\label{sec:1}
The main motivation to work out QQM-coupling in the context of
the constituent quark-model (CQM) is that in the extended-softcore (ESC) 
baryon-baryon interactions, see {\it i.e.} \cite{Rij06,Rijk15,NRY19a}, 
the quark-pair creation (QPC)
model is very successful to explain the meson-baryon-baryon (MBB) coupling
constants.

A major success of the non-relativistic (additive) quark model (CQM) 
has been the description of the magnetic moments of the baryons with
 $m_Q= M_N/3$. 

\begin{figure}
 \begin{center}
 \resizebox{5.25cm}{!}
 {\includegraphics[225,675][425,925]{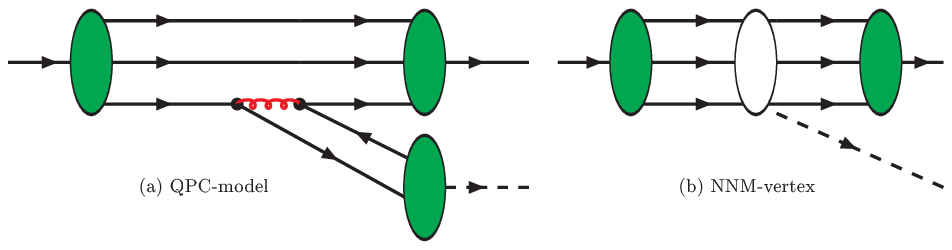}}
 \caption{Meson-nucleon-nucleon coupling.}                         
 \label{fig.ts11}
  \end{center}
  \end{figure}

Also in these notes the description of the nucleon with Dirac-spinors
in the SU(6) semi-relativistic "constituent" quark-model (CQM) as formulated by
LeYouanc, {\it et al} \cite{Yao73} is employed.
In Fig.~\ref{fig.ts11} the QPC-mechanism for NNM-coupling is illustrated.
From the sub-figure (a) it is clear that the basis is the assumption that 
the mesons couple in first instance to the quarks. Then, with folding this
leads to the NNM-coupling illustrated in sub-figure (b).
In this paper we show that the quark-quark-meson QQM) interaction can be 
chosen such that in the folding with the 3-quark nucleon wave function the correct 
$1/\sqrt{M'M}$ expansion of the NN-potentials can be obtained.

In QCD two non-perturbative effects occur: confinement and chiral
symmetry breaking. The SU(3)$_L$xSU(3)$_R$ chiral symmetry is spontaneously 
broken to an SU(3)$_v$ symmetry at some scale $\Lambda_{\chi SB} \approx 1$ GeV
\cite{Nambu60,NJL61,Man84}.
Below this scale there is an octet of pseudoscalar Nambu-Goldstone-bosons: 
$(\pi, K, \eta)$.
The confinement scale $\Lambda_{QCD} \approx 100-330$ MeV. The complex QCD-vacuum
structure can be described as an BPST instanton/anti-instanton liquid giving the
valence quarks a dynamical or constituent effective mass $\approx M_N/3$ 
\cite{BPST75,DY-PE86}. This corresponds to the CQM \cite{Man84}, 
and explains the success of the program proposed in this paper.

In these notes we consider the nucleon-nucleon (NN) potential
from meson-exchange between the (single) quarks in impulse-approximation, 
and folding these with the nucleon quark wave functions.
(In the CQM the 3-quark model wave functions for the SU(3) octet baryons 
are, with respect to flavor and color, 
properly antisymmetrized gaussian quark wave functions reflecting the
ground state of an effective harmonic oscillator binding force.)

We employ the description of the nucleon with Dirac-spinors
in the SU(6)-version of the CQM, see \cite{Yao73}.
In this study we evaluate the NN-meson vertices 
and analyze whether the expansion of these vertices in Pauli-invariants 
is in accordance with the similar expansion used in NN-models
using meson exchange at the nucleon level.

\noindent For elastic scattering with the (external) nucleons on the mass shell,
Lorentz invariance and parity conservation imply that there are
6 independent amplitudes \cite{SNRV71}, i.e. the NN-amplitude can be expressed
in terms of the free nucleon Dirac-spinors as follows

\onecolumngrid
\begin{eqnarray*}
&& {\cal M} = \sum_{i=1}^6 M_i(s,t)\ \left[\bar{u}_{N_1'}(p_1')\bar{u}_{N_2'}(p_2')\  
 O_i\ u_{N_1}(p_1)\ u_{N_2}(p_2)\right] 
\end{eqnarray*}
where a complete set independent
(t-channel) Lorentz-invariants can be chosen as
\begin{eqnarray*}
 O_1 = 1 \otimes 1 \hspace{8mm} && O_2= \gamma_5 \otimes \gamma~_5 \\
 O_3 = \gamma_\mu\otimes\gamma^\mu  \hspace{4mm} &&
   O_4 = \gamma_5\gamma_\mu \otimes \gamma~_5\gamma^\mu \\
 O_5 = \sigma_{\mu\nu}\otimes\sigma^{\mu\nu}  \hspace{0.5mm} &&
   O_6 = i\left\{\gamma_\mu K^\mu\otimes 1 - 1\otimes\gamma_\mu P^\mu\right\},   
\end{eqnarray*}
\twocolumngrid
where $P = p_1+p_1^\prime, K= p_2+p_2^\prime$. We note that 
$O_i = \Gamma_{1,i}\otimes \Gamma_{2,i}$ and that in the meson-exchange 
contribution to the NN-amplitude the NNM-vertex is of the form
 $\bar{u}(p',s') \Gamma u(p,s)$. The Lorentz structure of the NN-amplitude 
 and NNM-vertices 
given above is general and independent of the internal structure of the nucleon.
Therefore, the QQM-exchange vertices folded with the nucleon quark wave functions
has to reproduce at the nucleon level the same structure. 
This observation is the key to the procedure followed in these
notes to define the QQM-vertices.\\

\noindent {\bf Conjecture :}
{\blue The phenomenological expansion of
the vertices in powers of $1/\sqrt{M^\prime M}$ should not depend on the 
internal structure
of the nucleons. So, the ratio's of the central-, spin-spin-, tensor-, 
and spin-orbit operators should be independent internal structure 
of the nucleon}. 

\onecolumngrid
\vspace{2mm}

At the nucleon level, in Pauli-spinor space, the vertices have 
the general structure:              
\begin{eqnarray*}
 \bar{u}(p',s') \Gamma\ u(p,s) &=& \chi^{\prime \dagger}_{s'}\left\{ \Gamma_{bb}
+ \Gamma_{bs}\frac{\bm{\sigma}\cdot{\bf p}}{E+M} 
- \frac{\bm{\sigma}\cdot{\bf p}'}{E'+M'} \Gamma_{sb}
- \frac{\bm{\sigma}\cdot{\bf p}'}{E'+M'} \Gamma_{ss} \frac{\bm{\sigma}\cdot{\bf p}}{E+M} 
 \right\}\ \chi_s \nonumber\\ &\approx& 
 \chi^{\prime \dagger}_{s'}\left\{ \Gamma_{bb}
 + \Gamma_{bs} \frac{(\bm{\sigma}\cdot{\bf p})}{2\sqrt{M'M}}
 - \frac{(\bm{\sigma}\cdot{\bf p}')}{2\sqrt{M'M}} \Gamma_{sb}
 - \frac{(\bm{\sigma}\cdot{\bf p}')\ \Gamma_{ss}\ (\bm{\sigma}\cdot{\bf p})}
  {4M'M} \right\}\ \chi_s \nonumber\\ &\equiv &  
 \sum_{l} c^{(l)}_{NN}\ O_l({\bf p}',{\bf p},\bm{\sigma})\ (\sqrt{M'M})^{\alpha_l}\ \ 
(l=bb,bs,sb,ss),
\end{eqnarray*}
where 
$ O_l({\bf p}',{\bf p},\bm{\sigma})$ denotes the set of operators 
$ 1,\ \bm{\sigma},\ {\bf p},\ {\bf p}',\ \bm{\sigma}\cdot{\bf p},\ 
 \bm{\sigma}\cdot{\bf p}',\ 
 \bm{\sigma}\cdot{\bf p}'\times{\bf p}$, etc.\\
The question is how this structure is reproduced using the coupling of the 
mesons to the quarks directly, {\it i.e.} whether 
for the constants $c^{(l)}_{CQM}=c^{(l)}_{NN}$.
In fact, we want to demonstrate that for the CQM, i.e. 
$m_Q=\sqrt{M'M}/3$, the ratio's $c^{(l)}_{CQM}/c^{(l)}_{NN}$ 
are constant for each type of meson.
Then, by scaling the expansion coffients can be made equal.\\

\twocolumngrid
 
\noindent Therefore, we expect these ratio's are essentially the same as for the 
expansion of the NNM-vertices with Pauli-spinor invariants.\\
\noindent {\bf We found this to be possible for most of the terms,
up to order $1/M'M$, in the CQM where 
 $m_Q=M_N/3$, for all couplings: pseudo-scalar (P), scalar (S), vector (V) and 
axial-vector (A) coupling.}\\

{\it In the scalar, vector, and axial-vector vertex there appear    
"spurious" terms $\propto 1/R_N^2$. 
This is only the case for the central and spin term of the scalar/vector and
axial-vector respectively.                                                  
In view of the "conjecture", these terms should not be present, and must be eliminated.
We demonstrate that such "spurious" terms can be eliminated by introducing 
a momentum exchange between the "active" quark, {\it i.e.} the quark line with the meson vertex,
 and the two "spectator" quarks.
(In the simplest model without such a momentum exchange, this amounts to the introduction 
of a gaussian momentum distribution at the QQM-vertex.)
}.\\

In this study we evaluate the QQM vertices 
and analyze whether the expansion of these vertices in Pauli-invariants 
matches with the similar expansion used in NN-models
using meson exchange at the nucleon level.
To accomplish this we add the following vertices at the quark level:
(i) for the vector-mesons a zero in the scalar derivative part, and (ii) 
in the case of the axial-vector coupling an additional pseudoscalar derivative 
interaction.
To work out these ideas concretely, we 
use the description of the nucleon with Dirac-spinors
in the SU(6)-version of the CQM, see \cite{Yao73}.
In the CQM the rational for this is that since $M_N=3m_Q$ the quark kinetic and
potential energies cancel each other, which means that for the quark
energies $E_i \approx m_i$.


{\it As a final note: The QQM- and NNM-vertices are for potentials ${\cal V}$
in the Lippmann-Schwinger equation. For the relation with the 
(kinematically relativistic) Thompson, Kadyshevsky etc. equations, 
see Ref.~\cite{NRS77}}.
 
The content of these notes is as follows. 
In section~\ref{sec:17} the QCD basis of the CQM based on the instanton-model
of the QCD-vacuum is briefly reviewed.
In section \ref{sec:2} we review the 
quark wave functions and the overlap integrals. In section~\ref{sec:3}-\ref{sec:6}
 we treat scalar-exchange, pseudo-scalar-, vector-, and axial-vector-meson exchange.
In section~\ref{sec:interm} a method is given to remove "spurious" terms from the 
NN-vertices $\Gamma_{CQM}$. To complete this it is necessary to use a (gaussian)
QQM cut-off.
In section~\ref{sec:8} we formulate our conclusions. 
In Appendix~\ref{app:O1} the overlap integral for meson exchange is worked out.
Similarly in Appendix~\ref{app:O2}, where a momentum correlation is included between 
the quark with the meson-vertex and the remaining quark-pair, henceforth referred to
as the "active" quark and "spectator" quarks respectively. It is shown that
with such arrangement the "spurious" terms are eliminated, and can explain the  
procedure introduced in section~\ref{sec:interm}.
In Appendix~\ref{app:A} we discuss the quark summation. In Appendix~\ref{app:B}
we list the Pauli-spinor invariants for the nucleon-nucleon potentials. In 
Appendix~\ref{app:C} the extended-soft-core (ESC) quark-quark (QQ) 
OBE-interactions in 
momentum and configuration space are given for reference of the vertex 
structures with Pauli-invariants.
In Appendix~\ref{app:sc-vertex2} the lower vertex for the scalar-meson 
QQ-coupling is worked out for comparison with the upper vertex.
In Appendix~\ref{app:ten} tensor-meson exchange is analyzed and compared with 
scalar- and vector-meson exchange.

\twocolumngrid

\section{Constituent Quarks and Meson-exchange}
\label{sec:17}
The spectra of the nucleons, $\Delta$ resonances and the hyperons $\Lambda, \Sigma, \Xi$ 
are, for example, described in detail by the Glozman-Riska model \cite{Gloz96a}.
This is a modern version of the constituent quark model (CQM) \cite{Greenberg64} 
based on the Nambu-Goldstone spontaneous chiral-symmetry breaking (SCSB) 
with quarks interacting by the exchange of the SU(3)$_F$ octet of pseudoscalar
mesons \cite{Gloz96a}. The pseudoscalar octet are the Goldstone bosons associated 
with the hidden (approximate) chiral symmetry of QCD.
The confining potential is chosen to be harmonic, as is rather common in 
constituent quark models. This is in line with the harmonic wave functions we used 
in the derivation of
the connection between the meson-baryon and meson-quark couplings \cite{Rij.nnqq14}.
The $\eta'$, which is dominantly an SU(3) singlet, decouples from the original 
pseudoscalar nonet because of the U$_A$(1) anomaly \cite{Weinberg75,GtH76}. 
According to the two-scale picture of Manohar and Georgi \cite{Man84} the effective
degrees for the 3-flavor QCD at distances beyond that of SCSB 
($\Lambda_{\chi SB}^{-1} \approx 0.2-0.3$ fm),
but within that of the confinement scale $\Lambda_{QCD}^{-1} \approx 1$ fm,
should be the constituent quarks and chiral meson fields.
The two non-perturbative effects in QCD are confinement and chiral symmetry
breaking. The SU(3)$_L\otimes$SU(3)$_R$ chiral symmetry is spontaneously broken to 
an SU(3)$_v$ symmetry at a scale $\Lambda_{\chi SB} \approx 1$ GeV. 
The confinement scale is $\Lambda_{QCD} \approx 100-300$ MeV, which roughly
corresponds to the baryon radius $\approx$ 1 fm.
Due to the complex structure of the QCD vacuum, which can be understood as a
liquid of BPST instantons and anti-instantons \cite{BPST75,DY-PE84a,DY-PE84b,DY-PE86}, 
the valence quarks acquire a dynamical or constituent mass \cite{Weinberg75,Man84,
SHUR84,DY-PE84b,DY-PE86}. 
The interaction between the instanton and the anti-instanton is a dipole-interaction
\cite{SHUR82a}, similar to ordinary molecules: weak attraction at large distances
and strong repulsion at small ones. With the empirical value of the gluon
condensate \cite{SVZ79} as input the instanton density and radius become 
\cite{SHUR82a}
$n_c= 8\cdot 10^{-4}\ {\rm GeV}^{-4}$, and $\rho_c= (600\ {\rm MeV})^{-1}  
\approx 0.3\ {\rm fm}$ respectively.
Also, with these parameters the non-perturbative vacuum expectation value for 
the quark fields is
$\langle vac|\bar{\psi}\psi|vac\rangle \approx -10^{-2}\ {\rm GeV}^3$ and the quark effective
mass $\approx 200$ MeV, which is much larger than the almost massless (u,d) 
"current quarks". In the calculation of light quarks in the 
instanton vacuum \cite{DY-PE86}  the effective quark mass $m_Q(p=0)= 345$ MeV was 
calculated, which is remarkably close to the constituent mass $M_N/3$.
-----------------------------------------------------------------------------\\

\noindent In this paper we extend the meson-exchange between quarks by proposing to include,
besides the pseudoscalar, all meson nonets: vector, axial-vector, scalar etc.
{\it Since all these meson nonets can be considered as quark-antiquark bound states,
there is no reason to exclude any of these mesons from the quark-quark interactions.}
{\it Furthermore, our preferred value for the constituent quark mass has 
a solid basis in the instanton-liquid model of the QCD vacuum.}


\onecolumngrid

\section{Quark Wave Functions of the Nucleons}             
\label{sec:2}
 
\subsection{Kinematics and Dirac spinors}                     
\label{sec:2.1}
We consider a nucleon having a momentum $P$ and label the 3 quarks
by $a,b,c$. The quark momenta are denoted by $p_{a}, p_{b},p_{c}$.

The spatial part of the composite nucleon wave function is taken to be
\cite{Yao73}
\begin{eqnarray}
\psi(p_{a},p_{b},p_{c}) &=& \psi(p_{1},p_{2},p_{3})=
\left(\frac{\sqrt{3} R_N^{2}}{\pi}\right)^{3/2}\ \exp\left[
 -\frac{R^{2}}{6} \sum_{i<j}\left({\bf p}_{i}-{\bf p}_{j}\right)^{2}\right]
\label{eq:2.2} \end{eqnarray}
The normalization constant in (\ref{eq:2.2})\
we denote by ${\cal N} \equiv \left(\sqrt{3}R_N^2/\pi\right)^{3/2}$.


\noindent  
In the constituent quark model (CQM) the nucleon (baryon) mass is given by the
sum of quark masses, {\it i.e.} $M_N= 3m_Q$, the quark energies satisfy
$E_Q = m_Q+T_Q+U_Q$, the kinetic($T_Q$) and potential ($U_Q$) energies 
cancel approximately $T_Q+U_Q \approx 0$. Therefore, the constituent 
quark spinors are \cite{Yao73}
\begin{equation}
 u^{(0)}_i({\bf p}_i) = \sqrt{\frac{E_i+m_i}{2 m_i}}
 \left[\begin{array}{c} 
 1 \\ 
\frac{\bm{\sigma}_i\cdot{\bf p}_i}{E_i+m_i}
 \end{array}\right] \otimes\ \chi_i \approx    
 \left[\begin{array}{c} 
 1 \\ 
\frac{\bm{\sigma}_i\cdot{\bf p}_i}{2m_i}
 \end{array}\right] \otimes\ \chi_i,   
\label{eq:2.3} \end{equation}
where ${\bf p_i}_i$ denotes the three-momentum of the quarks in e.g. the CM-system.


\subsection{Overlap Integrals, Vertex functions}              
\label{sec:2.2}
We consider the nucleon-nucleon graph for meson-exchange between
the constituent quarks of the two nucleons.
In the following we will use, instead of indices a,b,c, the indices  i=1,2,3.

 \begin{figure}
 \begin{center}
 \resizebox{7.25cm}{!}
 {\includegraphics[225,575][425,825]{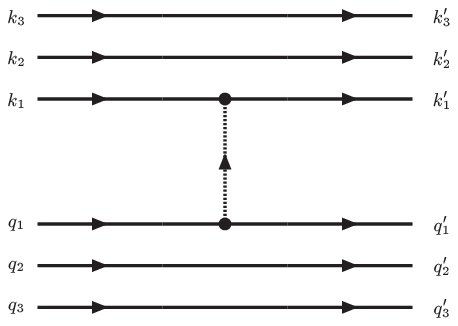}}
\caption{Nucleon internal quark momenta with meson-exchange}   
\label{fig:2.2}
  \end{center}
  \end{figure}                     








In Fig.~\ref{fig:2.2} we have given the momenta for the initial
and final nucleons, and the assigned momenta of the quarks.
From momentum conservation we have
\begin{eqnarray}
 p_{1} = k_{1}+k_{2}+k_{3}\ &,& \  p_{2}=q_{1}+q_{2}+q_{3}\ ,
 \nonumber \\ && \nonumber \\
 p'_{1} = k'_{1}+k'_{2}+k'_{3}\ &,& \ p'_{2}=q'_{1}+q'_{2}+q'_{3}\ ,
\label{eq:2.6} \end{eqnarray}
For meson-exchange with $p_{1}-p'_{1}=p'_{2}-p_{2} \equiv k$, we have
for the matrix-element of the potential
\begin{eqnarray}
\langle p'_{1}p'_{2}|V|p_{1}p_{2}\rangle &=& 
\int\prod_{i=1,3}\ d^{3}k_{i}\ \delta\left({\bf p}_{1}-\sum_{i}{\bf k}_{i}
\right)\cdot
\int\prod_{j=1,3}\ d^{3}k'_{j}\ \delta\left({\bf p'}_{1}-\sum_{j}{\bf k'}_{j}
\right)\cdot
\nonumber \\  &\times&
\int\prod_{i=1,3}\ d^{3}q_{i}\ \delta\left({\bf p}_{2}-\sum_{i}{\bf q}_{i}
\right)\cdot
\int\prod_{j=1,3}\ d^{3}q'_{j}\ \delta\left({\bf p'}_{2}-\sum_{j}{\bf q'}_{j}
\right)
\cdot \nonumber \\ &\times&
\tilde{\psi}_{p'_{1}}^{*}\left({\bf k'}_{1},{\bf k'}_{2},{\bf k'}_{3}\right)\
\tilde{\psi}_{p'_{2}}^{*}\left({\bf q'}_{1},{\bf q'}_{2},{\bf q'}_{3}\right)
\cdot
\tilde{\psi}_{p_{1}}\left({\bf k}_{1},{\bf k}_{2},{\bf k}_{3}\right)\
\tilde{\psi}_{p_{2}}\left({\bf q}_{1},{\bf q}_{2},{\bf q}_{3}\right)\
\cdot \nonumber \\ &\times&
 \delta^{3}\left({\bf k'}_{2}-{\bf k}_{2}\right)\ 
 \delta^{3}\left({\bf k'}_{3}-{\bf k}_{3}\right)\ 
 \delta^{3}\left({\bf q'}_{2}-{\bf q}_{2}\right)\ 
 \delta^{3}\left({\bf q'}_{3}-{\bf q}_{3}\right)\ 
\cdot \nonumber \\ &\times&
\frac{\gamma\left(k; k'_{1},k_{1}\right)\ \gamma\left(k; q'_{1},q_{1}\right)}
{{\bf k}^{2}+m_{M}^{2}}\cdot 
 \delta^{3}\left({\bf k}-{\bf k}'_{1}+{\bf k}_{1}\right)\ 
 \delta^{3}\left({\bf k}+{\bf q}'_{1}-{\bf q}_{1}\right)\ .
\label{eq:2.7} \end{eqnarray}
In (\ref{eq:2.7}) the $\gamma$'s denote the vertex functions.
Using the gaussian wave function of equation (\ref{eq:2.2}), the overlap
integral in Eq.~(\ref{eq:2.7}) can be evaluated in a straightforward manner.
For details see Appendix~\ref{app:O1}.

\noindent For doing later integrals with explicit terms for the QQ-potential, it is
useful to write the expression (\ref{app:O1.11}) with separated vertex factors:
\begin{eqnarray}
\langle p'_{1}p'_{2}|V|p_{1}p_{2}\rangle &=& \left(\frac{1}{8}\right)^3
\left(\frac{2\pi}{R_{N}^{2}}\right)^3\ {\cal N}^4\ \exp\left[-\frac{R_N^2}{3}\left(
{\bf q}^2+{\bf k}^2\right)\right]\ 
\cdot\nonumber\\[0.3cm] && \times
 \int d^3Q\ \exp\left[-\frac{R_{N}^{2}}{6}
 \left\{\frac{9}{4}\left({\bf Q}^2-\frac{4}{3}{\bf q}\cdot{\bf Q}\right) 
\right\}\vphantom{\frac{A}{A}}\right]\cdot \nonumber\\[0.3cm] && \times 
 \int d^3S\ \exp\left[-\frac{R_{N}^{2}}{6}
 \left\{\frac{9}{4}\left({\bf S}^2+\frac{4}{3}{\bf q}\cdot{\bf S}\right) 
\right\}\vphantom{\frac{A}{A}}\right]\cdot \nonumber\\[0.3cm] && \times 
 V_{QQ}({\bf Q},{\bf S};{\bf k}, {\bf q})\ 
 \times \delta\left({\bf k}'_1+{\bf q}'_1-{\bf k}_1-{\bf q}_1\right),              
\label{eq:2.18a} \end{eqnarray}
where the QQ-potential is
\begin{eqnarray}
 V_{QQ}({\bf Q},{\bf S};{\bf k}, {\bf q}) &=& 
\frac{\gamma\left(k; k'_{1},k_{1}\right)\ \gamma\left(k; q'_{1},q_{1}\right)}
{{\bf k}^{2}+m_{M}^{2}}  
\label{eq:2.18b} \end{eqnarray}
with the momenta, see Appendix~\ref{app:O1}, defined as
\begin{subequations}
\begin{eqnarray}
 {\bf k}'_1 = \frac{1}{2}\left({\bf Q} + {\bf k}\right) &,&          
 {\bf k}_1=\frac{1}{2}\left({\bf Q} - {\bf k}\right), \\          
 {\bf q}'_1 = \frac{1}{2}\left({\bf S} - {\bf k}\right) &,&          
 {\bf q}_1=\frac{1}{2}\left({\bf S} + {\bf k}\right).             
\label{eq:2.18c} \end{eqnarray}
\end{subequations}
More explicitly for spin-J mesons ( m,n =1, ...., 2J+1)
the numerator in (\ref{eq:2.18b} stands for
\begin{equation}
\gamma\left(k; k'_{1},k_{1}\right)\ \gamma\left(k; q'_{1},q_{1}\right) \rightarrow
\gamma^{\{\mu_m\}}\left(k; k'_{1},k_{1}\right)\ 
 {\cal P}_{\{\mu_m\},\{\nu_n\}}(k)
\gamma^{\{\nu_n\}}\left(k; q'_{1},q_{1}\right).
\label{eq:2.18d} \end{equation}
 
\noindent The basic $d^3Q$ and $d^3S$ integrals are  
\begin{subequations}\label{eq:2.19}
\begin{eqnarray}
 I_0({\bf q}) &=& \int d^3Q\ \exp\left[-\frac{R_{N}^{2}}{6}
 \left\{\frac{9}{4}\left({\bf Q}^2-\frac{4}{3}{\bf q}\cdot{\bf Q}\right) 
 \right\}\right]
 = \left(\frac{8\pi}{3R_N^2}\right)^{3/2}\ 
 \exp\left[\frac{1}{6} {\bf q}^2\ R_N^2\right], \\
 J_0({\bf q}) &=& \int d^3S\ \exp\left[-\frac{R_{N}^{2}}{6}
 \left\{\frac{9}{4}\left({\bf S}^2+\frac{4}{3}{\bf q}\cdot{\bf Q}\right) 
 \right\}\right]
 = \left(\frac{8\pi}{3R_N^2}\right)^{3/2}\ 
 \exp\left[\frac{1}{6} {\bf q}^2\ R_N^2\right]. 
 \end{eqnarray} \end{subequations}
Then,
\begin{subequations}\label{eq:2.20a}
\begin{eqnarray}
 I_i({\bf q}) &=& \int d^3Q\ Q_i\ \exp\left[-\frac{R_{N}^{2}}{6}
 \left\{\frac{9}{4}\left({\bf Q}^2-\frac{4}{3}{\bf q}\cdot{\bf Q}\right) 
 \right\}\right] = \frac{2}{R_N^2} \bm{\nabla}_i\ I_0({\bf q}) 
 = +\frac{2}{3}\ q_i\ I_0({\bf q}), \\
 J_i({\bf q}) &=& \int d^3S\ S_i\ \exp\left[-\frac{R_{N}^{2}}{6}
 \left\{\frac{9}{4}\left({\bf S}^2+\frac{4}{3}{\bf q}\cdot{\bf S}\right) 
 \right\}\right] = -\frac{2}{R_N^2} \bm{\nabla}_i\ I_0({\bf q}) 
 = -\frac{2}{3}\ q_i\ I_0({\bf q}), 
 \end{eqnarray} \end{subequations}
\begin{subequations}\label{eq:2.20b}
and
\begin{eqnarray}
 I_{i,j}({\bf q}) &=& \int d^3Q\ Q_i Q_j\ \exp\left[-\frac{R_{N}^{2}}{6}
 \left\{\frac{9}{4}\left({\bf Q}^2-\frac{4}{3}{\bf q}\cdot{\bf Q}\right) 
 \right\}\right] = \left(\frac{2}{R_N^2}\right)^2 
\bm{\nabla}_i \bm{\nabla}_j\ I_0({\bf q}) \nonumber\\ &=&
  \left[\frac{4}{3 R_N^2} \delta_{i,j} + \frac{4}{9}\ q_i q_j\right]
 I_0({\bf q}), \\ 
 J_{i,j}({\bf q}) &=& \int d^3S\ S_i S_j\ \exp\left[-\frac{R_{N}^{2}}{6}
 \left\{\frac{9}{4}\left({\bf S}^2+\frac{4}{3}{\bf q}\cdot{\bf S}\right) 
 \right\}\right] = \left(\frac{2}{R_N^2}\right)^2 
\bm{\nabla}_i \bm{\nabla}_j\ I_0({\bf q}) \nonumber\\ &=&
  \left[\frac{4}{3 R_N^2} \delta_{i,j} + \frac{4}{9}\ q_i q_j\right]
 I_0({\bf q}).    
 \end{eqnarray} \end{subequations}
The meson-nucleon vertex $\Gamma$ is, analogously with V, given by
\begin{eqnarray}
\langle p'_{1}|\Gamma|p_{1}\rangle &=& {\cal N}^2\left(\frac{1}{8}\right)^{3/2}
\left(\frac{2\pi}{R_{N}^{2}}\right)^{3/2}\ 
 \exp\left[-\frac{R_N^2}{6}\left(
{\bf q}^2+{\bf k}^2\right)\right]\ 
\cdot\nonumber\\[0.3cm] && \times
 \int d^3Q\ \exp\left[-\frac{R_{N}^{2}}{6}
 \left\{\frac{9}{4}\left({\bf Q}^2-\frac{4}{3}{\bf q}\cdot{\bf Q}\right) 
\right\}\vphantom{\frac{A}{A}}\right]\cdot 
 \gamma({\bf Q};{\bf k}, {\bf q})\ 
 \delta^{(3)}\left({\bf p}'_1-{\bf p}_1-{\bf k}\right) \nonumber\\ &\sim&
 {\cal N}^2 \left(\frac{2}{3}\right)^{3/2}
\left(\frac{\pi}{R_{N}^{2}}\right)^{3}\cdot
 \exp\left[-\frac{1}{6} R_N^2{\bf k}^2\right]\ 
 \bar{\Gamma}({\bf q}, {\bf k})\ 
 \delta^{(3)}\left({\bf p}'_1-{\bf p}_1-{\bf k}\right).                      
\label{eq:2.21} \end{eqnarray}
Here the last expression shows that the vertex has a gaussian local 
form factor.
 
 \begin{figure}
 \begin{center}
 \resizebox{7.25cm}{!}
 {\includegraphics[225,700][425,950]{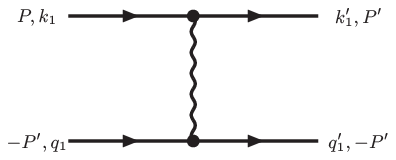}}
\caption{$V_{QQ}$ Scalar-exchange in the CM-frame}                 
\label{fig:3.1}
  \end{center}
  \end{figure}                     






\section{Scalar-exchange}                              
\label{sec:3}
The coupling of the scalar meson to the quarks we assume to be of the
form
\begin{equation}
 {\cal H}_S = \left[g_1 \bar{\psi}\psi - g_2 \Box(\bar{\psi}\psi)/(2\mu^2)
 \right]\ \sigma.
\label{eq:3a.1} \end{equation}
The corresponding vertex is
\begin{eqnarray}
 \Gamma_{QQ} &=& \bar{u}_Q(p')\left[g_1 +g_2\left(M^2-p'\cdot p\right)/\mu^2
 \right]\ u_Q(p) 
\label{eq:3a.2} \end{eqnarray}
Now,
\begin{equation}
 M^2-p'\cdot p = M^2-E'E+{\bf p}'\cdot{\bf p} \approx -{\bf k}^2/2.
\label{eq:3a.3} \end{equation}
Taking a common form factor for the two couplings, (\ref{eq:3a.3})
implies a {\bf zero in the potential} 
\begin{equation}
 0 = \left(g_1- \frac{{\bf k}^2}{2\mu^2} g_2\right)^2 
 \approx g_1^2-\frac{{\bf k}^2}{\mu^2} g_1 g_2,\ \ 
{\bf k}^2(0) = \mu^2\ \frac{g_1}{g_2},
\label{eq:3a.4} \end{equation}
which marks an approximate simple zero.
In ESC-models $g_1=g_2$ and ${\bf k}^2(0)= m_\sigma^2$, so 
$\mu=m_\sigma \approx 2m_Q$.\\
\noindent {\it On the quark level, the inclusion of the zero implies a change
in the coefficients of the ${\bf k}^2$-term which are of order $1/M^2$}.


\subsection{Folding Amplitude Scalar-exchange}            
\label{sec:3c}
The Dirac-spinor part of the scalar-meson QQ-vertex is
\begin{eqnarray}
 \left[\bar{u}_{i}({\bf k'}_{i}) u_{i}({\bf k}_{i}) 
  \right] &=& 
 \sqrt{\frac{E'_{i}+m'_{i}}{2m'_{i}}\ \frac{E_{i}+m_{i}}{2m_{i}} }\cdot
 \chi^{\prime \dagger}_{i}\cdot 
 \left[ 1 -
 \frac{\mbox{\boldmath $\sigma$}_{i}\cdot{\bf k'}_{i}}{E'_{i}+m_{i}}
 \frac{\mbox{\boldmath $\sigma$}_{i}\cdot{\bf k}_{i}}{E_{i}+m_{i}}\right]
 \nonumber\\ &\approx& 
 \chi^{\prime \dagger}_{i}\left[ 1 -
 \frac{{\bf k}'_i\cdot{\bf k}_i}{4m_i^2}                                        
 -\frac{i}{4m_i^2}\bm{\sigma}_i\cdot{\bf k}'_i\times{\bf k}_i\right]\ \chi_i
 \nonumber\\ &=& 
 \chi^{\prime \dagger}_{i}\left[ 1 -
 \frac{{\bf Q}_i^2-{\bf k}^2}{16m_i^2}                                        
 +\frac{i}{8m_i^2}\bm{\sigma}_i\cdot{\bf Q}_i\times{\bf k}\right]\ \chi_i.
\label{eq:3.41} \end{eqnarray}
Here is used that for the CQM $E_i \approx m_i$. The performance of the 
${\bf Q}$-integral in (\ref{eq:3.41}) gives
\begin{eqnarray}
 \left[\bar{u}_{i}({\bf k'}_{i}) u_{i}({\bf k}_{i}) 
  \right] &\Rightarrow& \chi^{\prime \dagger}_{i}
 \left[ 1 -\left(\frac{1}{4m_i^2R_N^2}+\frac{{\bf q}^2}{36m_i^2}\right)
 + \frac{{\bf k}^2}{16m_i^2}             
 +\frac{i}{12m_i^2}\bm{\sigma}_i\cdot{\bf q}\times{\bf k}\right]\ \chi_i
\label{eq:3.42} \end{eqnarray}
Summing over the quarks leads to the vertex
\begin{eqnarray}
 \Gamma_{CQM} &=&
 \sum_{i=1-3}\left[\bar{u}_{i}({\bf k'}_{i}) u_{i}({\bf k}_{i}) 
  \right] \Rightarrow
 3\left[ 1 -\left(\frac{1}{4m_Q^2R_N^2}+\frac{{\bf q}^2}{36m_Q^2}\right)
 + \frac{{\bf k}^2}{16m_Q^2}             
 +\frac{i}{36m_Q^2}\sum_i \bm{\sigma}_i\cdot{\bf q}\times{\bf k}\right]
\label{eq:3.43} \end{eqnarray}
The CQM replacement $m_Q \approx \sqrt{M'M}/3$ leads to
\begin{eqnarray}
 \Gamma_{CQM} &=&
 3\left[ \left(1 -\frac{1}{4m_Q^2R_N^2}\right) -\frac{{\bf q}^2}{4M'M}
 + \frac{9{\bf k}^2}{16M'M}             
 +\frac{i}{4M'M}\sum_i \bm{\sigma}_N\cdot{\bf q}\times{\bf k}\right], 
\label{eq:3.44} \end{eqnarray}
where we used $\sum_i \bm{\sigma}_i=\bm{\sigma}_N$.
This assumes that the spin of the nucleon is given by the total
spin of the quarks \cite{remark-spin}


\noindent This result should be compared with the 
$\left[\vphantom{\frac{A}{A}} \ldots \right]$-part
of the vertex computed at the nucleon-level, 
$\bm{\Delta}^2=({\bf P}'-{\bf P})^2= {\bf k}^2$,
\begin{subequations}\label{eq:3.45}
\begin{eqnarray}
 \Gamma_{NN} &\equiv& \bar{u}({\bf P}')\ u({\bf P}) =
 \sqrt{\frac{E'+M'}{2M'}}\ \sqrt{\frac{E+M}{2M}}\dot\nonumber\\ && \times
 \chi^{\prime\ \dagger}\left[1- \frac{{\bf p}'\cdot{\bf p}}{(E'+M')(E+M)}
 -\frac{ i {\bf p}'\times{\bf p}\cdot\bm{\sigma}}{(E'+M')(E+M)}
 \right] \chi \nonumber\\ &\approx& 
 \sqrt{\frac{E'+M'}{2M'}}\ \sqrt{\frac{E+M}{2M}}\                           
 \chi^{\prime\ \dagger}\left[1- \frac{{\bf q}^2}{4M'M}
 +\frac{\bm{\Delta}^2}{16M'M}
 +\frac{i}{4M'M} {\bf q}\times\bm{\Delta}\cdot\bm{\sigma}
 \right] \chi \\ &\Rightarrow& 
 \left[1- \frac{{\bf q}^2}{4M'M} +\frac{\bm{\Delta}^2}{16M'M}
 +\frac{i}{4M'M} {\bf q}\times\bm{\Delta}\cdot\bm{\sigma}
 \right].                        
\end{eqnarray}\end{subequations}
The last expression for $\Gamma_{NN}$ is the correspondence of $\Gamma_{CQM}$
in (\ref{eq:3.44}). This because in the transition from the potential
V to the Lippmann-Schwinger potential ${\cal V}$
there occurs a factor $(E'+E)/(M'+M)$ \cite{NRS77}. Now,
\begin{equation}  
\frac{E'+E}{M'+M} \approx  1+\frac{{\bf q}^2+\bm{\Delta}^2/4}{2M'M},\  
 \frac{(E'+M')(E+M)}{4M'M} \approx                  
1-\frac{{\bf q}^2+\bm{\Delta}^2/4}{2M'M},    
\label{eq:3.46} \end{equation}
showing that the product is $1+ O((M'M)^{-2} \approx 1$.\\

To bring $\Gamma_{CQM}$ and $\Gamma_{NN}$ in agreement the following:
\begin{enumerate}
\item[a.] The factor 3 is accounted for by scaling
the quark-meson coupling, {\it i.e. } $g^{(S)}_Q= g^{(S)}/3$.
\item[b.] The "spurious" term $ 1/(4m_Q^2/R_N^2)=9/(4M'MR_N^2)$
can be removed by introducing a gaussian distribution for ${\bf K}$, see 
subsection~\ref{sec:interm}.
\item[c.] Compared to $\Gamma_{NN}$ the quark vertex $\Gamma_{CQM}$ has
an extra $8{\bf k}^2/(16M'M)$-term. This term can be canceled by tuning the 
$g_2$-coupling. For that purpose we set
\begin{equation}
 -g_2\frac{9}{16M'M} = g_1\frac{8}{16M'M},\ g_2/g_1 = -8/9 \approx -g_1.
\label{eq:3.47} \end{equation}
\end{enumerate}
With these remarks it is shown that, although not identical, the QQ- and NN-vertex
are (approximately) equivalent as far as the NN-potential is concerned.
It also shows that the combination of scalar
and vector exchange are necessary to bring this about.\\

 This is consistent with the remarks after 
 Eq.~(\ref{eq:3a.4}), i.e. $g_2=g_1$ and $\mu=2m_i$.\\

\subsection{Scalar Form-factor Zero in QQ- and NN-vertex}\label{sec:zero}
\noindent {\it Furthermore we remark that in the ESC-models we use a simple 
(first-order) zero for the scalar-meson exchange potential. 
Taking a zero at the vertex,
as suggested by the analysis here, would imply a double zero.
To match the practice in the ESC-models one can use the zero at the
vertex partly for the proper generation of the ${\bf k}^2$-term and
partly for the simple zero in the potential, i.e. we expand}
\begin{eqnarray*}
 \left(1-\frac{{\bf k}^2}{2U^2}\right)^2 = 
 \left(1-\frac{{\bf k}^2}{U^2}\right) + O({\bf k}^4/U^4).
\end{eqnarray*}
Including the ESC-zero, the scalar-vertex becomes
\begin{eqnarray}
 \Gamma_S &=& \bar{u}(p^\prime)\left[g_1\left(1-\frac{{\bf k}^2}{2U_S^2}\right)
 -g_2 \frac{{\bf k}^2}{8m_Q^2}\right]\ u(p) \nonumber\\
 &\approx& g_1 \bar{u}(p^\prime)\left[1-\left(\frac{1}{2U_S^2}
 +\frac{1}{8m_Q^2}\right)\ {\bf k}^2\right]\ u(p),
\label{eq:3.48} \end{eqnarray}
which implies a zero in the scalar-quark coupling at
\begin{eqnarray*}
 {\bf k}^2 &=& \frac{4m_Q^2 U_S^2}{U_S^2+4m_Q^2} \equiv 2 U^2_Q.
\end{eqnarray*}
With $U_S = 3M_N/3$ and $m_Q = M_N/3$ we get 
$U_Q \approx \sqrt{\frac{7}{32}} U_S \approx U_S/2$.\\
Considering scalar-exchange between a quark-line and a nucleon-line we
have, up to terms of order ${\bf k}^4$,
\begin{eqnarray}
 \Gamma_Q\cdot \Gamma_N &\approx& 
1-\left(\frac{1}{2U_Q^2}+\frac{1}{2U_S^2}\right) {\bf k}^2 
 = 1-\left(\frac{U_Q^2+U_S^2}{2U_Q^2 U_S^2}\right) {\bf k}^2 
 \nonumber\\ &\equiv& 1 - \frac{{\bf k}^2}{U_{Q+S}^2}, \ \
 U_{Q+S}^2 = 2 U_S^2 \left[1+\frac{U_S^2}{U_Q^2}\right]^{-1}.   
\label{eq:3.49} \end{eqnarray}
We have
\begin{eqnarray*}
 \frac{U_S^2}{U_Q^2} &=& 1+ U_S^2/(4m_Q^2) \approx 32/7,    
\end{eqnarray*}
 where we used that $U_S = 750\ {\rm MeV} \approx (3/4) M_N$, and $m_Q = M_N/3$.
This means that $U_{Q+S} \approx U_S/\sqrt{3}$.\\

\noindent {\it Although this method can be chosen for the scalar- and
and axial-meson coupling, it is not available for the vector-mesons.
Therefore, the use of an extra coupling to match the ${\bf k}^2$ terms
is preferable.}

\subsection{Removal spurious central term}\label{sec:interm}
Instead of the $\delta^{3}({\bf K}-{\bf k})$-function we introduce a
distribution of the momentum ${\bf K}$ exchange. Such a distribution might be
caused by momentum exchange between the quark-line with the meson-vertex and
the other two quarks in the nucleon, see Appendix~\ref{app:O2} for an explicit
demonstration.
To produce a $\Gamma_{CQM}$ without a "spurious" term, we consider the integrals
\begin{eqnarray}
K_1({\bf k}^2) &=& N_1 \int d^3K\ \exp\left[-\alpha {\bf K}^2
+\beta {\bf K}\cdot{\bf k}\right]\ e^{-\gamma {\bf k}^2} 
 = N_1 \left(\frac{\pi}{\alpha}\right)^{3/2} 
 \exp\left[-\left(\gamma-\frac{\beta^2}{4\alpha}\right) {\bf k}^2\right],
\label{eq:interm.1}\end{eqnarray}
\begin{eqnarray}
K_2({\bf k}^2) &=& N_1 \int d^3K\ {\bf K}^2\ \exp\left[-\alpha {\bf K}^2
+\beta {\bf K}\cdot{\bf k}\right]\ e^{-\gamma {\bf k}^2} 
 = \left[\frac{3}{2}\alpha^{-1} + \frac{\beta^2}{4\alpha^2}{\bf k}^2\right]\
K_1({\bf k}^2),
\label{eq:interm.2}\end{eqnarray}
\begin{eqnarray}
K_{3,i}({\bf k}^2) &=& N_1 \int d^3K\ K_i\ \exp\left[-\alpha {\bf K}^2
+\beta {\bf K}\cdot{\bf k}\right]\ e^{-\gamma {\bf k}^2} = 
 (\beta/2\alpha)\ {\bf k}_i\ K_1({\bf k}^2), 
\label{eq:interm.3}\end{eqnarray}
and require
\begin{eqnarray}
(i)\  K_1({\bf k}^2) = \exp\left(-\frac{1}{6} R_N^2 {\bf k}^2\right) &,& 
(ii)\  K_2({\bf k}^2) = \left(\frac{4}{R_N^2}
 +\frac{\beta^2}{2\alpha^2}{\bf k}^2\right)\ K_1{\bf k}^2)\
 ,\ K_{3,i}({\bf k}^2) = {\bf k}_i\ K_1({\bf k}^2).
\label{eq:interm.4}\end{eqnarray}
These conditions give $N_1=\left(\pi/\alpha\right)^{-3/2}$ and the 
equations
\begin{eqnarray}
&& a)\  \gamma-\frac{\beta^2}{4\alpha} = \frac{1}{6} R_N^2\ ,\
 b)\ \frac{3}{2} \alpha^{-1} = \frac{1}{4} R_N^2\ ,\ 
 c)\ \frac{\beta}{2\alpha} =1.
\label{eq:interm.5}\end{eqnarray}
It follows that $\alpha=(3/8) R_N^2$, $\beta= (3/4) R_N^2$, and
$\gamma = (13/24) R_N^2$, and 
\begin{eqnarray}
 K_1({\bf k}^2) &=& N_1 \int d^3K\ \exp\biggl[-\frac{R_N^2}{6}\biggl\{\frac{9}{4} {\bf K}^2
 -\frac{9}{2} {\bf K}\cdot{\bf k} +\frac{13}{4} {\bf k}^2\biggr\}\biggr].
\label{eq:interm.6}\end{eqnarray}
Using this expression the meson-nucleon vertex (\ref{eq:2.21}) becomes
\footnote{
Comparing with (\ref{eq:2.21}) shows the ${\bf K}$-distribution change
$\delta^{(3)}({\bf K}-{\bf k}) \rightarrow (\pi\epsilon)^{-3/2} \exp\left[
 -\left({\bf K}-{\bf k}\right)^2/\epsilon \right],\ \ 
 \epsilon= 8/(3R_N^2).$  
}
\begin{eqnarray}
\langle p'_{1}|\Gamma|p_{1}\rangle &=& {\cal N}^2\left(\frac{1}{8}\right)^{3/2}
\left(\frac{2\pi}{R_{N}^{2}}\right)^{3/2}\ 
\left(\frac{3R_{N}^{2}}{8\pi}\right)^{3/2}\ 
 \exp\left[-\frac{R_N^2}{6}\left(
{\bf q}^2+{\bf k}^2\right)\right]\ 
\cdot\nonumber\\[0.3cm] && \times
 \int d^3Q\ \exp\left[-\frac{R_{N}^{2}}{6}
 \left\{\frac{9}{4}\left({\bf Q}^2-\frac{4}{3}{\bf q}\cdot{\bf Q}\right) 
\right\}\vphantom{\frac{A}{A}}\right] 
\cdot\nonumber\\[0.3cm] && \times
 \int d^3K\ \exp\left[-\frac{R_{N}^{2}}{6}
 \left\{\frac{9}{4}\left({\bf K}^2-2{\bf k}\cdot{\bf K}+{\bf k}^2\right)
\right\}\vphantom{\frac{A}{A}}\right] 
\cdot\nonumber\\[0.3cm] && \times
 \gamma({\bf Q},{\bf K};{\bf k}, {\bf q})\ 
 \delta^{(3)}\left({\bf p}'_1-{\bf p}_1-{\bf k}\right).                   
\label{eq:interm.7}\end{eqnarray}

\noindent With this result we obtain
\begin{eqnarray}
\Gamma_{CQM} &=& 3 \biggl[1-\frac{{\bf q}^2}{36m_Q^2}+\frac{{\bf k}^2}{16m_Q^2}
 +\frac{i}{36m_Q^2} \sum_i \bm{\sigma}_i\cdot{\bf q}\times{\bf k}
\biggr].
\label{eq:interm.8}\end{eqnarray}
{\it With this method we reproduce the central and spin-orbit term in 
Eqn.~(\ref{eq:3.43}) without the $1/(4m_i^2R_N^2)$ term!}\\




\section{Pseudoscalar-exchange }                              
\label{sec:4}
We determine the $QQ$ ps-exchange amplitude. Below, again i=1 is understood.\\
For the upper vertex in 
Fig.~\ref{fig:3.1}, line '1', we evaluate following spinor matrix-element
\begin{eqnarray}
 \left[\bar{u}_{i}({\bf k'}_{i}) \gamma_5 u_{i}({\bf k}_{i}) 
  \right] &=& 
 \sqrt{\frac{E'_{i}+m'_{i}}{2m'_{i}}\ \frac{E_{i}+m_{i}}{2m_{i}} }\cdot
 \chi^{\prime \dagger}_{i}\cdot \left[ 
 \frac{\mbox{\boldmath $\sigma$}_{i}\cdot{\bf k}_{i}}{E_{i}+m_{i}}
 -\frac{\mbox{\boldmath $\sigma$}_{i}\cdot{\bf k'}_{i}}{E'_{i}+m_{i}}
 \right]\ \chi_i
 \nonumber\\ &\approx& 
 -\chi^{\prime \dagger}_{i}\left[ \frac{\bm{\sigma}_i\cdot{\bf k}}{2m_i}
 \right]\ \chi_i,
\label{eq:4.11} \end{eqnarray}
again because in the CQM $E_i \approx m_i$. 
Summing over the quarks gives
\begin{eqnarray}
 \widetilde{\Gamma}_{CQM,5} &=&
 \sum_{i=1-3}\left[\bar{u}_{i}({\bf k'}_{i}) \gamma_5 u_{i}({\bf k}_{i}) 
  \right] = -3
 \chi^{\prime \dagger}_{N}\left[ \frac{\bm{\sigma}\cdot{\bf k}}{2m_i}
 \right]\ \chi_N
\label{eq:4.12} \end{eqnarray}
It is clear that this vertex is proportional to that for the OBE-coupling
of the pseudoscalar meson.
So, also for $m_i=\sqrt{M'M}/3$, i.e. the so-called "constituent"
quarks, $\bm{\Gamma}_{QQ}$ in equivalent with $\bm{\Gamma}_{NN}$.\\            
From $g_p=(2m_q/m_\pi) f_{pv}$, $g_P= (2M_N/m_\pi) f_{PV}$, and $g_p=g_P/3$ we 
find
\begin{eqnarray*}
 f_{pv} = \frac{1}{3} \frac{M_N}{m_q}\ f_{PV},
\end{eqnarray*}
which for $m_q = M_N/3$ the relation $f_{pv} = f_{PV}$.


\section{Folding Vector-exchange Vertex}            
\label{sec:5}
The coupling of the vector mesons ($ J^{PC}= 1^{--}$) to the
quarks is given by the interaction Hamiltonian 
\begin{eqnarray}
   {\cal H}_{V QQ} &=&
 g_{v} \left(\bar{\psi} \gamma^{\mu} \psi\right)\ V_{\mu} +  
\frac{f_{v}}{4{\cal M}}
 \left(\bar{\psi}\ \sigma^{\mu\nu}\ \psi\right)\ 
\left(\partial_{\mu} V_{\nu} -\partial_{\nu} V_{\mu}\right) 
 \nonumber\\ &=& \left[ (\bar{\psi}\gamma^\mu\psi) F_{1,v}
+\frac{i}{2}\left(\bar{\psi} \stackrel{\leftrightarrow}{\partial^\mu}
\psi\right)\ F_{2,v}\right]\cdot V_\mu, 
\label{eq:5.10} \end{eqnarray}
where $a \stackrel{\leftrightarrow}{\partial^\mu} b =
 a\cdot\partial^\mu b- \partial^\mu a\cdot b$. The relation
between the different coupling constants is
$ F_{1,v}= g_v+\frac{m_Q}{\cal M} f_v\ ,\ F_{2,v}=-\frac{f_v}{\cal M}$, 
and reversely $g_v= F_{1,v}+m_Q F_{2,v}, f_v= -{\cal M}\ F_{2,e}$.

\subsection{Direct-coupling}            
\label{sec:5a}
\noindent {\bf 1}.\ $\Gamma^0_{1,CQM}$-vertex: The QQ-meson vertices are 
\begin{eqnarray}
 \left[\bar{u}_{i}({\bf k'}_{i}) \gamma^0 u_{i}({\bf k}_{i}) 
  \right] &=& 
 \sqrt{\frac{E'_{i}+m'_{i}}{2m'_{i}}\ \frac{E_{i}+m_{i}}{2m_{i}} }\cdot
 \chi^{\prime \dagger}_{i}\cdot 
 \left[ 1 +
 \frac{\mbox{\boldmath $\sigma$}_{i}\cdot{\bf k'}_{i}}{E'_{i}+m_{i}}
 \frac{\mbox{\boldmath $\sigma$}_{i}\cdot{\bf k}_{i}}{E_{i}+m_{i}}\right]
 \nonumber\\ &\approx& 
 \chi^{\prime \dagger}_{i}\left[ 1 +
 \frac{{\bf k}'_i\cdot{\bf k}_i}{4m_i^2}                                        
 +\frac{i}{4m_i^2}\bm{\sigma}_i\cdot{\bf k}'_i\times{\bf k}_i\right]\ \chi_i
 \nonumber\\ &=& 
 \chi^{\prime \dagger}_{i}\left[ 1 +
 \frac{{\bf Q}_i^2-{\bf k}^2}{16m_i^2}                                        
 -\frac{i}{8m_i^2}\bm{\sigma}_i\cdot{\bf Q}_i\times{\bf k}\right]\ \chi_i.
\label{eq:5.11} \end{eqnarray}
Notice that the $1/m_i^2$ terms are the same as for scalar-exchange apart from
the sign. Therefore, from the expression (\ref{eq:3.44}) we now have
\begin{eqnarray}
 \Gamma^0_{1,CQM} &=&
 3\left[ \left(1 +\frac{1}{4m_Q^2R_N^2}\right) +\frac{{\bf q}^2}{4M'M}
 - \frac{9{\bf k}^2}{16M'M}             
 -\frac{i}{4M'M}\sum_i \bm{\sigma}_N\cdot{\bf q}\times{\bf k}\right], 
\label{eq:5.12} \end{eqnarray}

The direct coupling to the nucleons gives
\begin{eqnarray}
 \Gamma^0_{1,NN} &=&
 \left[\bar{u}_N({\bf p'} \gamma^0 u_N({\bf p}) 
  \right] = 
 \sqrt{\frac{E'+M'}{2M'}\ \frac{E+M}{2M} }\cdot
 \chi^{\prime \dagger}_N\cdot 
 \left[ 1 +
 \frac{\mbox{\boldmath $\sigma$}\cdot{\bf p'}}{E'+M'}
 \frac{\mbox{\boldmath $\sigma$}\cdot{\bf p}}{E+M}\right]
 \nonumber\\ &\approx& 
 \sqrt{\frac{E'+M'}{2M'}\ \frac{E+M}{2M} }\cdot
 \chi^{\prime \dagger}_N\left[ 1 +
 \frac{{\bf p}'\cdot{\bf p}}{4M'M}                                        
 +\frac{i}{4M'M}\bm{\sigma}\cdot{\bf p}'\times{\bf p}\right]\ \chi_N
 \nonumber\\ &=& 
 \sqrt{\frac{E'+M'}{2M'}\ \frac{E+M}{2M} }\cdot
 \chi^{\prime \dagger}_N\left[ 1 + \frac{{\bf q}^2-{\bf k}^2/4}{4M'M}       
 -\frac{i}{4M'M}\bm{\sigma}\cdot{\bf q}\times{\bf k}\right]\ \chi_N.
 \nonumber\\ &\Rightarrow& 
 \left[ 1 + \frac{{\bf q}^2-\bm{\Delta}^2/4}{4M'M}       
 -\frac{i}{4M'M}\bm{\sigma}\cdot{\bf q}\times{\bf k}\right].
\label{eq:5.17} \end{eqnarray}
To bring $\Gamma^0_{1,CQM}$ and $\Gamma^0_{1,NN}$ in agreement the following:
\begin{enumerate}
\item[a.] The factor 3 is accounted for by scaling
the quark-meson coupling, {\it i.e. } $g^{(V)}_Q= g^{(V)}/3$.
\item[b.] The term $ 1/(4m_Q^2/R_N^2)=9/(4M'MR_N^2) \approx 0.1$ for 
$R_N \approx 1$ fm, giving a $10\%$ amplified of the central term.. 
\item[c.] Compared to $\Gamma^0_{NN}$ the quark vertex $\Gamma^0_{QQ}$ has
an extra $-8{\bf k}^2/(16M'M)$-term. This term can be canceled by introducing
an extra QQV-interaction, similar to (\ref{eq:3a.1}),
\begin{equation}
 \Delta{\cal H}_V^{1)} =  f'_{1,v} 
 \left[\Box(\bar{\psi}\gamma^\mu\psi)(2\mu^2) \right]\ V_\mu, 
\label{eq:5.10a} \end{equation}
and determine for $\mu=0$ the coupling from the condition
\begin{equation}
 f'_{1,v}\frac{9}{8M'M} = F_{1,v}\frac{8}{16M'M},\ f'_{1,v}/F_{1,v}=4/9.
\label{eq:5.18} \end{equation}
\end{enumerate}
Terms for $\mu=m$ are of order $\sim 1/M^3$ which we neglect.
With these remarks it is shown that, as in the scalar case, the QQ- and NN-vertex
are (approximately) equivalent as far as the NN-potential is concerned.
It also shows that the combination of scalar
and vector exchange are necessary to bring this about. With the inclusion of the
$g'_2$-contribution, the QQV-vertex becomes
\begin{eqnarray}
 \Gamma^0_{1,CQM} &=&
 3\left[ 1 +\left(\frac{1}{4m_Q^2R_N^2}+\frac{{\bf q}^2-{\bf k}^2/4}{4M'M}\right)
 -\frac{i}{4M'M}\sum_i \bm{\sigma}_N\cdot{\bf q}\times{\bf k}\right], 
\label{eq:5.19} \end{eqnarray}
So, the ${\bf q}^2$-, the ${\bf k}^2$-, and spin-orbit
term are the same as for the coupling of the vector meson on the nucleon
level. {\it The central term in $\Gamma^{(0)}_{QQ}$ has an extra 
 $9/[4M'M R_N^2]$-term, which is a slight violation of the Idea/conjecture
as formulated in the Introduction, similar to the scalar-meson case.
As demonstrated in subsection~\ref{sec:interm} such "spurious" terms can be 
eliminated by introducing a ${\bf K}$-distribution. Henceforth, we omit such terms.
}\\


\noindent {\bf 2}.\ $\bm{\Gamma}_{1,QQ}$-vertex: The QQ-meson vertices are 
\begin{eqnarray}
 \left[\bar{u}_{i}({\bf k'}_{i}) \bm{\gamma}_i u_{i}({\bf k}_{i}) 
  \right] &=& 
 \sqrt{\frac{E'_{i}+m'_{i}}{2m'_{i}}\ \frac{E_{i}+m_{i}}{2m_{i}} }\cdot
 \chi^{\prime \dagger}_{i}\cdot 
 \left[\vphantom{\frac{A}{A}}
 \frac{\bm{\sigma}_i \bm{\sigma}_{i}\cdot{\bf k}_{i}}{E_{i}+m_{i}}
 + \frac{\bm{\sigma}_{i}\cdot{\bf k}'_i\ \bm{\sigma}_i}{E'_{i}+m_{i}}
 \right]\ \chi_i
 \nonumber\\ &\approx& 
 \chi^{\prime \dagger}_{i}\cdot 
 \left[\vphantom{\frac{A}{A}}
 \frac{{\bf Q}_i}{2m_i}+\frac{i}{2m_i}\left(\bm{\sigma}_i\times{\bf k}\right)
 \right]\ \chi_i
 \nonumber\\ &\Rightarrow&
 \chi^{\prime \dagger}_{i}\cdot 
 \left[\vphantom{\frac{A}{A}}
 \frac{{\bf q}}{3m_i}+\frac{i}{2m_i}\left(\bm{\sigma}_i\times{\bf k}\right)
 \right]\ \chi_i
\label{eq:5.21} \end{eqnarray}
Summing over the quarks leads to
\begin{subequations}\label{eq:5.22} 
\begin{eqnarray}
 \bm{\Gamma}_{1,CQM} &=&
 \sum_{i=1-3}\left[\bar{u}_{i}({\bf k'}_{i}) \bm{\gamma}_i u_{i}(\bar{\bf k}_{i}) 
  \right] =  \chi^{\prime \dagger}_{i}\cdot 
 \left[\vphantom{\frac{A}{A}}
 \frac{{\bf q}}{m_i}+\frac{i}{2m_i}\left(\bm{\sigma}_N\times{\bf k}\right)
 \right]\ \chi_i  \\ &\Rightarrow& 
 3 \left[\vphantom{\frac{A}{A}}
 \frac{{\bf q}}{M}+\frac{i}{2M}\left(\bm{\sigma}_N\times{\bf k}\right) \right].  
 \end{eqnarray}\end{subequations}
The direct coupling to the nucleons gives
\begin{eqnarray}
 \bm{\Gamma}_{1,NN} &=&
 \left[\bar{u}_{N}({\bf p'}) \bm{\gamma} u_{N}({\bf p}) 
  \right] = 
 \sqrt{\frac{E'+M'}{2M'}\ \frac{E+M}{2M} }\cdot
 \chi^{\prime \dagger}_{N}\cdot \left[\vphantom{\frac{A}{A}}
 \frac{\bm{\sigma} \bm{\sigma}\cdot{\bf p}}{E+M}
 + \frac{\bm{\sigma}\cdot{\bf p}'\ \bm{\sigma}}{E'+M'}
 \right]\ \chi_N
 \nonumber\\ &\approx& 
 \sqrt{\frac{E'+M'}{2M'}\ \frac{E+M}{2M} }\cdot
 \chi^{\prime \dagger}_{N}\cdot 
 \left[\vphantom{\frac{A}{A}}
 \frac{{\bf q}}{M}
 +\frac{i}{2M}\left(\bm{\sigma}\times{\bf k}\right)
 \right]\ \chi_N
 \nonumber\\ &\Rightarrow& 
 \chi^{\prime \dagger}_{N}\cdot 
 \left[\vphantom{\frac{A}{A}}
 \frac{{\bf q}}{M}
 +\frac{i}{2M}\left(\bm{\sigma}\times{\bf k}\right)
 \right]\ \chi_N
\label{eq:5.23} \end{eqnarray}
Again, we see that for $m_i=\sqrt{M'M}/3$, i.e. the so-called "constituent"
quarks, $\bm{\Gamma}_{QQ}$ matches with $\bm{\Gamma}_{NN}$.\\             

\subsection{Derivative-coupling via Gordon-decomposition}            
\label{sec:5c}
It remains to established the relation of the vertices
\begin{equation}
 \Gamma^\mu_{2,NN} = (p^\prime+p)^\mu\ \left[\bar{u}(p^\prime u(p\right]\ \
{\rm and}\ \ 
\widetilde{\Gamma}^\mu_{2,QQ} = \sum_{i=1-3} (k^{\prime \mu}+k^\mu)
 \left[\bar{u}({\bf k}^\prime_i) u({\bf k}_i)\right].
\label{eq:5.42} \end{equation}
 
\noindent {\bf 1}.\ $\Gamma^0_2$-vertex: From the analysis of the scalar
coupling, see Eqns.~(\ref{eq:3.41})-(\ref{eq:3.44}), the QQ-meson vertices are 
\begin{eqnarray}
 \sum_{i=1-3} (k^\prime_{i,0}+k_{i,0})\ 
 \left[\bar{u}_{i}({\bf k'}_{i})\ u_{i}({\bf k}_{i}) \right] &\Rightarrow& 
 6m_Q\ \left[ 1 -\left(\frac{1}{4m_Q^2R_N^2}+\frac{{\bf q}^2}{36m_Q^2}\right)
 + \frac{{\bf k}^2}{16m_Q^2}             
 +\frac{i}{36m_Q^2}\sum_i \bm{\sigma}_i\cdot{\bf q}\times{\bf k}\right]
\label{eq:5.43} \end{eqnarray}
The CQM replacement $m_Q \approx \sqrt{M'M}/3, (M'+M)/6$ leads to
\begin{eqnarray}
 \Gamma^0_{2,CQM} &\approx& 
 (M'+M)\left[ 1 -\left(\frac{1}{4m_Q^2R_N^2}+\frac{{\bf q}^2}{4M'M}\right)
 + \frac{9{\bf k}^2}{16M'M}             
 +\frac{i}{4M'M}\sum_i \bm{\sigma}_N\cdot{\bf q}\times{\bf k}\right], 
\label{eq:5.44} \end{eqnarray}
This vertex has to be compared with that at the NN-level:
\begin{eqnarray}
 \Gamma^0_{2,NN} &\approx& (M'+M) 
\left[1- \frac{{\bf q}^2}{4M'M} +\frac{{\bf k}^2}{16M'M}
 +\frac{i}{4M'M} {\bf q}\times {\bf k}\cdot\bm{\sigma}
 \right].
\label{eq:5.45} \end{eqnarray}
So, the situation is again similar to the 
scalar case. The remedy to obtain agreement for the ${\bf k}^2$-term
is the as in that case by introducing a zero in the coupling, or
by adding an extra QQV-interaction, similar to (\ref{eq:3a.1}) and
(\ref{eq:5.10a}),
\begin{equation}
 \Delta{\cal H}_V^{2)} =  f'_{2,v} \left[\Box(i\bar{\psi}
\stackrel{\leftrightarrow}{\partial_\mu}\psi)(2\mu^2)\right]\ V^\mu, 
\label{eq:5.10b} \end{equation}
and determine for $\mu=0$ the coupling from the condition 
\begin{equation}
 f'_{2,v}\frac{9}{8M'M} = F_{2,v}\frac{8}{16M'M},\ f'_{2,v}/F_{2,v}=4/9.
\label{eq:5.18b} \end{equation}
Terms for $\mu=m$ are of order $\sim 1/M^3$ which again we neglect.\\

\noindent {\bf 2}.\ $\bm{\Gamma}_2$-vertex: For this term we neglect the
$1/m_Q^2 \sim 1/M^\prime M$-terms as in the NN-potential derivation,
and therefore we get
\begin{eqnarray}
 \bm{\Gamma}_{2,CQM} &=& 
 \sum_{i=1-3}({\bf k}^\prime_i+{\bf k}_i)\ 
 \left[\bar{u}_{i}({\bf k'}_{i})\ u_{i}({\bf k}_{i}) \right] 
  \nonumber\\ &\Rightarrow& 2 \sum_i {\bf Q}_i 
 \left[\bar{u}_{i}({\bf k'}_{i})\ u_{i}({\bf k}_{i}) \right] 
  \Rightarrow 2{\bf q}, 
\label{eq:5.46} \end{eqnarray}
showing that without scaling, as in the case of $\Gamma^0_{2,QQ}$, the
NN-vertex is produced.\\

\noindent So, with the results of the scalar and vector 
couplings, i.e. $\Gamma = 1, \gamma^\mu$, utilizing the Gordon-decomposition,
the relation between QQM- and NNM-derivative couplings is most
easily demonstrated.

\subsection{Full quark-vector coupling}                              
\label{sec:5d}
At the quark-level the additional interaction is
\begin{equation}
 {\cal H}_V^{(2)} = -h_v\left[\frac{\Box}{4m_Q^2}\left(i\bar{q}(x)
 \stackrel{\leftrightarrow}{\partial}_\mu q(x)\right)\right]\cdot \phi_V^\mu.
\label{eq:5.51} \end{equation}
Since adaption is necessary in the direct and derivative term, 
we get for the full correction $h_v=g'_v+f'_v= (4/3)(g_V+f_V)/{\cal M}$,
with ${\cal M}= m_Q/3$. Here, the $({\bf p}'+{\bf p})$ term in
(\ref{eq:5.51}) is 

\noindent Since the adaption is in the direct and derivative term, 
we get for the full vector vertex
\begin{eqnarray}
\bar{u}(p^\prime) \Gamma_v^\mu\ u(p) &=& \bar{u}(p^\prime)\left[
 G_{m,v} \gamma^\mu + \frac{1}{{\cal M}} G_{e,v}\ (p^\prime+p)^\mu\right]\ u(p),
\label{eq:5.52} \end{eqnarray}
with 
\begin{eqnarray}
 G_{m,v} &=& g_v + f_v\ ,\ G_{e,v} = 
 -f_v\left[1 + \frac{\kappa^\prime}{\kappa}\frac{k^2}{8m_Q^2}\right],
\label{eq:5.53} \end{eqnarray}
where $f_v = \kappa_v\ g_v$. Now $k^2 \approx -{\bf k}^2$ so that $G_{e,v}$ 
exhibits a {\it zero} at ${\bf k}^2 = 8m_Q^2 (\kappa_v/\kappa_v^\prime)$.


\section{Folding Axial-vector-exchange Vertex}                              
\label{sec:6}
The coupling of the axial-vector mesons ($ J^{PC}= 1^{++}$, 1$^{st}$ kind) to the
quarks is given by the interaction Hamiltonian 
\begin{equation}
 {\cal H}_A = g_A[\bar{\psi}\gamma_\mu\gamma_5\psi] \phi^\mu_A + \frac{if_A}{{\cal M}}
 [\bar{\psi}\gamma_5\psi]\ \partial_\mu\phi_A^\mu.
\label{eq:6.0} \end{equation}

\noindent {\bf 1}.\ $\Gamma_5^0$-vertex: The QQ-meson vertices are 
\begin{eqnarray}
 \left[\bar{u}_{i}({\bf k'}_{i}) \gamma^0_i\gamma_5 u_{i}({\bf k}_{i}) 
  \right] &=& 
 \sqrt{\frac{E'_{i}+m'_{i}}{2m'_{i}}\ \frac{E_{i}+m_{i}}{2m_{i}} }\cdot
 \chi^{\prime \dagger}_{i}\cdot \left[ 
 \frac{\mbox{\boldmath $\sigma$}_{i}\cdot{\bf k}_{i}}{E_{i}+m_{i}}
 +\frac{\mbox{\boldmath $\sigma$}_{i}\cdot{\bf k'}_{i}}{E'_{i}+m_{i}}
 \right]\ \chi_i
 \nonumber\\ &\approx& 
 \chi^{\prime \dagger}_{i}\left[ \frac{\bm{\sigma}_i\cdot{\bf Q}_i}{2m_i}
 \right]\ \chi_i \Rightarrow
 \chi^{\prime \dagger}_{i}\left[ \frac{\bm{\sigma}_i\cdot{\bf q}}{3m_i}
 \right]\ \chi_i
\label{eq:6.1} \end{eqnarray}

Summing over the quarks gives
\begin{eqnarray}
 \Gamma_{5,CQM}^0 &=&
 \sum_{i=1-3}\left[\bar{u}_{i}({\bf k'}_{i}) \gamma^0_i\gamma_5 
 u_{i}({\bf k}_{i}) \right] = 
 \chi^{\prime \dagger}_{N}\left[ \frac{\bm{\sigma}_N\cdot{\bf q}}{3m_i}
 \right]\ \chi_N
 \Rightarrow \left[ \frac{\bm{\sigma}_N\cdot{\bf q}}{\sqrt{M'M}}\right].
\label{eq:6.2} \end{eqnarray}
It is clear that this vertex is proportional to that for the OBE-coupling
of the axial-vector meson.\\

\noindent {\bf 2}.\ $\bm{\Gamma}_5$-vertex: The QQ-meson vertices are 
\begin{eqnarray}
 &&\left[\bar{u}_{i}({\bf k'}_{i}) \bm{\gamma}_i\gamma_5 u_{i}({\bf k}_{i}) 
  \right] = 
 \sqrt{\frac{E'_{i}+m'_{i}}{2m'_{i}}\ \frac{E_{i}+m_{i}}{2m_{i}} }\cdot
 \chi^{\prime \dagger}_{i}\cdot 
 \left[ \bm{\sigma}_i + \frac{\bm{\sigma}_{i}\cdot{\bf k'}_{i}\ 
 \bm{\sigma}_i\ \bm{\sigma}_i\cdot{\bf k}_i}
{(E'_{i}+m_{i})(E_{i}+m_{i})}\right]\ \chi_i
 \nonumber\\ && \hspace{1cm} \approx 
 \chi^{\prime \dagger}_{i}\left[ \bm{\sigma}_i+\frac{1}{4m_i^2}\left\{
 \vphantom{\frac{A}{A}} {\bf k}'_i (\bm{\sigma}_i\cdot{\bf k}_i)
 +{\bf k}_i (\bm{\sigma}_i\cdot{\bf k}'_i) 
 -({\bf k}'_i\cdot{\bf k}_i) \bm{\sigma}_i
 -i({\bf k}'_i\times{\bf k}_i)\right\} \right]\ \chi_i
 \nonumber\\ && \hspace{1cm} =       
 \chi^{\prime \dagger}_{i}\left[ \bm{\sigma}_i+\frac{1}{16m_i^2}\left\{
 \vphantom{\frac{A}{A}} 2{\bf Q}_i (\bm{\sigma}_i\cdot{\bf Q}_i)
 -2{\bf k} (\bm{\sigma}_i\cdot{\bf k}) 
 -({\bf Q}_i^2-{\bf k}^2) \bm{\sigma}_i
 +2i({\bf Q}_i\times{\bf k})\right\} \right]\ \chi_i
 \nonumber\\ && \hspace{1cm} \Rightarrow
 \chi^{\prime \dagger}_{i}\left[ \bm{\sigma}_i+
 \frac{1}{2m_i^2}\frac{1}{3R_N^2}\ \bm{\sigma}_i
 -\frac{1}{4m_i^2}\frac{1}{R_N^2}\bm{\sigma}_i \right.\nonumber\\ && \left.
 \hspace{1.2cm} +\frac{1}{16m_i^2}\left\{
 \vphantom{\frac{A}{A}} \frac{8}{9}{\bf q} (\bm{\sigma}_i\cdot{\bf q})
 -2{\bf k} (\bm{\sigma}_i\cdot{\bf k}) 
 -(\frac{4}{9}{\bf q}^2-{\bf k}^2) \bm{\sigma}_i
 +\frac{4i}{3}({\bf q}\times{\bf k})\right\} \right]\ \chi_i\nonumber\\
\label{eq:6.11} \end{eqnarray}
Summing over the quarks gives
\begin{eqnarray}
 \bm{\Gamma}_{5,CQM} &=&
 \sum_{i=1,3}\left[\bar{u}_{i}({\bf k'}_{i}) \bm{\gamma}_i\gamma_5 
 u_{i}({\bf k}_{i}) \right] = 
 \chi^{\prime \dagger}_{N}\left[ 
 \left(1-\frac{1}{12(m_iR_N)^2}\right) \bm{\sigma}
 \right.\nonumber\\ && \left.\hspace{1.2cm} +\frac{1}{16m_i^2}\left\{
 \vphantom{\frac{A}{A}} \frac{8}{9}{\bf q} (\bm{\sigma}\cdot{\bf q})
 -2{\bf k} (\bm{\sigma}\cdot{\bf k}) 
 -(\frac{4}{9}{\bf q}^2-{\bf k}^2)\ \bm{\sigma}
 +4i({\bf q}\times{\bf k})\right\} \right]\ \chi_N.
\label{eq:6.12} \end{eqnarray}


The direct coupling to the nucleons gives
\begin{eqnarray}
&& \bm{\Gamma}_{5,NN} =
 \left[\bar{u}_{N}({\bf p'}) \bm{\gamma}\gamma_5 u_{N}({\bf p}) 
  \right] = 
 \sqrt{\frac{E'+M'}{2M'}\ \frac{E+M}{2M} }\cdot
 \chi^{\prime \dagger}_{N}\ \left[\vphantom{\frac{A}{A}}
 \bm{\sigma} + \frac{(\bm{\sigma}\cdot{\bf p'})\ 
 \bm{\sigma}\ (\bm{\sigma}\cdot{\bf p})}
{(E'+M')(E+M)}\right]\ \chi_N
 \nonumber\\ && \hspace{5mm} \approx 
 \sqrt{\frac{E'+M'}{2M'}\ \frac{E+M}{2M} }\cdot
 \chi^{\prime \dagger}_{N}\left[ \bm{\sigma}+\frac{1}{4M'M}\left\{
 \vphantom{\frac{A}{A}} {\bf p}' (\bm{\sigma}\cdot{\bf p})
 +{\bf p} (\bm{\sigma}\cdot{\bf p}') 
 -({\bf p}'\cdot{\bf p}) \bm{\sigma}
 -i({\bf p}'\times{\bf p})\right\} \right]\ \chi_N
 \nonumber\\ && \hspace{5mm} \Rightarrow 
 \left[ \bm{\sigma}+\frac{1}{4M'M}\left\{
 \vphantom{\frac{A}{A}} 2{\bf q} (\bm{\sigma}\cdot{\bf q})
 -\frac{1}{2}{\bf k} (\bm{\sigma}\cdot{\bf k}) 
 -\left({\bf q}^2-{\bf k}^2/4\right)\ \bm{\sigma}
 +i({\bf q}\times{\bf k})\right\} \right].        
\label{eq:6.16} \end{eqnarray}
Similarly to the scalar- and vector-meson, the last expression is to
be compared to $\bm{\Gamma}_{5,CQM}$ in (\ref{eq:6.12}).\\
For "constituent" quarks with $m_i=M/3$ the result (\ref{eq:6.12}) reads
\begin{eqnarray}
 \bm{\Gamma}_{5,CQM} &\Rightarrow&
 \chi^{\prime \dagger}_{N}\left[ 
 \left(1-\frac{3}{4(MR_N)^2}\right) \bm{\sigma}
 \right.\nonumber\\ && \left.\hspace{1.2cm} +\frac{1}{4M'M}\left\{
 \vphantom{\frac{A}{A}} 2 {\bf q} (\bm{\sigma}\cdot{\bf q})
 -\frac{9}{2}{\bf k} (\bm{\sigma}\cdot{\bf k}) 
 -({\bf q}^2-\frac{9}{4}{\bf k}^2)\ \bm{\sigma}
 +9i({\bf q}\times{\bf k})\right\} \right]\ \chi_N
 \nonumber\\ &=&                        
 \chi^{\prime \dagger}_{N}\left[ 
 \left(1-\frac{3}{4(MR_N)^2} +\frac{{\bf k}^2}{2M'M} \right) \bm{\sigma} 
 -\frac{{\bf k}(\bm{\sigma}\cdot{\bf k})}{M'M}
 \right.\nonumber\\ && \left.\hspace{1.2cm} +\frac{1}{4M'M}\left\{
 \vphantom{\frac{A}{A}} 2 {\bf q} (\bm{\sigma}\cdot{\bf q})
 -\frac{1}{2}{\bf k} (\bm{\sigma}\cdot{\bf k}) 
 -({\bf q}^2-{\bf k}^2/4)\ \bm{\sigma}
 +9i({\bf q}\times{\bf k})\right\} \right]\ \chi_N
\label{eq:6.17} \end{eqnarray}

\noindent {\bf 3}.\ $\bm{\Gamma}_5$-vertex(continued A): 
Next, we impose for the quarks the conservation of the axial current. The current is
\begin{equation}
 J^a_\mu = g_a \bar{\psi} \gamma_\mu \gamma_5\psi +\frac{i f_a}{\cal M}
 \partial_\mu(\bar{\psi}\gamma_5\psi),
\label{eq:6.21} \end{equation}
and $\partial\cdot J^A=0$ imposes the relation
\begin{equation}
 f_a = \left(\frac{m_{A_1}^2}{2m_Q {\cal M}}\right)^{-1} g_a.
\label{eq:6.22} \end{equation}
Taking $m_{A_1} = \sqrt{2} m_\rho \approx 2\sqrt{2} m_Q$ the 
axial current becomes
\begin{equation}
 J^a_\mu = g_a \left[\bar{\psi} \gamma_\mu \gamma_5\psi +\frac{i}{4m_Q}
 \partial_\mu(\bar{\psi}\gamma_5\psi)\right].
\label{eq:6.23} \end{equation}
The $f_a$-contributions to the axial-vertex are
\begin{subequations}
\begin{eqnarray}
 \mu=0 &:& \sim (E'-E) \sim (M'M m_Q)^{-1} \approx 0, \\
 \mu=i &:& -\frac{1}{4m_Q}\ {\bf k}\ 
 \left[\bar{u}({\bf k}'_i)\gamma_5 u({\bf k}_i)\right]
 \Rightarrow +
 \sqrt{\frac{E'_{i}+m'_{i}}{2m'_{i}}\ \frac{E_{i}+m_{i}}{2m_{i}} }\cdot
 \chi^{\prime \dagger}_{i}\left[\frac{1}{8m_i^2}\ {\bf k}\
 (\bm{\sigma}_i\cdot{\bf k})\right] \chi_i.
\label{eq:6.24} \end{eqnarray}
\end{subequations}
\noindent Taking this $f_a$-contributions into account we obtain
for "constituent" quarks:
\begin{subequations}
\begin{eqnarray}
&& \bm{\Gamma}_{5,NN} \Rightarrow
 \chi^{\prime \dagger}_{N}\left[\  \vphantom{\frac{A}{A}} \bm{\sigma} +
 \frac{1}{4M'M}\left\{
 \vphantom{\frac{A}{A}} 2{\bf q} (\bm{\sigma}\cdot{\bf q})
 -\left({\bf q}^2-{\bf k}^2/4\right)\ \bm{\sigma}
 +i({\bf q}\times{\bf k})\right\} \right]\ \chi_N, \\
&& \bm{\Gamma}_{5,CQM} \Rightarrow
 \chi^{\prime \dagger}_{N}\left[ 
 \left(1-\frac{3}{4(MR_N)^2}+\frac{{\bf k}^2}{2M'M}\right) \bm{\sigma}
 \right.\nonumber\\ && \left.\hspace{1.2cm} +\frac{1}{4M'M}\left\{
 \vphantom{\frac{A}{A}} 2 {\bf q} (\bm{\sigma}\cdot{\bf q})
 -({\bf q}^2-{\bf k}^2/4)\ \bm{\sigma}
 +9i({\bf q}\times{\bf k})\right\} \right]\ \chi_N
\label{eq:6.27} \end{eqnarray}
\end{subequations}
Here, we omitted the factor $\sqrt{(E'+M')(E+M)/4M'M}$ for the same reason 
as for the scalar- and vector-meson.

\noindent {\bf Remark} $\bm{\Gamma}_{5,CQM}$: (i) for $R_N \approx 1 fm$ the term 
$3/4(M R_N)^2 \approx 3/100 \ll 1$ 
and may be neglected, (ii) the ${\bf k}^2/(2M'M)$ term can
be removed by taking into account the zero in the vertices (see above), 
and (iii) the ${\bf k}(\bm{\sigma}\cdot{\bf k})/M'M$-term has been removed by 
adding an $f_a$-coupling
at the quark-level in a way compatible with axial-current conservation. \\
The change in the zero is as follows: we write the zero in the form
\begin{eqnarray*}
&&\left(1-{\bf k}^2/U^2\right)\left(1+{\bf k}^2/2M_N^2\right) \approx
1-{\bf k}^2/\bar{U}^2,\ \bar{U} = U/\sqrt{1-U^2/2M_N^2}. 
\end{eqnarray*}

\noindent {\bf So, there (only) remains the problem with the spin-orbit terms!}
For the solution see the next paragraphs.  \\


\noindent {\bf 4}.\ $\bm{\Gamma}_5$-vertex(continued B): 
We note that $\bm{\Gamma} = \sum_{i=1}^3\bar{u}_i \bm{\gamma}_i\gamma_5 u_i
 ~ \langle \bar{u}_N \bm{\Sigma}_N u_N\rangle$ for non-relativistic
quarks, i.e. it measures the contribution of the quarks to the
nucleon spin. In the parton model it appeared that a large portion of the 
nucleon spin has to come from gluonic and quark orbital angular momentum 
contributions \cite{Lead96}. In the ESC-model
we ascribe the meson-couplings to the quark-antiquark pair creation process.
To account for a modification for the axial-vector mesons we consider the
following additional phenomenological interaction at the quark level \cite{CC1}
\begin{eqnarray}
 \Delta{\cal H} &=& \frac{ig_a'}{{\cal M}^2}\ \epsilon^{\mu\nu\alpha\beta}\
 (\partial_\alpha\bar{\psi}) \gamma_\nu (\partial_\beta\psi)\ A_\mu
\nonumber\\ &\Rightarrow& \Delta\Gamma_5^\mu =
 \frac{ig_a'}{{\cal M}^2}\ \left[\bar{u}(p') \gamma_\nu u(p)\right]\ 
 \epsilon^{\mu\nu\alpha\beta} p'_\alpha p_\beta.
\label{eq:6.31} \end{eqnarray}
Now, we assume that ${\cal M} \sim M_N$. Then, if $\nu=n=1,2,3$ the vertex
is $\propto 1/M^3 \approx 0$. So, the only important contribution is given
for $\nu=0$. In this case, summing over the (valence) quarks,
\begin{eqnarray}
\Delta\bm{\Gamma}_{5,CQM}^\mu &=& \sum_{i=1}^3\Delta\bm{\Gamma}_{5,i}^\mu =
 -\frac{ig_a'}{{\cal M}^2}\ 
\sum_{i=1}^3\left[\bar{u}(k_i') \gamma_{i,0} u(k_i)\right]\ 
 ({\bf Q}_i\times{\bf k}) + O(1/M^3) \nonumber\\ &\Rightarrow &
 -\frac{2ig_a'}{M' M }\ 
 \sqrt{\frac{E'+M'}{2M'}\ \frac{E+M}{2M} }\cdot
 \left[\chi_N^{\prime \dagger} \chi_N\right]\ 
 ({\bf q}\times{\bf k}).                                       
\label{eq:6.32} \end{eqnarray}
By choosing $g'_a = g_a$, where $g_a$ is the axial coupling constant at
the quark level, the axial-vertex becomes   
$\bm{\Gamma}_{5,CQM} \sim \bm{\Gamma}_{5,NN}$.\\

\noindent {\bf 5}.\ {\bf Orbital Angular Momentum interpretation}:
In the parton model it appeared that a large portion of the 
nucleon spin comes from orbital quark motion and gluonic 
contributions \cite{Lead96}. The orbital angular momentum of the quarks
is present for the non-forward matrix element, i.e. ${\bf p} \neq {\bf p}'$.
Therefore we consider the 
following form of the additional interaction at the quark level \cite{CC2}
\begin{eqnarray}
 \Delta{\cal H}' &=& g_a^{\prime\prime}\ 
\epsilon^{\mu\nu\alpha\beta}\
 \left[\bar{\psi}(x) {\cal L}_{\nu\alpha\beta}\psi(x)\right]\ A_\mu,
\label{eq:6.51} \end{eqnarray}
where \cite{BD65}
\begin{equation}
 {\cal L}_{\nu\alpha\beta} = 
 i\gamma_\nu\left(x_\alpha\frac{\partial}{\partial x^\beta}
-x_\beta\frac{\partial}{\partial x^\alpha}\right)
\label{eq:6.52} \end{equation}
is the orbital part of ${\cal M}_{\nu\alpha\beta}$, the angular momentum 
density operator. The vertex for the NNA$_1$-coupling is given by
\begin{eqnarray}
 \langle p',s'|\Delta H'| p,s;k,\rho\rangle &=& \int d^4x 
 \langle p',s'|\Delta {\cal H}'| p,s;k,\rho\rangle \sim  
 \varepsilon_\mu(k,\rho)\ \epsilon^{\mu\nu\alpha\beta}\cdot\nonumber\\ 
 && \times \int d^4x\ e^{-ik\cdot x}\ \langle p',s'|i\bar{\psi}(x)
 \gamma_\nu \left(x_\alpha\nabla_\beta -x_\beta\nabla_\alpha\right)
 \psi(x)| p,s\rangle  
\label{eq:6.53} \end{eqnarray}
As pointed out in the previous paragraph the dominant contribution comes from 
$\nu=0$. For this we have to evaluate the integral
\begin{equation}
 J_{ab} = i \int d^4x\ e^{-ik\cdot x}\ \langle p',s'|i\psi^\dagger(x)
 \left(x_a\nabla_b -x_b\nabla_a\right) \psi(x)| p,s\rangle  
\label{eq:6.54} \end{equation}
Since we have only quarks, focusing on quark i=1, the quark field operator is 
\begin{equation}
 \psi_i(x) \Rightarrow  \sum_s\int\frac{d^3k_i}{(2\pi)^{3/2}}
 \sqrt{\frac{m_Q}{{\cal E}(k_i)}}\ b(k_i,s_i)\ u(k_i,s_i)\ 
 e^{-ik_i\cdot x}\
e^{-\alpha({\bf k}_i^2-{\bf k}_i\cdot{\bf p}/2)}
\label{eq:6.55} \end{equation}
where $\alpha = 2R_N^2/3$. Using this in (\ref{eq:6.54}) we get 
\begin{eqnarray}
 J_{ab} &=& \left[u^\dagger(k'_i,s')\ u(k_i,s)\right]\ 
 \int d^4x\ e^{i(k'_i-k_i-k)\cdot x}\ 
 \left(x_a k_{i,b} -x_b k_{i,a}\right) 
 e^{-\alpha({\bf k}_i^{\prime 2}+{\bf k}_i^2)}
 e^{\alpha({\bf k}'_i\cdot{\bf p}'+{\bf k}_i\cdot{\bf p})/2}
\nonumber\\ 
 &=& \left[u^\dagger(k'_i,s')\ u(k_i,s)\right]\ 
 \int d^4x\ e^{i(k'_i-k_i-k)\cdot x}\ 
 \left(x_a k_{i,b} -x_b k_{i,a}\right) 
 e^{-\alpha({\bf Q}^2-2{\bf q}\cdot{\bf Q})/2}
\nonumber\\ 
 &=& (2\pi)\delta(E'-E-k^0)\left[u^\dagger(k'_i,s')\ u(k_i,s)\right]\ 
 \int d^3x\ e^{-i({\bf p}'-{\bf p}-{\bf k})\cdot {\bf x}}\ 
 \left(x_a k_{i,b} -x_b k_{i,a}\right) 
 e^{-\alpha({\bf Q}^2-2{\bf q}\cdot{\bf Q})/2}
\nonumber\\ 
 &=& -(2\pi i)\delta(E'-E-k^0)\left[u^\dagger(k'_i,s')\ u(k_i,s)\right]\ 
 \int d^3x\ 
 \left[\left(\nabla_{p,a} k_{i,b} -\nabla_{p,b} k_{i,a}\right) 
e^{-i({\bf p}'-{\bf p}-{\bf k})\cdot {\bf x}}\right]\
 e^{-\alpha({\bf Q}^2-2{\bf q}\cdot{\bf Q})/2}
\nonumber\\ &=& 
 +(2\pi)^4 i\delta(E'-E-k^0)\delta^{(3)}({\bf p}'-{\bf p}-{\bf k})\
 \cdot(\alpha/2)\
 \left[u^\dagger(k'_i,s')\ u(k_i,s)\right]\cdot\nonumber\\ && \times
 \left(Q_a k_{i,b} -Q_b k_{i,a}\right)\
 e^{-\alpha({\bf Q}^2-2{\bf q}\cdot{\bf Q})/2}
\nonumber\\ &\Rightarrow& 
 +(2\pi)^4 i\delta(E'-E-k^0)\delta^{(3)}({\bf p}'-{\bf p}-{\bf k})\
 \cdot(\alpha/3)\
 \left[u^\dagger(k'_i,s')\ u(k_i,s)\right]\cdot\nonumber\\ && \times
 \left(q_a k_{i,b} -q_b k_{i,a}\right)\
 e^{-\alpha({\bf q}^2-2{\bf q}\cdot{\bf Q})/2}
\label{eq:6.56} \end{eqnarray}
Substitution in (\ref{eq:6.53}) gives
\begin{eqnarray}
 \langle p',s'|\Delta H'| p,s;k,\rho\rangle &\approx&  
 +(2\pi)^4 i\delta(E'-E-k^0)\delta^{(3)}({\bf p}'-{\bf p}-{\bf k})\
 g_a^{\prime\prime}(\alpha/3)\ \varepsilon_m(k,\rho)\cdot\nonumber\\ && \times
 \left[u^\dagger(k'_i,s')\ u(k_i,s)\right]\
 \epsilon_{mab} \left(q_a k_{i,b} -q_b k_{i,a}\right)\
 e^{-\alpha({\bf q}^2-2{\bf q}\cdot{\bf Q})/2}
\nonumber\\ &\Rightarrow& 
 +(2\pi)^4 i\delta^{(4)}(p'-p-k)\
 g_a^{\prime\prime}(2\alpha/3)\ \varepsilon_m(k,\rho)\cdot\nonumber\\ 
&& \times \left[u^\dagger(k'_i,s')\ u(k_i,s)\right]\
 \bm{\varepsilon}(k,\rho)\cdot{\bf q}\times{\bf k}\
 e^{-\alpha({\bf q}^2-2{\bf q}\cdot{\bf Q})/2}
\label{eq:6.57} \end{eqnarray}
This leads to
\begin{eqnarray}
\Delta\bm{\Gamma}_{5,CQM}^{\prime m} &\propto& ig_a^{\prime\prime}(4R_N^2/3)\
 \sqrt{\frac{E'+M'}{2M'}\ \frac{E+M}{2M} }\cdot
 \left[\chi_N^{\prime \dagger} \chi_N\right]\ 
 ({\bf q}\times{\bf k}).                                       
\label{eq:6.58} \end{eqnarray}
which is equivalent to the result (\ref{eq:6.32}) for
\begin{equation}
 g_a^{\prime\prime} = -\frac{3g_a'}{2(MR_N)^2}
 = -\frac{3g_a}{8(MR_N)^2}.
\label{eq:6.59} \end{equation}
{\bf Therefore, we can give the extra quark-coupling for the axial-vector 
vertex the interpretation as representing the orbital angular momentum 
of the three quarks in a nucleon (baryon) in the non-forward matrix element.
In this sense it is related to the "spin-crisis" \cite{Lead96}.}

\begin{center}
\hspace{-1cm}\fbox{ \begin{minipage}[b][ 2.5cm][l]{16.5cm}
\vspace{5mm}

\noindent The "spin-crisis" in the quark-parton model revealed the importance
of the orbital angular momentum and the gluonic content of the nucleon. 
At low energy the similar "crisis"
shows up quite naturally in the axial-vector coupling. Taking the orbital 
angular momentum of the quarks 
into account nicely connects the "constituent" quark model
with the axial-vector vertex at the nucleon level.
Interesting would be to analyze this phenomenon in the IMF.
\end{minipage} }\\
\end{center}


\section{Conclusions and Discussion}                                               
\label{sec:8}   
We have shown that for all meson-nucleon-nucleon couplings the Pauli-expansion
structure of the vertices can be reproduced by the "constituent" quark model.
For the scalar, the vector, and axial-vector mesons it required extra couplings
at the quark level in order to achieve this compatibility: 
(a) In the central part for scalar and vector mesons an extra interaction
is necessary on the quark level to produce the correct ${\bf k}^2/M'M$
terms at the nucleon level; (b) 
Using $\delta^3({\bf K}-{\bf k})$ at the meson vertex leads to 
"spurious" $1/R_N^2$-terms in the central parts for (i) the
scalar- and vector-meson vertex, and (ii) the axial-vector vertex. 
As demonstrated in subsection~\ref{sec:interm} such terms can eliminated by
the introduction of a  gaussian like distribution in ${\bf K}$. Therefore, these
terms are omitted.
This leads, at least for terms up to $1/M'M$, to the conclusion:
\begin{center}
\hspace{-0mm}\fbox{ \begin{minipage}[b][ 2.0cm][l]{16.5cm}
\vspace{5mm}
{\blue 
The Idea/conjecture made in the Introduction, asserting that based 
on the Lorentz structure the ratio's of spin-spin, 
tensor-, and spin-orbit-vertices and the central potentials as given 
by the nucleon-level potentials are independent of the internal structure,
can be realized completely in the CQM.}
\end{minipage} }\\
\end{center}
For the axial-vector coupling we have to introduce 
next to the usual $\gamma_\mu\gamma_5$-coupling a new coupling related to
the orbital angular-momentum contents related to the transverse motion of the 
quarks in the nucleon.
This is in line with the quark-parton model, where the so-called "spin-crisis"
can be solved by invoking such and/or a {\it gluonic} contribution 
to the spin of the nucleon.

[In passing we note that an important non-zero {\it gluonic} contribution 
would be in line with the soft-core NN-models
 (OBE and ESC) contain the pomeron-exchange potential which also has a 
 {\it gluonic} interpretation \cite{NRS78,PTP185.a}.
The same is true for the multi-pomeron repulsion in nuclear matter 
 \cite{PTP185.a,YFYR13}.]

The "constituent" quark model (CQM) is understood in a fundamental way
by spontaneous dynamical chiral-symmetry breaking. The instanton solutions
in QCD lead to a complex vacuum structure, which can be described by 
the instanton-liquid model. The pseudoscalar Nambu-Goldstone bosons are
ordinary $Q\bar{Q}$-states with a small mass due to the strong instanton
induced attraction. For other $Q\bar{Q}$-states there is not such a strong
attraction giving vector- and scalar-meson masses of about $2m_Q \approx 750$ MeV.
Strong-coupling QCD comes close to an understanding of the phenomenology
of the CQM \cite{Mil89}.
Another approach to derive the CQM is that of the Light-Front QCD of Wilson and 
collaborators \cite{Wil94}. 

In connection with the latter approach we note that putting ${\bf p}$ and
${\bf p}'$ in the xy-plane and going to the infinite-momentum-frame
(I.M.F.) along the z-axis, translates directly our results for the
meson-vertices to the quark-parton model. There, our impulse approximation
makes perfect sense and our results may be considered as realistic.
Thus, one would expect that the CQM-vertices correspond neatly to those
at the nucleon-level. However, also here one expects 
to find an orbital contribution to the spin due to the transverse motion 
of the quarks, in view of the "spin-crisis".\\
Applications to quark-quark and quark-nucleon potentials have been made 
and used in mixed quark-nuclear matter studies \cite{YYR22,YYR23,YYR24}.\\
Finally, the results of this paper can readily be extended to baryons.


\begin{flushleft}
\rule{16cm}{0.5mm}
\end{flushleft}


\appendix 


\section{Overlap Integral I}              
\label{app:O1}   
We consider the nucleon-nucleon graph for meson-exchange between
the constituent quarks of the two nucleons.
In the following we will use, instead of indices a,b,c, the indices  i=1,2,3.

 \begin{figure}
 \begin{center}
 \resizebox{7.25cm}{!}
 {\includegraphics[225,575][425,825]{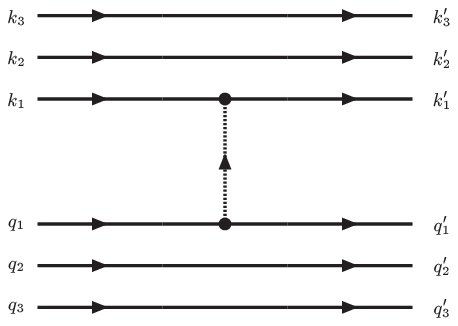}}
\caption{Internal quark momenta for meson-exchange}   
\label{fig:app.O1}
  \end{center}
  \end{figure}                     








In Fig.~\ref{fig:app.O1} we have given the momenta for the initial
and final nucleons, and the assigned momenta of the quarks.
From momentum conservation we have
\begin{eqnarray}
 p_{1} = k_{1}+k_{2}+k_{3}\ &,& \  p_{2}=q_{1}+q_{2}+q_{3}\ ,
 \nonumber \\ && \nonumber \\
 p'_{1} = k'_{1}+k'_{2}+k'_{3}\ &,& \ p'_{2}=q'_{1}+q'_{2}+q'_{3}\ ,
\label{app:O1.1} \end{eqnarray}
For meson-exchange with $p_{1}-p'_{1}=p'_{2}-p_{2} \equiv k$, we have
for the matrix-element of the potential
\begin{eqnarray}
\langle p'_{1}p'_{2}|V|p_{1}p_{2}\rangle &=& 
\int\prod_{i=1,3}\ d^{3}k_{i}\ \delta\left({\bf p}_{1}-\sum_{i}{\bf k}_{i}
\right)\cdot
\int\prod_{j=1,3}\ d^{3}k'_{j}\ \delta\left({\bf p'}_{1}-\sum_{j}{\bf k'}_{j}
\right)\cdot
\nonumber \\  &\times&
\int\prod_{i=1,3}\ d^{3}q_{i}\ \delta\left({\bf p}_{2}-\sum_{i}{\bf q}_{i}
\right)\cdot
\int\prod_{j=1,3}\ d^{3}q'_{j}\ \delta\left({\bf p'}_{2}-\sum_{j}{\bf q'}_{j}
\right)
\cdot \nonumber \\ &\times&
\tilde{\psi}_{p'_{1}}^{*}\left({\bf k'}_{1},{\bf k'}_{2},{\bf k'}_{3}\right)\
\tilde{\psi}_{p'_{2}}^{*}\left({\bf q'}_{1},{\bf q'}_{2},{\bf q'}_{3}\right)
\cdot
\tilde{\psi}_{p_{1}}\left({\bf k}_{1},{\bf k}_{2},{\bf k}_{3}\right)\
\tilde{\psi}_{p_{2}}\left({\bf q}_{1},{\bf q}_{2},{\bf q}_{3}\right)\
\cdot \nonumber \\ &\times&
 \delta^{3}\left({\bf k'}_{2}-{\bf k}_{2}\right)\ 
 \delta^{3}\left({\bf k'}_{3}-{\bf k}_{3}\right)\ 
 \delta^{3}\left({\bf q'}_{2}-{\bf q}_{2}\right)\ 
 \delta^{3}\left({\bf q'}_{3}-{\bf q}_{3}\right)\ 
\cdot \nonumber \\ &\times&
\frac{\gamma\left(k; k'_{1},k_{1}\right)\ \gamma\left(k; q'_{1},q_{1}\right)}
{{\bf k}^{2}+m_{M}^{2}}\cdot 
 \delta^{3}\left({\bf k}-{\bf k}'_{1}+{\bf k}_{1}\right)\ 
 \delta^{3}\left({\bf k}+{\bf q}'_{1}-{\bf q}_{1}\right)\ .
\label{app:O1.2} \end{eqnarray}
In (\ref{app:O1.2}) the $\gamma$'s denote the vertex functions.
Using the gaussian wave function of equation (\ref{eq:2.2}), we find for
the exponent, denoted by $f_{NN}$, taking into account that the momenta 
of the 'spectator quarks 2 and 3 do not change, the expression
\begin{eqnarray}
f_{NN} &=& \exp\left[-\frac{R_{N}^{2}}{6}\left\{\left(k_{1}-k_{2}\right)^{2}
+\left(k_{1}-k_{3}\right)^{2}+\left(k_{2}-k_{3}\right)^{2}
 \right.\right.\nonumber \\ & & \left.\left. \hspace{1.75cm}
+\left(q_{1}-q_{2}\right)^{2}+\left(q_{1}-q_{3}\right)^{2}
+\left(q_{2}-q_{3}\right)^{2}             
 \right.\right.\nonumber \\ & & \left.\left. \hspace{1.75cm}
+\left(k'_{1}-k_{2}\right)^{2}+\left(k'_{1}-k_{3}\right)^{2}
+\left(k_{2}-k_{3}\right)^{2}                    
 \right.\right.\nonumber \\ & & \left.\left. \hspace{1.75cm}
+\left(q'_{1}-q_{2}\right)^{2}+\left(q'_{1}-q_{3}\right)^{2}
+\left(q_{2}-q_{3}\right)^{2} \right\}\right] \nonumber \\       
      &=& \exp\left[-\frac{R_{N}^{2}}{6}\left\{
2\left(k_1^2+k_1^{\prime 2}\right)-2\left(k_2+k_3\right)\cdot
\left(k_1+k_1^\prime\right)  
 \right.\right.\nonumber \\ & & \left.\left. \hspace{1.75cm}
+2\left(k_2^2+k_3^2\right)+2\left(k_2-k_3\right)^2
 \right.\right.\nonumber \\ & & \left.\left. \hspace{1.75cm}
+2\left(q_1^2+q_1^{\prime 2}\right)-2\left(q_2+q_3\right)\cdot
\left(q_1+q_1^\prime\right)  
 \right.\right.\nonumber \\ & & \left.\left. \hspace{1.75cm}
+2\left(q_2^2+q_3^2\right)+2\left(q_2-q_3\right)^2 \right\}
\vphantom{\frac{A}{A}}\right]\ .
\label{app:O1.3} \end{eqnarray}
In (\ref{app:O1.3}) $k_1 \equiv {\bf k}_1$ etc. Introducing the 3-momenta
\begin{eqnarray}
 P_{23} = k_{2}+ k_{3}\ &,& \ R_{23}= q_{2}+q_{3}\ , \nonumber \\
 K_{23} = k_{2}- k_{3}\ &,& \ Q_{23}= q_{2}-q_{3}\ , 
\label{app:O1.4} \end{eqnarray}
for the 'spectator quarks' and the 3-momenta
\begin{eqnarray}
 {\bf k} = {\bf k}^\prime_1-{\bf k}_1\ & \ , \ & \ {\bf k}_1=\frac{1}{2} 
 \left({\bf Q} - {\bf k}\right) \nonumber \\
 {\bf Q} = {\bf k}_1+{\bf k}_1^\prime\ & \ , \ & \ {\bf k}_1^\prime=\frac{1}{2} 
 \left({\bf Q} + {\bf k}\right) \nonumber \\
 {\bf k} = {\bf q}_1-{\bf q}^\prime_1\ & \ , \ & \ {\bf q}_1=\frac{1}{2} 
 \left({\bf S} + {\bf k}\right) \nonumber \\
 {\bf S} = {\bf q}_1+{\bf q}_1^\prime\ & \ , \ & \ {\bf q}_1^\prime=\frac{1}{2} 
 \left({\bf S} - {\bf k}\right).
\label{app:O1.5}  \end{eqnarray}
For the 'active quarks', we can rewrite $f_{N}$ with the result
\begin{eqnarray}
f_{NN} &=& \exp\left[-\frac{R_{N}^{2}}{6}\left\{
\left({\bf Q}^2+{\bf k}^2\right)-2 {\bf P}_{23}\cdot{\bf Q}
 \right.\right.\nonumber \\ & & \left.\left. \hspace{1.75cm} +
\left({\bf P}_{23}^2+{\bf K}_{23}^2\right) + 2 {\bf K}_{23}^2
 \right.\right.\nonumber \\ & & \left.\left. \hspace{1.75cm} +
\left({\bf S}^2+{\bf k}^2\right)-2 {\bf R}_{23}\cdot{\bf S}
 \right.\right.\nonumber \\ & & \left.\left. \hspace{1.75cm} +
\left({\bf R}_{23}^2+{\bf Q}_{23}^2\right) + 2 {\bf Q}_{23}^2
\right\}\vphantom{\frac{A}{A}}\right]\ .
\label{app:O1.6a} \end{eqnarray}
In terms of the new variables defined in (\ref{app:O1.4}) and 
(\ref{app:O1.5}) the integration over the quark-momenta becomes
\begin{eqnarray}
&&\left(\frac{1}{8}\right)^{4} \int d^3Q\ d^3S\ d^3P_{23}\ d^3K_{23}\
     d^3R_{23}\ d^3Q_{23}\cdot \nonumber \\ 
&& \hspace{1cm} \times \delta^{(3)}\left(p_1+\frac{1}{2}\left(k-Q\right)
-P_{23}\right)\ \delta^{(3)}\left(p_2-\frac{1}{2}\left(k+S\right)-R_{23}\right)
\cdot \nonumber \\
&& \hspace{1cm} \times \delta^{(3)}\left(p'_1-\frac{1}{2}\left(k+Q\right)
-P_{23}\right)\ \delta\left(p'_2+\frac{1}{2}\left(k-S\right)-R_{23}\right)
\label{app:O1.6b}  \end{eqnarray}
From these $\delta$-function constraints one immediately gets 
\begin{eqnarray}
&& \delta^{(3)}\left(p'_1-p_1-k\right)\ \delta^{(3)}\left(p'_2-p_2+k\right) =
\nonumber \\ 
&& \delta^{(3)}\left(p'_1-p_1-k\right)\ \delta^{(3)}\left(p'_1+p'_2-p_1-p_2\right) 
\label{app:O1.7}  \end{eqnarray}
i.e. overall 3-momentum conservation and the fixing of ${\bf k}$ in
terms of the external momenta.

\noindent Next we go over to the CM-variables. We have
\begin{eqnarray}
 {\bf p}_1 &=& -{\bf p}_2 = {\bf p}\ \ , \ \ {\bf k} = {\bf p}'-{\bf p} \ \ , 
 \ \ {\bf p}= {\bf q}-\frac{1}{2}{\bf k}\ \ , \nonumber \\ 
 {\bf p}_1^\prime &=& -{\bf p}_2^\prime ={\bf p}^\prime \ \ , 
 \ \ {\bf q}=\frac{1}{2}({\bf p}+{\bf p}^\prime)
 \ \ , \ \ {\bf p}^\prime = {\bf q}+\frac{1}{2}{\bf k}\ .  
\label{app:O1.8}  \end{eqnarray}
Then using (\ref{app:O1.4}), we find for the expression between the curly
brackets in (\ref{app:O1.5}) the following expression
\begin{eqnarray}
\left\{\vphantom{\frac{A}{A}} \ldots \right\} &=& 
\left\{\vphantom{\frac{A}{A}} 2\left({\bf q}^{2}+{\bf k}^{2}\right)
+\frac{9}{4}\left({\bf Q}^2+{\bf S}^2\right)
-3{\bf q}\cdot\left({\bf Q}-{\bf S}\right)
\right.\nonumber \\ && \left.
+3\left({\bf K}_{23}^2+{\bf Q}_{23}^2\right) \vphantom{\frac{A}{A}}\right\}
\label{app:O1.9}  \end{eqnarray}
Now since the potential matrix elements will not depend on $K_{23}$ and
$Q_{23}$, apart from the appearance of these momenta in the exponential,
we can integrate these variables out, with the result:
\begin{equation}
\int d^3 K_{23}\ d^3 Q_{23}\ \exp\left[-\frac{R_{N}^{2}}{2}\left(
 {\bf K}_{23}^2+{\bf Q}_{23}^2\right)\right] \Rightarrow 
\left(\frac{2\pi}{R_{N}^{2}}\right)^3\ .
\label{app:O1.10}  \end{equation}

\noindent Collecting all results of the section, we find
\begin{eqnarray}
\langle p'_{1}p'_{2}|V|p_{1}p_{2}\rangle &=& \left(\frac{1}{8}\right)^4
\left(\frac{2\pi}{R_{N}^{2}}\right)^3\ {\cal N}^4 \int d^3Q\ d^3S\cdot 
\nonumber\\[0.3cm] && \times
 \exp\left[-\frac{R_{N}^{2}}{6}\left\{
 \frac{9}{4}\left({\bf Q}^2+{\bf S}^2\right) + 2\left({\bf q}^2+{\bf k}^2\right)
 -3{\bf q}\cdot\left({\bf Q}-{\bf S}\right) 
\right\}\vphantom{\frac{A}{A}}\right]\cdot \nonumber\\[0.3cm] && \times 
 V_{QQ}({\bf Q},{\bf S}; {\bf q},{\bf k})\ 
 \delta^{(3)}\left({\bf k}'_1+{\bf q}'_1-{\bf k}_1-{\bf q}_1\right),            
\label{app:O1.11} \end{eqnarray}
where 
$V_{QQ}({\bf Q},{\bf S}; {\bf q},{\bf k})$ denotes the QQ-potential which
contains the QQM-vertices and the meson propagator.

\begin{flushleft}
\rule{16cm}{0.5mm}
\end{flushleft}

\section{Overlap Integral II}              
\label{app:O2}   
\begin{figure}
 \begin{center}
 \resizebox{7.25cm}{!}
 {\includegraphics[225,625][425,875]{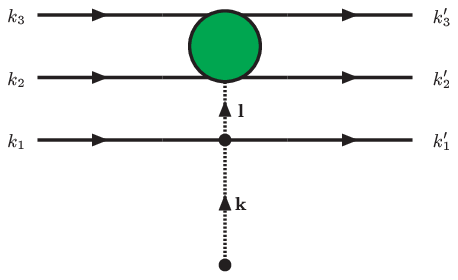}}
\caption{External and internal momenta meson-exchange
	 with diquark momentum correlation.}   
\label{fig:app.O2}
  \end{center}
  \end{figure}

We consider the nucleon-nucleon graph for meson-exchange between
the constituent quarks of the two nucleons.
In the following we will use, instead of indices a,b,c, the indices  i=1,2,3.
As shown in Fig.~\ref{fig:app.O2} we assume some momentum transfer between 
quark 1 and the pair quark 2 and quark 3.

In Fig.~\ref{fig:app.O2}, as in Fig.~\ref{fig:app.O1}, we have given the momenta for the initial
and final nucleons, and the assigned momenta of the quarks.
From momentum conservation we have
\begin{eqnarray}
 p = k_{1}+k_{2}+k_{3}\ &,& p' = k'_{1}+k'_{2}+k'_{3}.
\label{app:D1} \end{eqnarray}
For meson-exchange with $p'-p \equiv k$, we have for the QQM-vertex 
\begin{eqnarray}
\langle p'|\Gamma|p\rangle &=& 
\int\prod_{i=1,3}\ d^{3}k_{i}\ \delta\left({\bf p}-\sum_{i}{\bf k}_{i}
\right)\cdot
\int\prod_{j=1,3}\ d^{3}k'_{j}\ \delta\left({\bf p'}-\sum_{j}{\bf k'}_{j}
\right)\cdot
\nonumber \\  &\times&
\tilde{\psi}_{p'}^{*}\left({\bf k'}_{1},{\bf k'}_{2},{\bf k'}_{3}\right)\
\cdot
\tilde{\psi}_{p}\left({\bf k}_{1},{\bf k}_{2},{\bf k}_{3}\right)\cdot
\gamma({\bf k},{\bf l}; {\bf k'}_1,{\bf k}_1)
\cdot \nonumber \\ &\times&
 \delta^{3}\left({\bf k'}_{3}+{\bf k'}_{2}- {\bf k}_{3}-{\bf k}_{2}-{\bf l}\right)\ 
 \delta^{3}\left({\bf k}-{\bf l}-{\bf k}'_{1}+{\bf k}_{1}\right)\ 
\label{app:O2.2} \end{eqnarray}
Similar to Appendix~\ref{app:O1} we introduce the combinations
\begin{subequations}\label{app:O2.3} 
\begin{eqnarray}
 {\bf Q} = {\bf k'}_1+{\bf k}_1 &,& {\bf K} = {\bf k'}_1-{\bf k}_1, \\
 {\bf P}_{23} = {\bf k}_2+{\bf k}_3 &,& {\bf P'}_{23} = {\bf k'}_2+{\bf k'}_3,\\
 {\bf K'}_{23} = {\bf k'}_2-{\bf k'}_3 &,& {\bf K}_{23} = {\bf k}_2-{\bf k}_3.
\end{eqnarray} \end{subequations}
Furthermore, we use the customary definitions 
${\bf q} = ({\bf p'}+{\bf p})/2, {\bf k}={\bf p'}-{\bf p}$, and note that
${\bf K}={\bf k'}_1-{\bf k}_1={\bf k}-{\bf l}$, and ${\bf P'}_{23}={\bf P}_{23}+{\bf l}$.
The gaussian exponentials of the wave functions contain, see (\ref{eq:2.2}), 
\begin{eqnarray*}
h_N &=& (k_1-k_2)^2 + (k_1-k_3)^2 + (k_2-k_3)^2, \\
h'_N &=& (k'_1-k'_2)^2 + (k'_1-k'_3)^2 + (k'_2-k'_3)^2,   
\end{eqnarray*}
Using the definitions above and 
\begin{eqnarray*}
 {\bf P'}_{23} &=& {\bf P}_{23}+{\bf l}= {\bf P}_{23}+({\bf k}-{\bf K}) =
 {\bf p}+{\bf k}-\frac{1}{2}({\bf Q}+{\bf K}), \nonumber\\
 {\bf P}_{23} &=& {\bf p}-{\bf k}_1= {\bf p}-\frac{1}{2}({\bf Q}-{\bf K}),    
\end{eqnarray*}
one finds
\begin{eqnarray*}
 h_N &=& \frac{1}{4}\biggl[ \left({\bf Q}-{\bf K}-{\bf P}_{23}-{\bf K}_{23}\right)^2
 + \left({\bf Q}-{\bf K}-{\bf P}_{23}+{\bf K}_{23}\right)^2 + 4{\bf K}_{23}^2 \biggr] \\
 &=& \frac{1}{2}\biggl[({\bf Q}-{\bf K})^2+{\bf P}_{23}^2+3{\bf K}_{23}^2
 -2({\bf Q}-{\bf K})\cdot{\bf P}_{23}\biggr] \\
 &=& \frac{1}{2}\biggl[ \frac{9}{4} ({\bf Q}-{\bf K})^2
 -3({\bf Q}-{\bf K})\cdot\left({\bf q}-{\bf k}/2\right) 
 +({\bf q}^2+{\bf k}^2/4)+ 3 {\bf K}_{23}^2 \biggr], 
\end{eqnarray*}
and 
\begin{eqnarray*}
 h'_N &=& \frac{1}{4}\biggl[ \left({\bf Q}+{\bf K}-{\bf P'}_{23}-{\bf K'}_{23}\right)^2
 + \left({\bf Q}+{\bf K}-{\bf P'}_{23}+{\bf K'}_{23}\right)^2 + 4{\bf K'}_{23}^2 \biggr] \\
 &=& \frac{1}{2}\biggl[({\bf Q}+{\bf K})^2
 +{\bf P'}_{23}^2+3{\bf K'}_{23}-2({\bf Q}+{\bf K})\cdot{\bf P'}_{23}\biggr] \\
 &=& \frac{1}{2}\biggl[ \frac{9}{4} ({\bf Q}+{\bf K})^2
 -3({\bf Q}+{\bf K})\cdot\left({\bf q}+{\bf k}/2\right) 
 +({\bf q}^2+{\bf k}^2/4)+ 3 {\bf K'}_{23}^2 \biggr]. 
\end{eqnarray*}
Summing gives 
\begin{eqnarray*}
 h'_N+h_N &=&
  \frac{9}{4}({\bf Q}^2+{\bf K}^2) -3 {\bf Q}\cdot{\bf q}
 -\frac{3}{2}{\bf K}\cdot{\bf k+\left({\bf q}^2+\frac{1}{4}{\bf k}^2\right) }
 +\frac{3}{2}({\bf K'}_{23}^2+{\bf K}_{23}^2),   
\end{eqnarray*}
and, with performing the $d^3K'_{23}$ and $d^3K_{23}$ integrations, 
\begin{eqnarray}
 f_N = \exp\biggl[-\frac{R_N^2}{6}\left\{ h'_N+h_N\right\}\biggr] &\Rightarrow&
 \left(\frac{4\pi}{R_N^2}\right)^{3}\ \exp\biggl[-\frac{R_N^2}{6}\biggl\{
 \frac{9}{4}({\bf Q}^2+{\bf K}^2) -3{\bf Q}\cdot{\bf q}
 -\frac{3}{2}{\bf K}\cdot{\bf k}+\left({\bf q}^2+\frac{1}{4}{\bf k}^2\right) \biggr\}\biggr].
\label{app:O2.4}\end{eqnarray}
Note that for ${\bf K}={\bf k}$, after the $K'_{23}, K_{23}$ integrations:
\begin{eqnarray*}
 h'_N + h_N &\Rightarrow& \frac{9}{4}{\bf Q}^2 -3{\bf q}\cdot{\bf Q} +
 ({\bf q}^2+{\bf k}^2),
\end{eqnarray*}
which corresponds to the expression in Eqn.~(\ref{eq:2.21}).
Furthermore, the ${\bf k}, {\bf K}$ dependence differs from (\ref{eq:interm.6})
in the integrand by a factor
\begin{eqnarray*}
\gamma({\bf k},{\bf l}) &=&
\exp\biggl[-\frac{R_N^2}{6}\biggl\{-3{\bf K}\cdot{\bf k}+3{\bf k}^2\biggr\}\biggr]
=  \exp\biggl[-\frac{R_N^2}{2} ({\bf k}\cdot{\bf l})\biggr]
=  \exp\biggl[-\frac{R_N^2}{8}\biggl\{ 
({\bf k}+{\bf l})^2-({\bf k}-{\bf l})^2\bigg\}\biggr], 
\end{eqnarray*}
which has consequences in particular for the spin-orbit coupling,
 giving 1/3 instead of 1.
Including this factor in the vertex $\langle p' |\Gamma|p \rangle$ in
(\ref{app:O2.2}) makes it identical to (\ref{eq:interm.7}), and leads
to the expression for $\Gamma_{CQM}$ given Eqn.~(\ref{eq:interm.8}) {\red !}\\


\begin{center}
\hspace{-1cm}\fbox{ \begin{minipage}[b][ 8.5cm][l]{16.5cm}
\vspace{5mm}
\noindent {\bf Remark:} Consider the general gaussian integral:
\begin{eqnarray*}
J &=& \int\int d^3Q\ d^3K\ \exp\biggl[-\biggl\{\alpha {\bf Q}^2+\beta {\bf K}^2
+\gamma {\bf Q}\cdot{\bf K} +a {\bf Q}\cdot{\bf V}+ b {\bf K}\cdot{\bf W}\biggr\}\biggr]
\nonumber\\ &=& 
\left(\frac{\pi}{\alpha}\right)^{3/2}\ \int d^3K\ \exp\bigg[
-\left(\beta-\frac{\gamma^2}{4\alpha}\right)\ {\bf K}^2
-\left(b {\bf W}-\frac{a\gamma}{2\alpha} {\bf V}\right)\cdot {\bf K}\biggr]\cdot
\exp\left[+\frac{a^2}{4\alpha}\ {\bf V}^2\right] 
\nonumber\\ &=& 
\left(\frac{\pi}{\alpha}\right)^{3/2}\ 
\left(\frac{4\pi\alpha}{4\alpha\beta-\gamma^2}\right)^{3/2}
\exp\biggl[ \alpha\left(b {\bf W}-\frac{a\gamma}{2\alpha}{\bf V}\right)^2\bigg/ 
 (4\alpha\beta-\gamma^2)\biggr]\cdot
\exp\left[+\frac{a^2}{4\alpha}\ {\bf V}^2\right] 
\end{eqnarray*}
The factor in front $(4\alpha\beta-\gamma^2)^{-3/2}$ determines the possible 
"spurious" terms. One has
\begin{eqnarray*}
{\bf Q}^2 &:& -\frac{d}{d\alpha} \rightarrow 
6\beta \left(4\alpha\beta-\gamma^2\right)^{-5/2}, \\
{\bf K}^2 &:& -\frac{d}{d\beta} \rightarrow 
6\alpha \left(4\alpha\beta-\gamma^2\right)^{-5/2}.      
\end{eqnarray*}
{\it This implies that for a potential term $\propto \left({\bf Q}^2-{\bf K}^2\right)$
the "spurious" terms cancel when $\alpha=\beta$! }\\
The example worked out in this Appendix satisfies this condition. \\

\end{minipage} }\\
\end{center}
\section{Quark summation}                                           
\label{app:A}   
The nucleons are part of the irrep {\bf 56} of SU(6). These states have 
the following structure \cite{Close79}
\begin{equation}
 | N \rangle \sim \frac{1}{\sqrt{2}}\left(
 \phi_{M,S}\ \chi_{M,S} + \phi_{M,A}\ \chi_{M,A}\right) \equiv ({\bf 8,2}).
\label{app:1.1} \end{equation}
Here $\phi_{M,S}$ and $\phi_{M,A}$ denote the three-quark isospin
states with mixed symmetric and ant-symmetric character \cite{Close79}.
likewise for the spin states $\chi_{M,S}$ and $\chi_{M,A}$.\\
\noindent Since the total wave function is symmetric for the spin 
matrix elements one has
\begin{equation}
 \sum_{i=1}^3 \langle ... |\bm{\sigma}_i| ... \rangle \rightarrow
 3\langle ... |\bm{\sigma}_3| ... \rangle.   
\label{app:1.2} \end{equation}
To find the proper factor we evaluate the proton matrix element:
\begin{eqnarray}
 \langle P,+|\sum_{i=1}^3 \sigma_{i,z}| P,+ \rangle &=&
 3\langle P,+|\sigma_{3,z}| P,+ \rangle =
\frac{3}{2}\left\{ \langle \chi^P_{M,S}|\sigma_{3,z}|\chi^P_{M,S}\rangle
 +\langle \chi^P_{M,A}|\sigma_{3,z}|\chi^P_{M,A}\rangle \right\},
\label{app:1.3} \end{eqnarray}
where we used the orthonormality of the mixed states
\begin{eqnarray}
&& \langle \chi_{M,S}|\chi_{M,S}\rangle=1,\ \
\langle \chi_{M,A}|\chi_{M,A}\rangle=1,\ \
\langle \chi_{M,S}|\chi_{M,A}\rangle=0.     
\label{app:1.4} \end{eqnarray}
Explicit evaluation:
\begin{subequations}
\begin{eqnarray}
\langle \chi^P_{M,S}(+)|\sigma_{3,z}|\chi^P_{M,S}(+)\rangle &=&
\frac{1}{6}\langle (+-+) \oplus (-++) \ominus 2(++-)|\sigma_{3,z}
 |(+-+) \oplus (-++) \ominus 2(++-)\rangle
 \nonumber\\ &=& \frac{1}{6}[1+1-4]=-\frac{1}{3}, \\
\langle \chi^P_{M,A}(+)|\sigma_{3,z}|\chi^P_{M,A}(+)\rangle &=&
\frac{1}{2}\langle (+-+) \oplus (-++)|\sigma_{3,z}
 |(+-+) \oplus (-++) \rangle
 \nonumber\\ &=& \frac{1}{2}[1+1]=+1. 
\label{app:1.5} \end{eqnarray}
\end{subequations}
These results imply the relation
\begin{equation}
 \langle P,+|\sum_{i=1}^3 \bm{\sigma}_{i}| P,+ \rangle =
 \langle P,+|\bm{\sigma}_{N}| P,+ \rangle.     
\label{app:1.6} \end{equation}
It is now trivial to see that 
\begin{equation}
 \langle P,+|\sum_{i=1}^3 {\bf 1}_{op,i}| P,+ \rangle =
 3 \langle P,+| {\bf 1}_{op,N}| P,+ \rangle.     
\label{app:1.7} \end{equation}

\begin{flushleft}
\rule{16cm}{0.5mm}
\end{flushleft}


\section{Momentum-space Meson-Quark-Quark Vertices }
\label{app:OBE}
\subsection{Pauli-reduction Dirac-spinor $\Gamma$-matrix elements}
\label{app:Pauli}
The transition from Dirac spinors to Pauli spinors is given here,  
without approximations. We use the notations ${\cal E}= E+M$ and
${\cal E}'= E'+M'$, where $E=E(p,M)$ and $E'=E(p',M')$.
Also, we omit, on the right-hand side in the expressions below,
the final and initial Pauli spinors $\chi^{\prime \dagger}$ and
$\chi$ respectively, which are self-evident.
\begin{subequations}
\label{Pauli.1}
\begin{eqnarray}
   \bar{u}({\bf p}') u({\bf p}) &=&
 +\sqrt{\frac{{\cal E}'{\cal E}}{4M' M}}
  \left[ \left(1-\frac{{\bf p}'\cdot{\bf p}}{{\cal E}'{\cal E}}\right)
 -i\frac{ {\bf p}'\times{\bf p}\cdot\mbox{\boldmath $\sigma$}}{{\cal E}'{\cal E}} 
\right]
     ,    \label{Paus}\\
   \bar{u}({\bf p}')\gamma_5 u({\bf p}) &=&
 -\sqrt{\frac{{\cal E}'{\cal E}}{4M' M}}
 \left[ \frac{\mbox{\boldmath $\sigma$}\!\cdot\!{\bf p}'}{{\cal E}'}
      - \frac{\mbox{\boldmath $\sigma$}\!\cdot\!{\bf p}}{{\cal E}} \right]
     ,                      \label{Pau1p}\\
   \bar{u}({\bf p}')\gamma^{0} u({\bf p}) &=&
 +\sqrt{\frac{{\cal E}'{\cal E}}{4M' M}}
   \left[ \left(1+\frac{{\bf p}'\cdot{\bf p}}{{\cal E}'{\cal E}}\right)
 +i \frac{{\bf p}'\times{\bf p}\cdot\mbox{\boldmath $\sigma$}}{{\cal E}'{\cal E}}\right]
     ,    \label{Pauv0}\\
   \bar{u}({\bf p}')\mbox{\boldmath $\gamma$}\ u({\bf p}) &=& 
 +\sqrt{\frac{{\cal E}'{\cal E}}{4M' M}}
\left[ \left(\frac{{\bf p}'}{{\cal E}'}+\frac{{\bf p}}{{\cal E}}\right) 
 + i \left(\frac{\mbox{\boldmath $\sigma$}\times{\bf p}'}{{\cal E}'}
         - \frac{\mbox{\boldmath $\sigma$}\times{\bf p}}{{\cal E}}\right)
 \right] , \label{Pauvi}\\
   \bar{u}({\bf p}')\gamma_5\gamma^{0} u({\bf p}) &=&
 -\sqrt{\frac{{\cal E}'{\cal E}}{4M' M}}
 \left[ \frac{\mbox{\boldmath $\sigma$}\!\cdot\!({\bf p}'}{{\cal E}'}
      + \frac{\mbox{\boldmath $\sigma$}\!\cdot\!({\bf p}}{{\cal E}} \right]
     ,    \label{Paua0}\\
   \bar{u}({\bf p}')\gamma_5\mbox{\boldmath $\gamma$}\ u({\bf p}) &=& 
 -\sqrt{\frac{{\cal E}'{\cal E}}{4M' M}}
  \left[ \mbox{\boldmath $\sigma$} + \frac{ 
  (\mbox{\boldmath $\sigma$}\cdot{\bf p}')\ \mbox{\boldmath $\sigma$}\ 
  (\mbox{\boldmath $\sigma$}\cdot{\bf p})}{{\cal E}'{\cal E}}\right]  
 \nonumber\\ &=& 
 -\sqrt{\frac{{\cal E}'{\cal E}}{4M' M}}\left[
  \left(1-\frac{{\bf p}'\cdot{\bf p}}{{\cal E}'{\cal E}}\right)\mbox{\boldmath $\sigma$} 
 -i \frac{{\bf p}'\times{\bf p}}{{\cal E}'{\cal E}} 
 \right.\nonumber\\ && \left.
 +\frac{1}{{\cal E}'{\cal E}}\left(
 \mbox{\boldmath $\sigma$}\cdot{\bf p}\ {\bf p}' + 
  \mbox{\boldmath $\sigma$}\cdot{\bf p}'\ {\bf p}\right)\right] \approx 
  -\mbox{\boldmath $\sigma$} 
 , \label{Pauai} ,
\end{eqnarray}
\end{subequations}
\noindent
$\!\!\!$where we defined ${\bf k}={\bf p}'-{\bf p}$,
${\bf q}=({\bf p}'+{\bf p})/2$, and $\kappa_{V}=f_{V}/g_{V}$.\\

Using the the Gordon decomposition
\begin{eqnarray}
&& i\ \bar{u}(p')\ \sigma^{\mu\nu}(p'-p)_\nu u(p) = \bar{u}(p')\left\{\vphantom{\frac{A}{A}}
 (M'+M)\gamma^\mu - (p'+p)^\mu\right\}\ u(p)
\label{Gordon.app}\end{eqnarray}
one obtains for the complete vector-vertex 
\begin{subequations}
\label{Paul.3}
\begin{eqnarray}
 \bar{u}(p') \Gamma^\mu_V u(p) &\equiv& \bar{u}(p')\left[\gamma^\mu 
 + \frac{i}{2{\cal M}}\kappa_V \sigma^{\mu\nu}(p'-p)_\nu\right] u(p)                  
 \nonumber\\[3mm]           
 &=& \bar{u}(p')\left[\left(1+\frac{M'+M}{2{\cal M}}\kappa_V\right)\gamma^\mu 
 - \frac{\kappa_V}{2{\cal M}}(p'+p)_\mu\right] u(p) \Longrightarrow 
 \nonumber\\[3mm]           
 \mu=0 &:& +\sqrt{\frac{{\cal E}'{\cal E}}{4M' M}}\left[
  \left(1+\frac{M'+M}{2{\cal M}}\kappa_V\right)
 \left(1+\frac{\mbox{\boldmath $\sigma$}\cdot{\bf p}'\
 \mbox{\boldmath $\sigma$}\cdot{\bf p}}{{\cal E}'{\cal E}}\right)
 \right.\nonumber\\ && \left. 
 -\frac{\kappa_V}{2{\cal M}}(E'+E) 
 \left(1-\frac{\mbox{\boldmath $\sigma$}\cdot{\bf p}'\
 \mbox{\boldmath $\sigma$}\cdot{\bf p}}{{\cal E}'{\cal E}}\right)\right]  
, \label{PAU0}\\
  \mu=i &:& +\sqrt{\frac{{\cal E}'{\cal E}}{4M' M}}\left[
 \left(1+\frac{M'+M}{2{\cal M}}\kappa_V\right)
 \left\{\left(\frac{{\bf p}'}{{\cal E}'}+\frac{\bf p}{\cal E}\right)
 +i\left(\frac{\mbox{\boldmath $\sigma$}\times{\bf p}'}{{\cal E}'}
 - \frac{\mbox{\boldmath $\sigma$}\times{\bf p}}{{\cal E}}\right)\right\}
 \right.\nonumber\\ && \left.
 -\frac{\kappa_V}{2{\cal M}}({\bf p}'+{\bf p}) 
 \left(1-\frac{\mbox{\boldmath $\sigma$}\cdot{\bf p}'\
 \mbox{\boldmath $\sigma$}\cdot{\bf p}}{{\cal E}'{\cal E}}\right)\right]  
. \label{PAUi} 
\end{eqnarray}
\end{subequations}
 
\subsection{1/M-expansion $\Gamma$-matrix elements}

The exact transition from Dirac spinors to Pauli spinors is given in
Appendix~\ref{app:Pauli}. From the expressions in \ref{app:Pauli}, keeping only
terms up to order $1/M$, and setting the scaling mass ${\cal M}=M$, 
we find that the vertex operators in
Pauli-spinor space for the $N\!Nm$ vertices are given by
\begin{subequations}
\label{Pauliexp.3}
\begin{eqnarray}
   \bar{u}({\bf p}') u({\bf p}) &=&
      \left[ \left(1-\frac{{\bf p}'\cdot{\bf p}}{4M^2}\right)
 -\frac{i}{4M^2} {\bf p}'\times{\bf p}\cdot\mbox{\boldmath $\sigma$}\right]
     ,    \label{Gams}\\
   \bar{u}({\bf p}')\gamma_5 u({\bf p}) &=&
 -\frac{1}{2M}\!\left[ \mbox{\boldmath $\sigma$}\!\cdot\!({\bf p}'-{\bf p})\right]
  = -\frac{1}{2M}\!\left[ \mbox{\boldmath $\sigma$}\!\cdot\!{\bf k}\right]
     ,                      \label{Gam1pp}\\
   \bar{u}({\bf p}')\gamma^{0} u({\bf p}) &=&
      \left[ \left(1+\frac{{\bf p}'\cdot{\bf p}}{4M^2}\right)
 +\frac{i}{4M^2} {\bf p}'\times{\bf p}\cdot\mbox{\boldmath $\sigma$}\right]
     ,    \label{Gamv0}\\
   \bar{u}({\bf p}')\mbox{\boldmath $\gamma$}\ u({\bf p}) &=& 
 \frac{1}{2M}\left[ ({\bf p}'+{\bf p}) + i \mbox{\boldmath $\sigma$}\times({\bf p}'-{\bf p})
 \right] , \label{Gamvi}\\
   \bar{u}({\bf p}')\gamma_5\gamma^{0} u({\bf p}) &=&
 -\frac{1}{2M}\!\left[ \mbox{\boldmath $\sigma$}\!\cdot\!({\bf p}'+{\bf p})\right]
 = -\frac{1}{M}\!\left[ \mbox{\boldmath $\sigma$}\!\cdot\!{\bf q}\right]
     ,    \label{Gama0}\\
   \bar{u}({\bf p}')\gamma_5\mbox{\boldmath $\gamma$}\ u({\bf p}) &=& 
 -\left[ \mbox{\boldmath $\sigma$} + \frac{1}{4M^2} 
  (\mbox{\boldmath $\sigma$}\cdot{\bf p}')\ \mbox{\boldmath $\sigma$}\ 
  (\mbox{\boldmath $\sigma$}\cdot{\bf p})\right]  
 = -\left[\left(1-\frac{{\bf p}'\cdot{\bf p}}{4M^2}\right)\mbox{\boldmath $\sigma$}
 \right.\nonumber\\ && \left.
 -\frac{i}{4M^2} {\bf p}'\times{\bf p} +\frac{1}{4M^2}\left(
 \mbox{\boldmath $\sigma$}\cdot{\bf p}\ {\bf p}' + 
  \mbox{\boldmath $\sigma$}\cdot{\bf p}'\ {\bf p}\right)\right] \approx 
  -\mbox{\boldmath $\sigma$} 
 , \label{Gamai} ,
\end{eqnarray}
\end{subequations}
\noindent
$\!\!\!$where we defined ${\bf k}={\bf p}'-{\bf p}$,
${\bf q}=({\bf p}'+{\bf p})/2$, and $\kappa_{V}=f_{V}/g_{V}$.
In passing we note that the inclusion of the $1/M^2$-terms is necessary
in order to get spin-orbit potentials, like in the case of the OBE-potentials.\\

For the magnetic-coupling we use the Gordon decomposition
\begin{eqnarray}
&& i\ \bar{u}(p')\ \sigma^{\mu\nu}(p'-p)_\nu u(p) = \bar{u}(p')\left\{\vphantom{\frac{A}{A}}
 2M \gamma^\mu - (p'+p)^\mu\right\}\ u(p)
\label{Gordon}\end{eqnarray}
We get 
\begin{subequations}
\label{Pauliexp.4}
\begin{eqnarray}
&& i\ \bar{u}(p')\ \sigma^{\mu\nu}(p'-p)_\nu u(p) \Longrightarrow 
 \nonumber\\[3mm]           
&& \mu=0\ :\ -M \left[ \left(1-\frac{{\bf p}'\cdot{\bf p}}{4M^2}\right)
      +\frac{(p^{\prime 2}+p^2)}{2M^2}
 -\frac{i}{4M^2} {\bf p}'\times{\bf p}\cdot\mbox{\boldmath $\sigma$}\right]
, \label{Gamt0}\\
&& \mu=i\ :\ -\left[ \frac{1}{2}({\bf p}'+{\bf p}) -\frac{i}{2}\mbox{\boldmath $\sigma$}
 \times({\bf p}'-{\bf p})\right] . \label{Gamti}
\end{eqnarray}
\end{subequations}
For the vector-vertex with direct and derivative coupling one has
\begin{subequations}
\label{Pauliexp.5a}
\begin{eqnarray}
 \bar{u}(p') \Gamma^\mu_V u(p) &\equiv& \bar{u}(p')\left[\gamma^\mu 
 + \frac{i}{2M}\kappa_V \sigma^{\mu\nu}(p'-p)_\nu\right] u(p)                  
 \nonumber\\[3mm]           
 &=& \bar{u}(p')\left[(1+\kappa_V)\gamma^\mu 
 - \frac{\kappa_V}{2M}(p'+p)_\mu\right] u(p) \Longrightarrow 
 \nonumber\\[3mm]           
 \mu=0 &:& \left[ (1+\kappa_V)\left(1+\frac{{\bf p}'\cdot{\bf p}}{4M^2}
  +\frac{i}{4M^2} {\bf p}'\times{\bf p}\cdot\mbox{\boldmath $\sigma$}\right)
 \right.\nonumber\\[3mm] && \left.        
    -\kappa_V\frac{E_{p'}+E_p}{2M}\left(1-\frac{{\bf p}'\cdot{\bf p}}{4M^2}
  -\frac{i}{4M^2} {\bf p}'\times{\bf p}\cdot\mbox{\boldmath $\sigma$}\right)\right] \approx 
 \nonumber\\ && 
 \left[ 1+(1+2\kappa_V)\left\{\frac{{\bf p}'\cdot{\bf p}}{4M^2}
  +\frac{i}{4M^2} {\bf p}'\times{\bf p}\cdot\mbox{\boldmath $\sigma$}\right\}
    -\kappa_V\frac{{\bf p}^{\prime 2}+{\bf p}^2}{4M^2}\right] 
, \label{GAM0}\\
  \mu=i &:& \frac{1}{M}\left[ \frac{1}{2}({\bf p}'+{\bf p}) 
  +\frac{i}{2}(1+\kappa_V)\mbox{\boldmath $\sigma$}\times({\bf p}'-{\bf p})\right] 
. \label{GAMi} 
\end{eqnarray}
\end{subequations}
In terms of the magnetic and electric couplings, $g_V=G_M+G_E$ and
$f_V=-G_e$, we have 
$g_V\kappa_V= -G_E, g_V(1+\kappa_V)=G_M, g_V(1+2\kappa_V)=G_M-G_E$. 
This gives
\begin{subequations}
\label{Pauliexp.5b}
\begin{eqnarray}
 g_V\bar{u}(p') \Gamma^\mu_V u(p) &\equiv& g_V\bar{u}(p')\left[\gamma^\mu 
 + \frac{i}{2M}\kappa_V \sigma^{\mu\nu}(p'-p)_\nu\right] u(p)                  
 \nonumber\\[3mm]           
 \mu=0 &:& 
 \left[ (G_M+G_E) + (G_M-G_E)\left\{\frac{{\bf p}'\cdot{\bf p}}{4M^2}
  +\frac{i}{4M^2} {\bf p}'\times{\bf p}\cdot\mbox{\boldmath $\sigma$}\right\}
    +G_E\frac{{\bf p}^{\prime 2}+{\bf p}^2}{4M^2}\right] 
, \label{GAM10}\\
  \mu=i &:& \frac{1}{M}\left[ \frac{1}{2}({\bf p}'+{\bf p})(G_M+G_E) 
  +\frac{i}{2} G_M \mbox{\boldmath $\sigma$}\times({\bf p}'-{\bf p})\right] 
. \label{GAM11} 
\end{eqnarray}
\end{subequations}

\subsection{Meson-vertices in Pauli-spinor space}
The transition from Dirac spinors to Pauli spinors is reviewed in
Appendix C of \cite{Rij91}. Following this reference and keeping only
terms up to order $(1/M)^2$, we find that the vertex operators in
Pauli-spinor space for the $Q\!Qm$ vertices are given by

 
\begin{subequations}
\label{Pauliexp.1}
\begin{eqnarray}
   \bar{u}({\bf p}')\Gamma^{(1)}_{P}u({\bf p}) &=&
 -i\frac{f_{P}}{m_{\pi}}\!\left[ \mbox{\boldmath $\sigma$}_{1}\!\cdot\!{\bf k}
        \pm\frac{\omega}{2M}\mbox{\boldmath $\sigma$}_{1}\!\cdot\!
                 ({\bf p}'+{\bf p}) \right],            \label{Gam1p}\\
   \bar{u}({\bf p}')\Gamma^{(1)}_{V}u({\bf p}) &=&
      g_{V}\left[ 
      \left\{\left(1+\frac{{\bf p}'\cdot{\bf p}}{4M^2}\right)
 -\frac{i}{4M^2} {\bf p}'\times{\bf p}\cdot\mbox{\boldmath $\sigma$}
      \right\}\ \phi^{0}_{V} \right.\nonumber\\ && \left.
     - \frac{1}{2M}\biggl\{({\bf p}'+{\bf p})
          +i(1+\kappa_{V})\mbox{\boldmath $\sigma$}_{1}\!\times\!{\bf k}
     \biggr\}\!\cdot\!\mbox{\boldmath $\phi$}_{V}\right],    \label{Gam1v}\\
   \bar{u}({\bf p}')\Gamma^{(1)}_{A}u({\bf p}) &=&
      g_{A}\left[ -\frac{1}{2M} \left\{\bm{\sigma}\cdot({\bf p}'+{\bf p}) 
      \right\}\ \phi^{0}_{A} \right.\nonumber\\ && \left.
     + \biggl\{ \bm{\sigma} +\frac{1}{4M^2}(\bm{\sigma}\cdot{\bf p}')\
     \bm{\sigma}\ (\bm{\sigma}\cdot{\bf p})
     \biggr\}\!\cdot\!\mbox{\boldmath $\phi$}_{A}\right],    \label{Gam1a}\\
   \bar{u}({\bf p}')\Gamma^{(1)}_{S}u({\bf p}) &=& g_{S}
 \left[\left(1-\frac{{\bf p}'\cdot{\bf p}}{4M^2}\right)
 -\frac{i}{4M^2} {\bf p}'\times{\bf p}\cdot\mbox{\boldmath $\sigma$}
 \right] ,                                               \label{Gam1s}
\end{eqnarray}
\end{subequations}
\noindent
$\!\!\!$where we defined ${\bf k}={\bf p}'-{\bf p}$ and
$\kappa_{V}=f_{V}/g_{V}$.
In the pseudovector vertex, the upper (lower) sign stands for creation
(absorption) of the pion at the vertex.
In passing we note that the inclusion of the $1/M^2$-terms is necessary
in order to get spin-orbit potentials, like in the case of the OBE-potentials.\\

\noindent For the complete vector-meson coupling to the quarks we have,
writing $\Gamma_V = \Gamma_V^{(m)}+\Gamma_V^{(e)}$, 
\begin{subequations}
\label{eq:mevert1}     
\begin{eqnarray}
   \bar{u}({\bf p}')\Gamma^{(m)}_{V}u({\bf p}) &=&
      G_{m,v}\left[ 
      \left\{\left(1+\frac{{\bf p}'\cdot{\bf p}}{4M^2}\right)
 +\frac{i}{4M^2} {\bf p}'\times{\bf p}\cdot\mbox{\boldmath $\sigma$}
      \right\}\ \phi^{0}_{V} \right.\nonumber\\ && \hspace*{1cm} \left.
     + \frac{1}{2M}\biggl\{({\bf p}'+{\bf p})
          +i\mbox{\boldmath $\sigma$}_{1}\!\times\!{\bf k}
     \biggr\}\!\cdot\!\mbox{\boldmath $\phi$}_{V}\right], \\                
   \bar{u}({\bf p}')\Gamma^{(e)}_{V}u({\bf p}) &=&
      G_{m,e}\left[\frac{{\cal E}^\prime+{\cal E}}{{\cal M}}
      \left\{\left(1-\frac{{\bf p}'\cdot{\bf p}}{4M^2}\right)
 -\frac{i}{4M^2} {\bf p}'\times{\bf p}\cdot\mbox{\boldmath $\sigma$}
      \right\}\ \phi^{0}_{V} \right.\nonumber\\ && \hspace*{1cm} \left.
     + \frac{({\bf p}'+{\bf p})}{{\cal M}}
      \left\{\left(1-\frac{{\bf p}'\cdot{\bf p}}{4M^2}\right)
 -\frac{i}{4M^2} {\bf p}'\times{\bf p}\cdot\mbox{\boldmath $\sigma$}
   \right\}  \biggr\}\!\cdot\!\mbox{\boldmath $\phi$}_{V}\right] 
\nonumber\\ &\approx&      
      G_{m,e}\left[2\frac{M}{{\cal M}}
      \left\{\left(1+\frac{{\bf p}^{\prime 2}-{\bf p}'\cdot{\bf p}
 +{\bf p}^2}{4M^2}\right)
 -\frac{i}{4M^2} {\bf p}'\times{\bf p}\cdot\mbox{\boldmath $\sigma$}
      \right\}\ \phi^{0}_{V} 
     + \frac{({\bf p}'+{\bf p})}{{\cal M}}
   \biggr\}\!\cdot\!\mbox{\boldmath $\phi$}_{V}\right] 
\end{eqnarray}
\end{subequations}
The extra QQ axial-coupling has the vertex 
\begin{eqnarray}
   \bar{u}({\bf p}')\Gamma^{(o)}_{A}u({\bf p}) &=&
      \frac{g_a^\prime}{{\cal M}^2} \left[  \frac{1}{M}\left\{
 ({\bf p}^\prime\cdot{\bf p}-{\bf p}^2)\bm{\sigma}\cdot{\bf p}' +
 ({\bf p}^\prime\cdot{\bf p}-{\bf p}^{\prime 2})\bm{\sigma}\cdot{\bf p} 
      \right\}\ \phi^{0}_{A} 
 -2i {\bf p}^\prime\times{\bf p}\cdot\!\mbox{\boldmath $\phi$}_{A}\right].    
\label{Gaxia1} \end{eqnarray}
 

\section{Pauli-spinor Invariants for Nucleon-nucleon Potentials}
\label{app:B}   
Because of rotational invariance and parity conservation, the ${\cal V}$-matrix, which is
a $4\times 4$-matrix in Pauli-spinor space, can be expanded 
into the following set of in general 8 spinor invariants, see for example 
Ref.~\cite{SNRV71}. Introducing \cite{notation1}
\begin{equation}
  {\bf q}=\frac{1}{2}({\bf p}'+{\bf p})\ , \
  {\bf k}={\bf p}'-{\bf p}\ , \           
  {\bf n}={\bf p}\times {\bf p}',
\label{eq:30.30} \end{equation}
with, of course, ${\bf n}={\bf q}\times {\bf k}$,
we choose for the operators $P_{j}$ in spin-space
\begin{eqnarray}
&&  P_{1}=1,  \hspace{3mm} P_{2}= 
 \mbox{\boldmath $\sigma$}_1\cdot\mbox{\boldmath $\sigma$}_2,
 \nonumber\\[0.0cm]
&& P_{3}=(\mbox{\boldmath $\sigma$}_1\cdot{\bf k})(\mbox{\boldmath $\sigma$}_2\cdot{\bf k})
 -\frac{1}{3}(\mbox{\boldmath $\sigma$}_1\cdot\mbox{\boldmath $\sigma$}_2)
  {\bf k}^2,
 \nonumber\\[0.0cm]
&& P_{4}=\frac{i}{2}(\mbox{\boldmath $\sigma$}_1+
 \mbox{\boldmath $\sigma$}_2)\cdot{\bf n}, \hspace{3mm} 
 P_{5}=(\mbox{\boldmath $\sigma$}_1\cdot{\bf n})(\mbox{\boldmath $\sigma$}_2\cdot{\bf n}),
 \nonumber\\[0.0cm]
 && P_{6}=\frac{i}{2}(\mbox{\boldmath $\sigma$}_1-\mbox{\boldmath $\sigma$}_2)\cdot{\bf n}, 
 \nonumber\\[0.0cm]
 && P_{7}=(\mbox{\boldmath $\sigma$}_1\cdot{\bf q})(\mbox{\boldmath $\sigma$}_2\cdot{\bf k})
 +(\mbox{\boldmath $\sigma$}_1\cdot{\bf k})(\mbox{\boldmath $\sigma$}_2\cdot{\bf q}),
 \nonumber\\[0.0cm]
&& P_{8}=(\mbox{\boldmath $\sigma$}_1\cdot{\bf q})(\mbox{\boldmath $\sigma$}_2\cdot{\bf k})
 -(\mbox{\boldmath $\sigma$}_1\cdot{\bf k})(\mbox{\boldmath $\sigma$}_2\cdot{\bf q}).
\label{eq:30.31} \end{eqnarray}
Here we follow Ref.~\cite{MRS89}, where in contrast to Ref.~\cite{NRS78},
we have chosen $P_{3}$ to be a purely `tensor-force' operator.
The expansion in Pauli spinor-invariants reads
\begin{equation}
 {\cal V}({\bf p}',{\bf p}) = \sum_{j=1}^8\ \widetilde{V}_j({\bf p}^{\prime 2},{\bf p}^2,
 {\bf p}'\cdot{\bf p})\ P_j({\bf p}',{\bf p})\ .
\label{eq:30.32} \end{equation}

\begin{flushleft}
\rule{16cm}{0.5mm}
\end{flushleft}
\section{Extended-Soft-Core QQ-Potentials in Momentum Space}
\label{app:C}   
The potential of the ESC-model contains the contributions from 
(i) One-boson-exchanges, (ii) Uncorrelated 
Two-Pseudo-scalar exchange,  
and (iii) Meson-Pair-exchange. In this section we 
review the potentials and indicate the changes with respect to 
earlier papers on the OBE- and ESC-models.
The spin-1 meson-exchange is an important ingredient for the baryon-baryon force. 
In the ESC08-model we treat the vector-mesons and the axial-vector mesons 
according to the Proca- \cite{IZ80} and the B-field \cite{Nak72,NO90} formalism
respectively. For details, we refer to Appendix~\ref{app:C}.

\subsection{One-Boson-Exchange Interactions in Momentum Space}
\label{app.3a}
The OBE-potentials are the 
same as given in \cite{NRS78,MRS89}, with the exception of 
 (i) the zero in the scalar form factor, and 
 (ii) the axial-vector-meson potentials.
Here, we review the OBE-potentials briefly, and give those potentials
which are not included in the above references.
The local interaction Hamilton densities for the different couplings
are \cite{BD65} \\ \\
        a) Pseudoscalar-meson exchange $(J^{PC}=0^{-+})$
         \begin{equation}
         {\cal H}_{PV}= \frac{f_{PV}}{m_{\pi^{+}}}
         \bar{\psi}\gamma_{\mu}\gamma_{5}
                \psi\partial^{\mu}\phi_{P}. \label{eq:3.1}\end{equation}
This is the pseudovector coupling, and the
relation with the pseudoscalar coupling is 
$g_P = 2M_B/m_{\pi^+}$, where $M_B$ is the nucleon or hyperon mass.\\ \\     
        b) Vector-meson exchange $(J^{PC}=1^{--})$
       \begin{equation}
   {\cal H}_{V}=g_{V}\bar{\psi}\gamma_{\mu}\psi\phi_{V}^{\mu}
                +\frac{f_{V}}{4{\cal M}}\bar{\psi}\sigma_{\mu\nu}
                \psi (\partial^{\mu}\phi^{\nu}_{V}-\partial^{\nu}
                      \phi^{\mu}_{V}), \label{eq:3.2}\end{equation}
       where $\sigma_{\mu\nu}= i[\gamma_{\mu},\gamma_{\nu}]/2$,
       and the scaling mass ${\cal M}$, 
       will be taken to be the proton mass.\\ \\      
\noindent c)\ Axial-vector-meson exchange ( $J^{PC}=1^{++}$, 1$^{st}$ kind):
\begin{equation}
 {\cal H}_A = g_A[\bar{\psi}\gamma_\mu\gamma_5\psi] \phi^\mu_A + \frac{if_A}{{\cal M}}
 [\bar{\psi}\gamma_5\psi]\ \partial_\mu\phi_A^\mu.
\label{eq:OBE.1}\end{equation}
In ESC04 the $g_A$-coupling was included, but not the derivative $f_A$-coupling \cite{dercopax}.
Also, in ESC04  we used a local-tensor approximation (LTA) for the 
 $(\mbox{\boldmath $\sigma$}_1\cdot{\bf q})(\mbox{\boldmath $\sigma$}_2
  \cdot{\bf q})$ operator. Here, we improve on that considerably by avoiding such
 rather crude approximation. The details of our new treatment are given in 
Appendix~\ref{app:B}.
\\[0.2cm]
\noindent d)\ Axial-vector-meson exchange ( $J^{PC}=1^{+-}$, 2$^{nd}$ kind):
\begin{equation}
 {\cal H}_B = \frac{if_B}{m_B}
 [\bar{\psi}\sigma_{\mu\nu}\gamma_5\psi]\ \partial_\nu\phi_B^\mu\ .
\label{eq:OBE.2}\end{equation}
In ESC04 this coupling was not included. Like for the axial-vector mesons of the
1$^{st}$-kind we include an SU(3)-nonet with members $b_1(1235), h_1(1170), h_1(1380)$.
In the quark-model they are $Q\Bar{Q}(^1P_1)$-states.\\[0.2cm]
\noindent e)\ Scalar-meson exchange ($J^{PC}=0^{++}$):
\begin{equation}
 {\cal H}_S = g_S[\bar{\psi}\psi] \phi_S + \frac{f_S}{{\cal M}}
 [\bar{\psi}\gamma_\mu\psi]\ \partial^\mu\phi_S,
\label{eq:OBE.3}\end{equation}
which is the most general interaction. 
In ESC04 the possibility of the derivative $f_S$-coupling was not considered.
By partial integration it is clear that the derivative vertex is proportional to the 
baryon mass difference and therefore there can only be expected sizable effects for
$\kappa$-exchange. However, 
it is easily seen that for example for the $\Lambda N \leftrightarrow \Sigma N$ 
it leads to    
 a coupled-channel problem with a (non-real) hermitian 
potential. This can be handled in principle,
but complicates the solution and moreover this coupling is not needed. 
Therefore, we take $f_S=0$.\\[0.2cm]

\noindent f)\ Pomeron-exchange ($J^{PC}=0^{++}$):
       The vertices for this `diffractive'-exchange have the
       same Lorentz structure as those for scalar-meson-exchange.\\[0.2cm]
\noindent g)\ Odderon-exchange ($J^{PC}=1^{--}$):
\begin{equation}
 {\cal H}_O = g_O[\bar{\psi}\gamma_\mu\psi] \phi^\mu_O + \frac{f_O}{4{\cal M}}
 [\bar{\psi}\sigma_{\mu\nu}\psi] (\partial^\mu\phi^\nu_O-\partial^\nu\phi_O^\mu).
\label{eq:OBE.4}\end{equation}
Since the gluons are flavorless, Odderon-exchange is treated as an SU(3)-singlet.
Furthermore, since the Odderon represents a Regge-trajectory with an intercept
equal to that of the Pomeron, and is supposed not to contribute for small ${\bf k}^2$,
we include a factor ${\bf k}^2/{\cal M}^2$ in the coupling.\\[0.2cm]
 
Including form factors $f({\bf x}'-{\bf x})$ ,
the interaction Hamiltonian densities are modified to
\begin{equation}
        H_{X}({\bf x})=\int\!d^{3}x'\,
  f({\bf x}'-{\bf x}){\cal H}_{X}({\bf x}'),
\end{equation}
 for $X= P,\ V,\ A$, and $S$ ($P =$ pseudo-scalar, $V =$ vector,
 $A=$ axial-vector, and $S =$ scalar). The         
potentials in momentum space are the same as for point interactions,
except that the coupling constants are multiplied by the Fourier
transform of the form factors.
 
In the derivation of the $V_{i}$ we employ the same approximations as in 
\cite{NRS78,MRS89}, i.e.
\begin{enumerate}
\item   We expand in $1/M$: 
    $E(p) = \left[ {\bf k}^{2}/4 +
    {\bf q}^{2}+M^{2}\right]^{\frac{1}{2}}$\\
    $\approx M+{\bf k}^{2}/8M + {\bf q}^{2}/2M$
 and keep only terms up to first order in ${\bf k}^{2}/M$ and
 ${\bf q}^{2}/M$. This except for the form factors where
 the full ${\bf k}^{2}$-dependence is kept throughout
 the calculations. Notice that the gaussian form factors
suppress the high ${\bf k}^{2}$-contributions strongly.
\item   In the meson propagators
$       (-(p_{1}-p_{3})^{2}+m^{2})  
        \approx({\bf k}^{2}+m^{2})$ .
\item   When two different baryons are involved at a $BBM$-vertex
        their average mass is used in the
        potentials and the non-zero component of the momentum transfer
        is accounted for by using an effective mass in
        the meson propagator (for details see \cite{MRS89}).     
\end{enumerate}
 
Due to the approximations we get only a linear dependence on
${\bf q}^{2}$ for $V_{1}$. In the following, separating the local and the
non-local parts, we write
\begin{equation}
  V_{i}({\bf k}^{2},{\bf q}^{2})=
  V_{i a}({\bf k}^{2})+V_{i b}({\bf k}^{2})({\bf q}^{2}+\frac{1}{4}{\bf k}^2),
\label{vcdec} \end{equation}
where in principle $i=1,8$. 
 
The OBE-potentials are now obtained in the standard way (see e.g.\
\cite{NRS78,MRS89}) by evaluating the $BB$-interaction in Born-approximation.
We write the potentials $V_{i}$ of Eqs.~(\ref{vcdec}) in the form
\begin{equation}
  V_{i}({\bf k}\,^{2},{\bf q}\,^{2})=
   \sum_{X} \Omega^{(X)}_{i}({\bf k}\,^{2})
   \cdot \Delta^{(X)} ({\bf k}^{2},m^{2},\Lambda^{2}).
\label{nrexpv2} \end{equation}
Furthermore for $X=P,V$ 
\begin{equation}
   \Delta^{(X)}({\bf k}^{2},m^{2},\Lambda^{2})= e^{-{\bf k}^{2}/\Lambda^{2}}/  
                    \left({\bf k}^{2}+m^{2}\right),
\label{propm1} \end{equation}
and for $X=S,A$ a zero in the form factor
\begin{equation}
   \Delta^{(S)}({\bf k}^{2},m^{2},\Lambda^{2})= \left(1-{\bf k}^2/U^2\right)\
  e^{-{\bf k}^{2}/\Lambda^{2}}/  
  \left({\bf k}^{2}+m^{2}\right),
\label{propm2} \end{equation}
and for $X=D,O$
\begin{equation}
   \Delta^{(D)}({\bf k}^{2},m^{2},\Lambda^{2})=\frac{1}{{\cal M}^{2}}
   e^{-{\bf k}^{2}/(4m_{P,O}^{2})}.
\label{Eq:difdel}
\end{equation}
In the latter expression ${\cal M}$ is a universal
scaling mass, which is again taken to be the proton mass.
The mass parameter $m_{P}$ controls the ${\bf k}^{2}$-dependence of
the Pomeron-, $f$-, $f'$-, $A_{2}$-, and $K^{\star\star}$-potentials.
Similarly, $m_O$ controls the ${\bf k}^2$-dependence of the Odderon.\\

\noindent {\it In the following we give the OBE-potentials in momentum-space for the 
hyperon-nucleon systems. From these those for NN and YY can be deduced easily.
We assign the particles 1 and 3 to be hyperons, and particles 2 and 4 to be
nucleons. Mass differences among the hyperons and among the nucleons will be neglected.}\\


\onecolumngrid


\subsection{Non-strange Meson-exchange}
\label{app:OBE.a}
For the non-strange mesons the mass differences at the vertices are neglected,
we take at the $YYM$- and the $NNM$-vertex the average hyperon and the average
nucleon mass respectively. This implies that we do not include contributions
to the Pauli-invariants $P_7$ and $P_8$.
 For vector-, and diffractive OBE-exchange we  
refer the reader to Ref.~\cite{MRS89}, where the contributions to the different
$\Omega^{(X)}_{i}$'s for baryon-baryon scattering are given in detail.
\begin{enumerate}
 \item[(a)]   Pseudoscalar-meson exchange:
      \begin{subequations}
      \begin{eqnarray}
       \Omega^{(P)}_{2a} & = & -g^p_{13}g^p_{24}\left( \frac{{\bf k}^{2}}
           {12M_yM_n} \right) \ \ ,\ \ 
       \Omega^{(P)}_{3a}  =  -g^p_{13}g^p_{24}\left( \frac{1}
           {4M_yM_n}  \right), \label{eq1a} \\
       \Omega^{(P)}_{2b} & = & +g^p_{13}g^p_{24}\left( \frac{{\bf k}^{2}}
           {24M_y^2M_n^2} \right) \ \ ,\ \ 
       \Omega^{(P)}_{3b}  =  +g^p_{13}g^p_{24}\left( \frac{1}
           {8M_y^2M_n^2}  \right), \label{eq1b}.    
         \end{eqnarray}
\hspace{3mm} PV-formulas:
      \begin{eqnarray}
       \Omega^{(P)}_{2a} & = & -f^{pv}_{13}f^{pv}_{24}\left( \frac{{\bf k}^{2}}
           {3 m_{\pi^+}^2} \right) \ \ ,\ \ 
       \Omega^{(P)}_{3a}  =  -f^{pv}_{13}f^{pv}_{24}\left( \frac{1}
           { m_{\pi^+}^2} \right), \label{eq2a} \\
       \Omega^{(P)}_{2b} & = & +f^{pv}_{13}f^{pv}_{24}\left( \frac{{\bf k}^{2}}
           {6 m_{\pi^+}^2 M_y M_n} \right) \ \ ,\ \ 
       \Omega^{(P)}_{3b}  =  +f^{pv}_{13}f^{pv}_{24}\left( \frac{1}
           {2m_{\pi^+}^2 M_y^2M_n^2}  \right), \label{eq2b}.    
         \end{eqnarray}
      \end{subequations}
 \item[(b)]   Vector-meson exchange:
     \begin{eqnarray}  
       \Omega^{(V)}_{1a}&=&
   \left\{g^v_{13}g^v_{24}\left( 1-\frac{{\bf k}^{2}}{2M_yM_n}\right)
           -g^v_{13}f^v_{24}\frac{{\bf k}^{2}}{4{\cal M}M_n} 
      -f^v_{13}g^v_{24}\frac{{\bf k}^{2}}{4{\cal M}M_y}
 \vphantom{\frac{A}{A}}\right. \nonumber\\ && \left. \vphantom{\frac{A}{A}}
           +f^v_{13}f^v_{24}\frac{{\bf k}^{4}}
           {16{\cal M}^{2}M_yM_n}\right\},\ \                 
    \Omega^{(V)}_{1b} =  g^v_{13}g^v_{24}\left(
    \frac{3}{2M_yM_n}\right), \nonumber\\
  \Omega^{(V)}_{2a} &=& -\frac{2}{3} {\bf k}^{2}\,\Omega^{(V)}_{3a}, \ \ 
  \Omega^{(V)}_{2b}  =  -\frac{2}{3} {\bf k}^{2}\,\Omega^{(V)}_{3b}, 
 \nonumber\\
    \Omega^{(V)}_{3a}&=& \left\{
           (g^v_{13}+f^v_{13}\frac{M_y}{{\cal M}})
           (g^v_{24}+f^v_{24}\frac{M_n}{{\cal M}}) 
          -f^v_{13}f^v_{24}\frac{{\bf k}^{2}}{8{\cal M}^{2}} \right\}
            /(4M_yM_n), \nonumber\\                 
    \Omega^{(V)}_{3b}&=& -
           (g^v_{13}+f^v_{13}\frac{M_y}{{\cal M}})
           (g^v_{24}+f^v_{24}\frac{M_n}{{\cal M}}) 
            /(8M_y^2M_n^2), \nonumber\\                 
    \Omega^{(V)}_{4}&=&-\left\{12g^v_{13}g^v_{24}+8(g^v_{13}f^v_{24}+f^v_{13}g^v_{24})
           \frac{\sqrt{M_yM_n}}{{\cal M}} 
     - f^v_{13}f^v_{24}\frac{3{\bf k}^{2}}{{\cal M}^{2}}\right\}
            /(8M_yM_n)              \nonumber\\
       \Omega^{(V)}_{5}&=&- \left\{
           g^v_{13}g^v_{24}+4(g^v_{13}f^v_{24}+f^v_{13}g^v_{24})
           \frac{\sqrt{M_yM_n}}{{\cal M}}  
           +8f^v_{13}f^v_{24}\frac{M_yM_n}{{\cal M}^{2}}\right\}
          /(16M_y^{2}M_n^{2})        \nonumber\\
       \Omega^{(V)}_{6}&=&-\left\{(g^v_{13}g^v_{24}
           +f^v_{13}f^v_{24}\frac{{\bf k}^{2}}{4{\cal M}^{2}})
    \frac{(M_n^{2}-M_y^{2})}{4M_y^{2}M_n^{2}} 
      -(g^v_{13}f^v_{24}-f^v_{13}g^v_{24})
      \frac{1}{\sqrt{{\cal M}^{2}M_yM_n}}\right\}.
 \nonumber\\
 \label{eq2}\end{eqnarray}
 \item[(c)]   Scalar-meson exchange:  \hspace{2em}
      \begin{eqnarray} 
      \Omega^{(S)}_{1} & = & 
      -g^s_{13} g^s_{24} \left( 1+\frac{{\bf k}^{2}}{4M_yM_n} 
       -\frac{{\bf q}^2}{2M_yM_n}\right)
       \nonumber\\ &&\nonumber\\
      \Omega^{(S)}_{1b} & = & +g^s_{13} g^s_{24} \frac{1}{2M_yM_n}
  \ \ ,\ \ 
      \Omega^{(S)}_{4}= -g^s_{13} g^s_{24} \frac{1}{2M_yM_n}
       \nonumber\\ &&\nonumber\\
      \Omega^{(S)}_{5} &=& g^s_{13} g^s_{24}
        \frac{1}{16M_y^{2}M_n^{2} } 
  \ \ ,\ \ 
      \Omega^{(S)}_{6}= -g^s_{13} g^s_{24}
        \frac{(M_n^{2}-M_y^{2})}{4M_yM_n}.
       \label{Eq:scal} \end{eqnarray}
\item[(d)] Axial-vector-exchange $J^{PC}=1^{++}$:
      \begin{eqnarray} 
      \Omega^{(A)}_{2a} & = & -g^a_{13}g^a_{24}\left[
         1-\frac{2{\bf k}^2}{3M_yM_n}\right]
         +\left[\left(g_{13}^A f_{24}^A\frac{M_n}{{\cal M}}
         +f_{13}^A g_{24}^A \frac{M_y}{{\cal M}}\right)
         -f_{13}^A f_{24}^A \frac{{\bf k}^2}{2{\cal M}^2}\right]\
         \frac{{\bf k}^2}{6M_yM_n}
       \nonumber\\ && \nonumber\\
      \Omega^{(A)}_{2b} &=& 
        -g^a_{13}g^a_{24} \left(\frac{3}{2M_yM_n}\right) 
        \nonumber\\ && \nonumber\\
      \Omega^{(A)}_{3}&=&
        -g^a_{13}g^a_{24} \left[\frac{1}{4M_yM_n}\right]
         +\left[\left(g_{13}^A f_{24}^A\frac{M_n}{{\cal M}}
         +f_{13}^A g_{24}^A \frac{M_y}{{\cal M}}\right)
         -f_{13}^A f_{24}^A \frac{{\bf k}^2}{2{\cal M}^2}\right]\
         \frac{1}{2M_yM_n}
       \nonumber\\ && \nonumber\\
	\Omega^{(A)}_{4}  &=&
     -g^a_{13}g^a_{24}   \left[\frac{1}{2M_yM_n}\right] 
      \ \ ,\ \
      \Omega^{(A)}_{6} = 
     -g^a_{13}g^a_{24} \left[\frac{(M_n^{2}-M_y^{2})}{4M_y^2M_n^2}\right]
       \nonumber\\ && \nonumber\\
      \Omega^{(A)'}_{5} & = &
     -g^a_{13}g^a_{24}   \left[\frac{2}{M_yM_n}\right] 
         \label{eq:axi1} \end{eqnarray}
Here, we used the B-field description with $\alpha_r=1$, see Appendix~\ref{app:C}.
 The detailed treatment of the potential proportional to $P_5'$, i.e. 
 with $\Omega_5^{(A)'}$, is given in \cite{Rij04a}, Appendix~B.
\item[(e)] Axial-vector mesons with $J^{PC}=1^{+-}$: 
      \begin{eqnarray} 
       \Omega^{(B)}_{2a} & = & +f^B_{13}f^B_{24}\frac{(M_n+M_y)^2}{m_B^2}
       \left(1-\frac{{\bf k}^2}{4M_yM_n}\right)
       \left( \frac{{\bf k}^{2}}{12M_yM_n} \right),\ \ 
       \Omega^{(B)}_{2b}   =   +f^B_{13}f^B_{24}\frac{(M_n+M_y)^2}{m_B^2}
       \left( \frac{{\bf k}^{2}}{8M_y^2M_n^2} \right)     
     \nonumber\\ 
       \Omega^{(B)}_{3a} & = & +f^B_{13}f^B_{24}\frac{(M_n+M_y)^2}{m_B^2}
       \left(1-\frac{{\bf k}^2}{4M_yM_n}\right)
       \left( \frac{1}{4M_yM_n} \right),\ \             
       \Omega^{(B)}_{3b}   =   +f^B_{13}f^B_{24}\frac{(M_n+M_y)^2}{m_B^2}
       \left( \frac{3}{8M_y^2M_n^2} \right). \nonumber\\     
     \label{eq:bxi1} \end{eqnarray}
 \item[(f)]   Diffractive-exchange (pomeron, $f, f', A_{2}$): \\
         The $\Omega^{D}_{i}$ are the same as for scalar-meson-exchange
         Eq.(\ref{Eq:scal}), but with
         $\pm g_{13}^{S}g_{24}^{S}$ replaced by
         $\mp g_{13}^{D}g_{24}^{D}$, and except for the zero in the form factor.
\item[(g)] Odderon-exchange:              
         The $\Omega^{O}_{i}$ are the same as for vector-meson-exchange
         Eq.(ref{eq2}), but with
         $ g_{13}^{V}\rightarrow g_{13}^{O}$, 
         $ f_{13}^{V}\rightarrow f_{13}^{O}$ and similarly for the couplings
         with the 24-subscript.

\end{enumerate}

As in Ref.~\cite{MRS89} in the derivation of the expressions for $\Omega_i^{(X)}$, 
given above, $M_y$ and $M_n$ denote the mean hyperon and nucleon
mass, respectively \begin{math} M_y=(M_{1}+M_{3})/2 \end{math}
and \begin{math} M_n=(M_{2}+M_{4})/2 \end{math},
 and $m$ denotes the mass of the exchanged meson.
Moreover, the approximation                            
        \begin{math}
              1/ M^{2}_{N}+1/ M^{2}_{Y}\approx
              2/ M_nM_y,
        \end{math}
is used, which is rather good since the mass differences
between the baryons are not large.\\

\subsection{One-Boson-Exchange Interactions in Configuration Space I}
\label{app.IIIb}
In configuration space the BB-interactions are described by potentials
of the general form
\begin{subequations}
\begin{eqnarray}
 V &=& \left\{\vphantom{\frac{A}{A}} V_C(r) + V_\sigma(r)
\mbox{\boldmath $\sigma$}_1\cdot\mbox{\boldmath $\sigma$}_2
 + V_T(r) S_{12} + V_{SO}(r) {\bf L}\cdot{\bf S} + V_Q(r) Q_{12}
 \right.\nonumber\\ && \left.
 + V_{ASO}(r)\ \frac{1}{2}(\mbox{\boldmath $\sigma$}_1-
  \mbox{\boldmath $\sigma$}_2)\cdot{\bf L}
 -\frac{1}{2M_yM_n}\left(\vphantom{\frac{A}{A}} 
\mbox{\boldmath $\nabla$}^2 V^{n.l.}(r) + V^{n.l.}(r) 
 \mbox{\boldmath $\nabla$}^2\right)
\right\}\cdot {\cal P}, \\
 V^{n.l.} &=& \left\{\vphantom{\frac{A}{A}} \varphi_C(r) + \varphi_\sigma(r)
\mbox{\boldmath $\sigma$}_1\cdot\mbox{\boldmath $\sigma$}_2
 + \varphi_T(r) S_{12}\right\}\cdot {\cal P}, 
 \label{eq:3b.a}\end{eqnarray}
\end{subequations}
where for non-strange mesons ${\cal P}=1$, and
\begin{subequations}
\begin{eqnarray}
 S_{12} &=& 3 (\mbox{\boldmath $\sigma$}_1\cdot\hat{r})
 (\mbox{\boldmath $\sigma$}_2\cdot\hat{r}) -
 (\mbox{\boldmath $\sigma$}_1\cdot\mbox{\boldmath $\sigma$}_2), \\
 Q_{12} &=& \frac{1}{2}\left[\vphantom{\frac{A}{A}} 
 (\mbox{\boldmath $\sigma$}_1\cdot{\bf L})(\mbox{\boldmath $\sigma$}_2\cdot{\bf L})
 +(\mbox{\boldmath $\sigma$}_2\cdot{\bf L})(\mbox{\boldmath $\sigma$}_1\cdot{\bf L})
 \right], \\
 \phi(r) &=& \phi_C(r) + \phi_\sigma(r) 
 \mbox{\boldmath $\sigma$}_1\cdot\mbox{\boldmath $\sigma$}_2, 
 \label{eq:3b.b}\end{eqnarray}
\end{subequations}
For the basic functions for the Fourier transforms with gaussian form factors,
we refer to Refs.~\cite{NRS78,MRS89}.                           
For the details of the Fourier transform for the potentials with $P_5'$, which 
occur in the case of the axial-vector mesons with $J^{PC}=1^{++}$, we refer 
to Appendix~\ref{app:B}. 

\noindent (a)\ Pseudoscalar-meson-exchange:
\begin{subequations}
\begin{eqnarray}
  V_{PS}(r) &=& \frac{m}{4\pi}\left[ g^p_{13}g^p_{24}\frac{m^2}{4M_yM_n}
 \left(\frac{1}{3}(\mbox{\boldmath $\sigma$}_1\cdot\mbox{\boldmath $\sigma$}_2)\
 \phi_C^1 + S_{12} \phi_T^0\right)\right] {\cal P}, \\
  V_{PS}^{n.l.}(r) &=& \frac{m}{4\pi}\left[ g^p_{13}g^p_{24}\frac{m^2}{4M_yM_n}
 \left(\frac{1}{3}(\mbox{\boldmath $\sigma$}_1\cdot\mbox{\boldmath $\sigma$}_2)\
 \phi_C^1 + S_{12} \phi_T^0\right)\right] {\cal P}. 
 \label{eq:3b.1}\end{eqnarray}
\end{subequations}
\noindent (b)\ Vector-meson-exchange:          
\begin{subequations}
\begin{eqnarray}
&& V_{V}(r) = \frac{m}{4\pi}\left[\left\{ g^v_{13}g^v_{24}\left[ \phi_C^0 +
 \frac{m^2}{2M_yM_n} \phi_C^1 
\right]
\right.\right.\nonumber\\ && \left.\left.  \hspace{0cm} 
 +\left[g^v_{13}f^v_{24}\frac{m^2}{4{\cal M}M_n}
 +f^v_{13}g^v_{24}\frac{m^2}{4{\cal M}M_y}\right] \phi_C^1 + f^v_{13}f^v_{24}
\frac{m^4}{16{\cal M}^2 M_y M_n} \phi_C^2\right\}
  \right.\nonumber\\ && \left.  \hspace{0cm} 
 +\frac{m^2}{6M_yM_n}\left\{\left[ \left(g^v_{13}+f^v_{13}\frac{M_y}{{\cal M}}\right)\cdot
 \left(g^v_{24}+f^v_{24}\frac{M_n}{{\cal M}}\right)\right] \phi_C^1 
 +f^v_{13} f^v_{24}\frac{m^2}{8{\cal M}^2} \phi_C^2\right\}
 (\mbox{\boldmath $\sigma$}_1\cdot\mbox{\boldmath $\sigma$}_2)\
  \right.\nonumber\\ && \left.  \hspace{0cm} 
 -\frac{m^2}{4M_yM_n}\left\{\left[ \left(g^v_{13}+f^v_{13}\frac{M_y}{{\cal M}}\right)\cdot
 \left(g^v_{24}+f^v_{24}\frac{M_n}{{\cal M}}\right)\right] \phi_T^0 
 +f^v_{13} f^v_{24}\frac{m^2}{8{\cal M}^2} \phi_T^1\right\} S_{12}
  \right.\nonumber\\ && \left.  \hspace{0cm} 
 -\frac{m^2}{M_yM_n}\left\{\left[ \frac{3}{2}g^v_{13}g^v_{24}
 +\left(g^v_{13}f^v_{24}+f^v_{13}g^v_{24}\right)
 \frac{\sqrt{M_yM_n}}{{\cal M}}\right] \phi_{SO}^0 
 +\frac{3}{8}f^v_{13} f^v_{24}\frac{m^2}{{\cal M}^2} \phi_{SO}^1\right\} {\bf L}\cdot{\bf S}
  \right.\nonumber\\ && \left.  \hspace{0cm} 
 +\frac{m^4}{16M_y^2M_n^2}\left\{\left[ g^v_{13}g^v_{24}
 +4\left(g^v_{13}f^v_{24}+f^v_{13}g^v_{24}\right)
 \frac{\sqrt{M_yM_n}}{{\cal M}} 
 +8f^v_{13}f^v_{24}\frac{M_yM_n}{{\cal M}^2}\right]\right\} 
  \cdot\right.\nonumber\\ && \left.  \hspace{0cm} \times
\frac{3}{(mr)^2} \phi_T^0 Q_{12}
 -\frac{m^2}{M_yM_n}\left\{\left[ 
 \left(g^v_{13}g^v_{24}-f^v_{13}f^v_{24}\frac{m^2}{{\cal M}^2}\right)
  \frac{(M_n^2-M_y^2)}{4M_yM_n}
  \right.\right.\right.\nonumber\\ && \left.\left.\left.  \hspace{0cm} 
  -\left(g^v_{13}f^v_{24}-f^v_{13}g^v_{24}\right)\frac{\sqrt{M_yM_n}}{{\cal M}}\right] \phi_{SO}^0
 \right\}\cdot\frac{1}{2}\left(
 \mbox{\boldmath $\sigma$}_1-\mbox{\boldmath $\sigma$}_2\right)\cdot{\bf L}\right] {\cal P},
\\
&& V_{V}^{n.l.}(r) = \frac{m}{4\pi}\left[ \frac{3}{2} g^v_{13}g^v_{24}\ \phi_C^0 
  \right.\nonumber\\ && \left.  \hspace{0cm} 
 +\frac{m^2}{6M_yM_n}\left\{\left[ \left(g^v_{13}+f^v_{13}\frac{M_y}{{\cal M}}\right)\cdot
 \left(g^v_{24}+f^v_{24}\frac{M_n}{{\cal M}}\right)\right] \phi_C^1 \right\}
 (\mbox{\boldmath $\sigma$}_1\cdot\mbox{\boldmath $\sigma$}_2)\
  \right.\nonumber\\ && \left.  \hspace{0cm} 
 -\frac{m^2}{4M_yM_n}\left\{\left[ \left(g^v_{13}+f^v_{13}\frac{M_y}{{\cal M}}\right)\cdot
 \left(g^v_{24}+f^v_{24}\frac{M_n}{{\cal M}}\right)\right] \phi_T^0 \right\} S_{12}
\right] {\cal P}. 
 \label{eq:3b.2}\end{eqnarray}
\end{subequations}
Note: the non-local tensor and "associated" spin-spin terms are 
not included in ESC08c-model.\\

\noindent (c)\ Scalar-meson-exchange:          
\begin{eqnarray}
 V_{S}(r) &=& -\frac{m}{4\pi}\left[ g^s_{13}g^s_{24}\left\{\left[ \phi_C^0 
 -\frac{m^2}{4M_yM_n} \phi_C^1\right] + \frac{m^2}{2M_yM_n} \phi_{SO}^0\ {\bf L}\cdot{\bf S}
 +\frac{m^4}{16M_y^2M_n^2}
 \cdot\right.\right.\nonumber\\ && \left.\left. \times
\frac{3}{(mr)^2} \phi_T^0 Q_{12} 
 +\frac{m^2}{M_yM_n} \left[\frac{(M_n^2-M_y^2)}{4M_yM_n}\right] \phi_{SO}^0\cdot
 \frac{1}{2}\left(\mbox{\boldmath $\sigma$}_1-\mbox{\boldmath $\sigma$}_2\right)\cdot{\bf L}
  \right.\right.\nonumber\\ && \left.\left.  \hspace{0.0cm}
  +\frac{1}{4M_yM_n}\left(\mbox{\boldmath $\nabla$}^2 \phi_C^0 
 + \phi_C^0 \mbox{\boldmath $\nabla$}^2\right) \right\}\right] {\cal P}.
 \label{eq:3b.3}\end{eqnarray}
\noindent (d)\ Axial-vector-meson exchange $J^{PC}=1^{++}$:
\begin{eqnarray}
&& V_{A}(r) = -\frac{m}{4\pi}\left[ 
 \left\{ g^a_{13}g^a_{24}\left(\phi_C^0 +\frac{2m^2}{3M_yM_n} \phi_C^1\right)
 +\frac{m^2}{6M_yM_n}\left(g^a_{13}f^a_{24}\frac{M_n}{{\cal M}}
 +f^a_{13}g^a_{24}\frac{M_y}{{\cal M}}\right)\phi_C^1
\right.\right.\nonumber\\ && \left.\left.
 +f^a_{13}f^a_{24}\frac{m^4}{12M_yM_n{\cal M}^2}\phi_C^2\right\}
 (\mbox{\boldmath $\sigma$}_1\cdot\mbox{\boldmath $\sigma$}_2)
  -\frac{3}{4M_yM_n} g^a_{13}g^a_{24}\left(\mbox{\boldmath $\nabla$}^2 \phi_C^0 
 + \phi_C^0 \mbox{\boldmath $\nabla$}^2\right) 
 (\mbox{\boldmath $\sigma$}_1\cdot\mbox{\boldmath $\sigma$}_2)
 \right.\nonumber\\ && \left. 
 - \frac{m^2}{4M_yM_n}\left\{\left[g^a_{13}g^a_{24}-2\left(g^a_{13}f^a_{24}
 \frac{M_n}{{\cal M}}+f^a_{13}g^a_{24}\frac{M_y}{{\cal M}}\right)\right] \phi_T^0
 -f^a_{13}f^a_{24}\frac{m^2}{{\cal M}^2} \phi_T^1\right\} S_{12}
 \right.\nonumber\\ && \left. 
 +\frac{m^2}{2M_yM_n}g^a_{13}g^a_{24} \left\{\phi_{SO}^0\ {\bf L}\cdot{\bf S}
 +\frac{m^2}{M_yM_n} \left[\frac{(M_n^2-M_y^2)}{4M_yM_n}\right] \phi_{SO}^0\cdot
 \frac{1}{2}\left(\mbox{\boldmath $\sigma$}_1-\mbox{\boldmath $\sigma$}_2\right)\cdot{\bf L}
 \right\}\right] {\cal P}.     
 \label{eq:3b.4}\end{eqnarray}
\noindent (e)\ Axial-vector-meson exchange $J^{PC}=1^{+-}$:
\begin{subequations}
\begin{eqnarray}
 V_{B}(r) &=& -\frac{m}{4\pi}\frac{(M_n+M_y)^2}{m^2}\left[ 
 f^B_{13}f^B_{24}\left\{\frac{m^2}{12M_yM_n}\left(\phi_C^1+
 \frac{m^2}{4M_yM_n} \phi_C^2\right)
 (\mbox{\boldmath $\sigma$}_1\cdot\mbox{\boldmath $\sigma$}_2)
 \right.\right.\nonumber\\ && \left.\left. 
  -\frac{m^2}{8M_yM_n}\left(\mbox{\boldmath $\nabla$}^2 \phi_C^1 
 + \phi_C^1 \mbox{\boldmath $\nabla$}^2\right) 
 (\mbox{\boldmath $\sigma$}_1\cdot\mbox{\boldmath $\sigma$}_2)
 +\left[\frac{m^2}{4M_yM_n}\right] \phi^0_T\ S_{12}\right\}\right] {\cal P}, \\
 V_{B}^{n.l.}(r) &=& -\frac{m}{4\pi}\frac{(M_n+M_y)^2}{m^2}\left[ 
 f^B_{13}f^B_{24}\left\{
  \frac{3m^2}{4M_yM_n} \left(\frac{1}{3}  
  \mbox{\boldmath $\sigma$}_1\cdot\mbox{\boldmath $\sigma$}_2\ \phi_C^1
  + S_{12}\ \phi_T^0\right)\right\}\right] {\cal P}. 
 \label{eq:3b.5}\end{eqnarray}
\end{subequations}
\noindent (f)\ Diffractive exchange:           
\begin{eqnarray}
&& V_{D}(r) = \frac{m_P}{4\pi}\left[ g^D_{13}g^D_{24} 
 \frac{4}{\sqrt{\pi}}\frac{m_P^2}{{\cal M}^2}\cdot\left[\left\{
 1+\frac{m_P^2}{2M_yM_n}(3-2 m_P^2r^2) + \frac{m_P^2}{M_yM_n} {\bf L}\cdot{\bf S}
  \right.\right.\right.\nonumber\\ && \left.\left.\left.  \hspace{0.75cm} 
 +\left(\frac{m_P^2}{2M_yM_n}\right)^2 Q_{12}
 +\frac{m_P^2}{M_yM_n} \left[\frac{(M_n^2-M_y^2)}{4M_yM_n}\right]\cdot
 \frac{1}{2}\left(\mbox{\boldmath $\sigma$}_1-\mbox{\boldmath $\sigma$}_2\right)\cdot{\bf L}
 \right\}\ e^{-m_P^2r^2} 
  \right.\right.\nonumber\\ && \left.\left.  \hspace{1.5cm} 
  +\frac{1}{4M_yM_n}\left(\mbox{\boldmath $\nabla$}^2 e^{-m_P^2r^2} 
 + e^{-m_P^2r^2}\mbox{\boldmath $\nabla$}^2\right) \right]\right] {\cal P}.
 \label{eq:3b.6}\end{eqnarray}
\noindent (g)\ Odderon-exchange:                      
\begin{subequations}
\begin{eqnarray}
 V_{O,C}(r) &=& +\frac{g^O_{13}g^O_{24}}{4\pi}\frac{8}{\sqrt{\pi}}\frac{m_O^5}{{\cal M}^4}
 \left[\left(3-2m_O^2r^2\right) \right.\nonumber\\ && \left. 
 -\frac{m_O^2}{M_yM_n}\left( 15 - 20 m_O^2r^2+4 m_O^4r^4\right)
 \right]\exp(-m_O^2 r^2)\ , \\
 V_{O,n.l.}(r) &=& -\frac{g^O_{13}g^O_{24}}{4\pi}\frac{8}{\sqrt{\pi}}\frac{m_O^5}{{\cal M}^4}
 \frac{3}{4M_yM_n}\left\{\mbox{\boldmath $\nabla$}^2
 \left[(3-2m_O^2r^2)\exp(-m_O^2 r^2)\right]+ \right.\nonumber\\
 && \left. + \left[(3-2m_O^2r^2)\exp(-m_O^2 r^2)\right] 
 \mbox{\boldmath $\nabla$}^2 \right\}\ , \\
 V_{O,\sigma}(r) &=& 
-\frac{g^O_{13}g^O_{24}}{4\pi}\frac{8}{3\sqrt{\pi}}\frac{m_O^5}{{\cal M}^4}
 \frac{m_O^2}{M_yM_n}
 \left[15-20 m_O^2r^2+4 m_O^4 r^4\right]\exp(-m_O^2 r^2)\cdot \nonumber\\
 && \times\left(1+\kappa^O_{13}\frac{M_y}{\cal M}\right) 
 \left(1+\kappa^O_{24}\frac{M_n}{\cal M}\right) 
 , \\
 V_{O,T}(r) &=& -\frac{g^O_{13}g^O_{24}}{4\pi}\frac{8}{3\sqrt{\pi}}\frac{m_O^5}{{\cal M}^4}
 \frac{m_O^2}{M_yM_n}\cdot m_O^2 r^2
 \left[7-2 m_O^2r^2\right]\exp(-m_O^2 r^2)\cdot \nonumber\\
 && \times\left(1+\kappa^O_{13}\frac{M_y}{\cal M}\right) 
 \left(1+\kappa^O_{24}\frac{M_n}{\cal M}\right) 
 , \\
 V_{O,SO}(r) &=& -\frac{g^O_{13}g^O_{24}}{4\pi}\frac{8}{\sqrt{\pi}}\frac{m_O^5}{{\cal M}^4}
 \frac{m_O^2}{M_yM_n} \left[5-2 m_O^2r^2\right]\exp(-m_O^2 r^2)\cdot \nonumber\\
 && \times\left\{3+\left(\kappa^O_{13}+\kappa^O_{24}\right)\frac{\sqrt{M_yM_n}}{\cal M}\right\}
 , \\
 V_{O,Q}(r) &=& +\frac{g^O_{13}g^O_{24}}{4\pi}\frac{2}{\sqrt{\pi}}\frac{m_O^5}{{\cal M}^4}
 \frac{m_O^4}{M_y^2M_n^2} 
 \left[7-2 m_O^2r^2\right]\exp(-m_O^2 r^2)\cdot \nonumber\\
 && \times\left\{1+4\left(\kappa^O_{13}+\kappa^O_{24}\right)\frac{\sqrt{M_yM_n}}{\cal M}
 +8\kappa_{13}\kappa_{24}\frac{M_yM_n}{{\cal M}^2}\right\}
 , \\
 V_{O,ASO}(r) &=& -\frac{g^O_{13}g^O_{24}}{4\pi}\frac{4}{\sqrt{\pi}}\frac{m_O^5}{{\cal M}^4}
 \frac{m_O^2}{M_yM_n} \left[5-2 m_O^2r^2\right]\exp(-m_O^2 r^2)\cdot \nonumber\\
 &&  \times\left\{ \frac{M_n^2-M_y^2}{M_yM_n}
-4\left(\kappa^O_{24}-\kappa^O_{13}\right)
 \frac{\sqrt{M_yM_n}}{\cal M} \right\}.
 \label{eq:3b.7}\end{eqnarray}
\end{subequations}

\subsection{Strange Meson-exchange}
\label{app:OBE.d}
The rules for hypercharge nonzero exchange have been given in e.g.
Ref.~\cite{Rij04b}.
The potentials for non-zero hypercharge exchange ($K, K^*,\kappa, K_A, K_B)$ 
are obtained from the expressions   
given in the previous subsections for non-strange mesons by taking
care of the following points: 
(a) For strange meson exchange ${\cal P}=-{\cal P}_x {\cal P}_\sigma$.       
(b) In the latter case one has to replace both $M_n$ and $M_y$ by 
$\sqrt{M_yM_n}$, and reverse the sign of the antisymmetric spin orbit.
\begin{flushleft}
\rule{16cm}{0.5mm}
\end{flushleft}

\section{Folding Amplitude Scalar-exchange II}            
\label{app:sc-vertex2}
In this Appendix the lower vertex in Fig.~\ref{fig:2.2} is worked 
out for the scalar-meson coupling. This in order to check the
signs in the vertex function in comparison with the upper vertex.\\
The Dirac-spinor part of the scalar-meson QQ-vertex is
\begin{eqnarray}
 \left[\bar{u}_{i}({\bf q'}_{i}) u_{i}({\bf q}_{i}) 
  \right] &=& 
 \sqrt{\frac{E'_{i}+m'_{i}}{2m'_{i}}\ \frac{E_{i}+m_{i}}{2m_{i}} }\cdot
 \chi^{\prime \dagger}_{i}\cdot 
 \left[ 1 -
 \frac{\mbox{\boldmath $\sigma$}_{i}\cdot{\bf q'}_{i}}{E'_{i}+m_{i}}
 \frac{\mbox{\boldmath $\sigma$}_{i}\cdot{\bf q}_{i}}{E_{i}+m_{i}}\right]
 \nonumber\\ &\approx& 
 \chi^{\prime \dagger}_{i}\left[ 1 -
 \frac{{\bf q}'_i\cdot{\bf q}_i}{4m_i^2}                                        
 -\frac{i}{4m_i^2}\bm{\sigma}_i\cdot{\bf q}'_i\times{\bf q}_i\right]\ \chi_i
 \nonumber\\ &=& 
 \chi^{\prime \dagger}_{i}\left[ 1 -
 \frac{{\bf S}_i^2-{\bf k}^2}{16m_i^2}                                        
 -\frac{i}{8m_i^2}\bm{\sigma}_i\cdot{\bf S}_i\times{\bf k}\right]\ \chi_i.
\label{app:3.41} \end{eqnarray}
Here is used that for the CQM $E_i \approx m_i$. The performance of the 
${\bf Q}$-integral in (\ref{app:3.41}) gives
\begin{eqnarray}
 \left[\bar{u}_{i}({\bf q'}_{i}) u_{i}({\bf q}_{i}) 
  \right] &\Rightarrow& \chi^{\prime \dagger}_{i}
 \left[ 1 -\left(\frac{1}{4m_i^2R_N^2}+\frac{{\bf q}^2}{36m_i^2}\right)
 + \frac{{\bf k}^2}{16m_i^2}             
 +\frac{i}{12m_i^2}\bm{\sigma}_i\cdot{\bf q}\times{\bf k}\right]\ \chi_i
\label{app:3.42} \end{eqnarray}
Summing over the quarks leads to the vertex
\begin{eqnarray}
 \Gamma_{CQM} &=&
 \sum_{i=1-3}\left[\bar{u}_{i}({\bf q'}_{i}) u_{i}({\bf q}_{i}) 
  \right] \Rightarrow
 3\left[ 1 -\left(\frac{1}{4m_Q^2R_N^2}+\frac{{\bf q}^2}{36m_Q^2}\right)
 + \frac{{\bf k}^2}{16m_Q^2}             
 +\frac{i}{36m_Q^2}\sum_i \bm{\sigma}_i\cdot{\bf q}\times{\bf k}\right]
\label{app:3.43} \end{eqnarray}
The CQM replacement $m_Q \approx \sqrt{M'M}/3$ leads to
\begin{eqnarray}
 \Gamma_{CQM} &=&
 3\left[ \left(1 -\frac{1}{4m_Q^2R_N^2}\right) -\frac{{\bf q}^2}{4M'M}
 + \frac{9{\bf k}^2}{16M'M}             
 +\frac{i}{4M'M}\sum_i \bm{\sigma}_N\cdot{\bf q}\times{\bf k}\right], 
\label{app:3.44} \end{eqnarray}
where we used $\sum_i \bm{\sigma}_i=\bm{\sigma}_N$.
This assumes that the spin of the nucleon is given by the total
spin of the quarks \cite{remark-spin}.

\noindent {\it Notice that the $1/R_N^2$-term in (\ref{app:3.44})
has the same sign to that of (\ref{eq:3.44}). Hence, these terms 
would not cancel in the NN-potential in this simple treatment.}

\begin{flushleft}
\rule{16cm}{0.5mm}
\end{flushleft}

\section{Folding Tensor-exchange Vertex}            
\label{app:ten}
For the coupling of the tensor mesons ($ J^{PC}= 2^{++}$) to the
quarks, similar to that for the nucleons, we take
\begin{equation}
   {\cal H}_{fNN} = -
 \left[\frac{i}{2} \bar{\psi}\left(\gamma^{\mu}\partial^{\nu}+
 \gamma^{\nu}\partial^{\mu}\right)\psi\ F_{1} - \bar{\psi}
 \partial^{\mu}\partial^{\nu}\psi\ F_{2}\right]\cdot f_{\mu\nu}\ ,
\label{app:ten.1}\end{equation}
where $f_{\mu\nu}=f_{\nu\mu}$, i.e. symmetric, and
\begin{equation}
 F_{1}\ =\ \frac{G_{T,1}}{\cal M}\ ,\  {\rm and}\
 F_{2}\ =\ \frac{G_{T,2}}{{\cal M}^{2}}\ .
\label{app:ten.2}\end{equation}
The Sach form factors are in terms of the $G_{T,i}$ defined as
\begin{equation}
 G_{M} = G_{T,1}\ ,\
 G_{E} = G_{T,1}-\frac{(t-4M^{2})}{4M^{2}} G_{T,2}\ \approx
 G_{T,1}+\left(1+\frac{{\bf k}^{2}}{4M^{2}}\right) G_{T,2}\ .
\label{app:ten.3}\end{equation}
The latter are defined for general J in e.g. Rijken, Phd. Thesis
(Nijmegen, 1975). This is of importance when we apply the constraints
imposed by EXD, which relates the tensor-meson couplings to the
vector-meson couplings. As a matter of fact, EXD predicts that
${\cal M}F_{1}=F_{V,1}$ and ${\cal M} F_{2}= F_{V,2}$ for the pairs
$(A_{2},\rho)$ and $(f(1270),\omega)$.
 
Using the Gordon decomposition, the Pauli-
couplings $G^{T}_{i}$ are related to the Dirac-couplings $g_{T},f_{T}$
by
\begin{equation}
 g_{T} = G_{T,1}+ G_{T,2}\ ,\  f_{T} = -G_{T,2}\ ,
\label{app:ten.4}\end{equation}
and notice that
\begin{equation}
 F_{1}+ M F_{2} = g_{T}/{\cal M}\ ,\  M F_{2} = -f_{T}/{\cal M}\ ,
\label{app:ten.5}\end{equation}
which is strictly valid only for ${\cal M}= M$.
 
For the $QQf$-vertices this gives the factors
\begin{eqnarray}
  \bar{u}_{i}({\bf k_i}')\Gamma_{T}^{\mu\nu}u_{i}({\bf k}_i) & \sim & \frac{1}{4}
  \bar{u}_{i}({\bf k}'_i)\left[\vphantom{\frac{A}{A}}
  \left(\vphantom{\frac{A}{A}}(k_i+k'_i)^{\nu}\gamma^{\mu}
  +(k_i+k'_i)^{\mu}\gamma^{\nu}\right)\ F_{1} +
  (k_i+k'_i)^{\mu}(k+k')^{\nu}\ F_{2}\right] u_{i}({\bf k}_i)
  \nonumber \\ && \nonumber \\ & \sim& \frac{1}{2}\ (k_i+k'_i)^{\nu}\
  \bar{u}_{i}({\bf k}'_i)\left[\vphantom{\frac{A}{A}}
  \gamma^{\mu}F_{1}+\frac{1}{2}(k+k')^{\mu}\ F_{2}\right] u_{i}({\bf k}_i)
  \nonumber \\ && \nonumber \\
  \bar{u}_{j}({\bf q}_j')\Gamma_{T}^{\mu\nu}u_{j}({\bf q}_j) & \sim & \frac{1}{4}
  \bar{u}_{j}({\bf q}'_j) \left[\vphantom{\frac{A}{A}}
  \left(\vphantom{\frac{A}{A}}(q+q')^{\rho}\gamma^{\sigma} +
  (q_j+q'_j)^{\sigma}\gamma^{\rho}\right)\ F_{1}' +
  (q_j+q'_j)^{\rho}(q+q')^{\sigma}\ F_{2}'\right] u_{j}({\bf q}_j)
  \nonumber \\ && \nonumber \\  &\sim & \frac{1}{2}\ (q_j+q'_j)^{\sigma}\
  \bar{u}_{j}({\bf q}'_j) \left[\vphantom{\frac{A}{A}}
  \gamma^{\rho}F_{1}' + \frac{1}{2}(q_j+q'_j)^{\rho}\ F_{2}'\right]
  u_{j}({\bf q}_j)
\label{app:ten.6}\end{eqnarray}
Here the symbol $\sim$ indicates that factors coming
from the normalization $\sqrt{(E+m_Q)/2m_Q}$ of the Dirac-spinors
have been suppressed.
(In the present case, as also for vector- and scalar-exchange, they
cancel out when we pass to the level of the Lippmann-Schwinger
equation.)
The second form of the vertices is equivalent to the first form due
to the symmetry of the tnesor field $f_{\mu\nu}$. Notice that, apart from
the factor $(k'_1+k_1)^\nu$, the $\Gamma_T$ matrix element in (\ref{app:ten.6})
is identical to that for the vector meson. 

\noindent The propagator for the spin-2 mesons contains the projection operator
\begin{equation}
   {\cal P}_{\mu\nu;\rho\sigma}(k) = \frac{1}{2}\left(
   P_{\mu\rho} P_{\nu\sigma}+ P_{\mu\sigma} P_{\nu\rho}\right)
   -\frac{1}{3} P_{\mu\nu} P_{\rho\sigma}\ ,
\label{app:ten.15}\end{equation}
where $ P_{\mu\nu}(k)= -\eta_{\mu\nu}+k_{\mu}k_{\nu}/m^{2}$, with
 $ k= k'_i-k_i=p'-p = q-q'$.
On-mass-shell and equal quark masses 
the $k_{\mu}k_{\nu}$-terms in the $P_{\mu\nu}(k)$ do not
contribute, so
\begin{equation}
{\cal P}_{\mu\nu;\rho\sigma}(k) \Rightarrow \frac{1}{2}\left(
\eta_{\mu\rho} \eta_{\nu\sigma}+\eta_{\mu\sigma}\eta_{\nu\rho}\right)
 -\frac{1}{3} \eta_{\mu\nu} \eta_{\rho\sigma}\ .
\label{app:ten.16}\end{equation}
Therefore, we find three contributions to the QQ-potential
\begin{equation}
   V_{T,ij} = V_{T,ij}^{(1)} + V_{T,ij}^{(2)} + V_{T,ij}^{(3)}\ ,
\label{app:ten.17}\end{equation}
where, denoting $p := k_i,p' := k'_i$ and $q := q_j, q' := q'_j$,
\begin{eqnarray}
   V_{T,ij}^{(1)} &=& - \frac{1}{8}\ (p+p')\cdot(q+q')\
   \bar{u}_{i}({\bf p}')
  \left[\gamma^{\mu}\ F_{1} + \frac{1}{2} (p+p')^{\mu}\ F_{2}
  \right] u_{i}({\bf p})
 \cdot \nonumber \\ && \nonumber \\ &\times &
  \bar{u}_{j}({\bf q}')
  \left[\gamma_{\mu}\ F_{1}' + \frac{1}{2} (q+q')_{\mu}\ F_{2}'
  \right] u_{j}({\bf q})\
  \times \left[{\bf k}^{2}+m^{2}\right]^{-1}\ ,
 \nonumber \\ && \nonumber \\
   V_{T,ij}^{(2)} &=& - \frac{1}{8}\ \bar{u}_{i}({\bf p}')
  \left[\gamma\cdot(q+q')\ F_{1}+\frac{1}{2}(p+p')\cdot(q+q')\
  F_{2} \right] u_{i}({\bf p})
 \cdot \nonumber \\ && \nonumber \\ &\times &
  \bar{u}_{j}({\bf q}')
  \left[\gamma\cdot(p+p')\ F_{1}'+\frac{1}{2}(q+q')\cdot(p+p')\
  F_{2}' \right] u_{j}({\bf q})\
  \times \left[{\bf k}^{2}+m^{2}\right]^{-1}\ ,
 \nonumber \\ && \nonumber \\
   V_{T,ij}^{(3)} &=& +\frac{1}{48}
  \left[4M F_{1}+\left(4M^{2}-t\right) F_{2}\right]
  \left[4M'F_{1}'+\left(4M'^{2}-t\right) F_{2}'\right]
 \cdot \nonumber \\ && \nonumber \\ & & \times
  \left[ \bar{u}_{i}({\bf p}') u_{i}({\bf p})\right]
  \left[\bar{u}_{j}({\bf q}') u_{j}({\bf q})\right]
  \times \left[{\bf k}^{2}+m^{2}\right]^{-1}\ ,
\label{app:ten.20}\end{eqnarray}
where for the third contribution we used the
Dirac equation $\gamma\cdot p\ u(p)= m_Q u(p)$.
This last contribution is very akin to the scalar potential and the
result can be written down almost immediately using the results of
Phys.Rev. {\bf D} 17 (1978). \\
Working out the contribution for $\mu=0$, similar to that for the scalar- and
vector-meson one finds, with $F'_{1,2}=F_{1,2}$ and $m_i=m_j=m_Q$, 
\begin{eqnarray}
V^{(1)}_{T,ij} &\approx& -\frac{1}{2}m_i^2\
\biggl[ \biggl\{ 1+\frac{{\bf Q}_i^2-{\bf k}^2}{4m_i^2} 
+\frac{i}{8m_i^2}\bm{\sigma}_i\cdot{\bf Q}_i\times{\bf k}\biggr\}\ F_1
\nonumber\\ && +\biggl\{
\biggl\{ 1-\frac{{\bf Q}_1^2-{\bf k}^2}{16m_i^2} 
+\frac{i}{8m_i^2}\bm{\sigma}_i\cdot{\bf Q}_i\times{\bf k}\biggr\}\ m_i F_2 \biggr]
\cdot\nonumber\\ && \times
\biggl[ \biggl\{ 1+\frac{{\bf S}_j^2-{\bf k}^2}{4m_j^2} 
+\frac{i}{8m_j^2}\bm{\sigma}_j\cdot{\bf S}_j\times{\bf k}\biggr\}\ F_1
\nonumber\\ && +\biggl\{
\biggl\{ 1-\frac{{\bf S}_j^2-{\bf k}^2}{16m_j^2} 
+\frac{i}{8m_j^2}\bm{\sigma}_j\cdot{\bf S}_j\times{\bf k}\biggr\}\ m_j F_2 \biggr]
\nonumber\\ &\Rightarrow& -\frac{1}{2}m_Q^2
\biggl[(F_1+m_Q F_2) + (F_1-4m_Q F_2) \frac{{\bf Q}_i^2-{\bf k}^2}{16m_Q^2}\biggr]
\cdot\nonumber\\ && \times
\biggl[(F_1+m_Q F_2) + (F_1-4m_Q F_2) \frac{{\bf S}_i^2-{\bf k}^2}{16m_Q^2}\biggr]
\nonumber\\ &=& -\frac{1}{2} g_T^2 \left(\frac{m_Q^2}{{\cal M}^2}\right)
\biggl[1 + (1+5f_T/g_T)\
\biggl\{ \frac{{\bf Q}_i^2-{\bf k}^2}{16m_Q^2}
 +\frac{{\bf S}_i^2-{\bf k}^2}{16m_Q^2}\biggr\} +\ \ldots\ \biggr], 
\label{app:ten.21}\end{eqnarray}
where we neglect the quadratic terms of the product. Similarly, the second and third terms
give
\begin{eqnarray}
V^{(2)}_{T,ij} &\approx& -\frac{1}{2}\ g_T^2 \left(\frac{m_Q^2}{{\cal M}^2}\right)
\biggl[ 1-\frac{{\bf Q}_i^2-{\bf k}^2}{16m_i^2} 
+\frac{i}{8m_i^2}\bm{\sigma}_i\cdot{\bf Q}_i\times{\bf k}\biggr]      
\biggl[ 1-\frac{{\bf S}_j^2-{\bf k}^2}{16m_j^2} 
-\frac{i}{8m_j^2}\bm{\sigma}_j\cdot{\bf S}_j\times{\bf k}\biggr]      
\nonumber\\ &\Rightarrow& 
-\frac{1}{2} g_T^2 \left(\frac{m_Q^2}{{\cal M}^2}\right)          
\biggl[1-\frac{{\bf Q}_i^2-{\bf k}^2}{16m_Q^2}
 -\frac{{\bf S}_i^2-{\bf k}^2}{16m_Q^2} +\ \dots\ \biggr],  
\label{app:ten.22}\end{eqnarray}
and
\begin{eqnarray}
V^{(3)}_{T,ij} &\approx& +\frac{1}{3}\ g_T^2 \left(\frac{m_Q^2}{{\cal M}^2}\right)
\biggl[1-\frac{{\bf Q}_i^2-{\bf k}^2}{16m_Q^2}
 -\frac{{\bf S}_i^2-{\bf k}^2}{16m_Q^2} +\ \dots\ \biggr].  
\label{app:ten.23}\end{eqnarray}

\subsection{Cancellation $(m_QR_N)^{-2}$ terms in NN-potential}
\label{app:ten.b}
In the case one sticks to the $\delta^3({\bf K}-{\bf k})$ the "spurious"
contributions to the central potentials can be (almost) completely eliminated
by the inclusion of the tensor mesons, which is illustrated below.
From the vertices $\Gamma_{CQM}$ for scalar, vector, and tensor exchange
we get, with $\kappa_T=f_T/g_T$,
\begin{subequations}\label{app:ten.33}
\begin{eqnarray}
V_{NN,sc} &\sim& -g_S^2\biggl[ \left(1-\frac{1}{4m_Q^2R_N^2}\right)^2
 -\frac{{\bf q}^2+{\bf k}^2/4}{2M^2} + \ldots\ \biggr]\ ({\bf k}^2+m_S^2)^{-1}, \\
V_{NN,vc} &\sim& +g_V^2\biggl[ \left(1+\frac{1}{4m_Q^2R_N^2}\right)^2
 +\frac{{\bf q}^2+{\bf k}^2/4}{2M^2} + \ldots\ \biggr]\ ({\bf k}^2+m_V^2)^{-1}, \\
V_{NN,tn} &\sim& -\frac{2}{3}g_T^2\biggl[ 
 \left(1+\frac{(1+3\kappa_T/2)}{4m_Q^2R_N^2}\right)
 +(1+3\kappa_T/2)\frac{{\bf q}^2+{\bf k}^2/4}{4M^2} + \ldots\biggr]\ ({\bf k}^2+m_T^2)^{-1}, 
\end{eqnarray}\end{subequations}
where the couplings $g_S, g_V$, and $g_T$ are now NN coupling constants.
Neglecting the $R_N^{-4}$ terms, the volume integral of the "spurious" 
$1/R_N^2$ terms is proportional to
\begin{eqnarray}
 I_V &\sim & \left[ \frac{g_S^2}{m_S^2}+ \frac{g_V^2}{m_V^2}
 -\frac{1}{3}(1+3\kappa_T/2)\frac{g_T^2}{m_T^2}\right]
\end{eqnarray}\label{app:ten.34}
For $m_S = m_V= m_T/\sqrt{3}$, and $\kappa_T \approx \kappa_V \approx 3.7$,
 the vanishing of this part of $I_V$ 
implies $g_T^2 \approx (2/3)(g_S^2+g_V^2)$. From QPC-mechanism
$g_S \approx g_V := \bar{g}$ leading to 
$g_t \approx (\sqrt{2/3}) \bar{g} \approx \bar{q}$.
In the approximation of "contact-approximation this shows that the 
potentials from the "spurious" terms can be made to vanish.
 



\begin{thebibliography}{99}
\bibitem{Rij06} Th.A.\ Rijken, Phys.\ Rev.\ {\bf C 73} (2006) 044007.
\bibitem{Rijk15} M.M.\ Nagels, Th.A.\ Rijken, and Y.\ Yamamoto, 
 {\it Extended-soft-core Baryon-Baryon Model ESC08c,
    I. Nucleon-Nucleon Scattering },
    [arXiv:nucl-th/1408.4825].
\bibitem{NRY19a} M.M.\ Nagels, Th.A.\ Rijken, and Y.\ Yamamoto,
  Phys.\ Rev.\ {\bf C 99}, 044002 (2019).
\bibitem{Yao73}      
A.\ Le\ Yaouanc, L.\ Oliver, O.\ P\'{e}ne, and J.-C.\ Raynal, 
Phys.\ Rev.\ {\bf D 8} (1973) 2322; {\bf 9} (1974) 1415.
\bibitem{Nambu60} Y.\ Nambu, Phys.\ Rev.\ Lett.\ {\bf 60} (1960) 380.
\bibitem{NJL61}
 Y.\ Nambu and G.\ Jona-Lasinio, Phys.\ Rev.\ {\bf 122} (1961) 345, and
 {\it ibid} {\bf 124} (1961) 246.
\bibitem{Man84}
 A.\ Manohar and H.\ Georgi, Nucl.\ Phys.\ {\bf B234} (1984) 189.
\bibitem{BPST75}
A.A.\ Belavin, A.M.\ Polyakov, A.S.\ Schwarz, and Yu.S.\ Tyupkin,
Phys.\ Lett.\ {\bf 59B} (1975) 85.
\bibitem{DY-PE86}
D.I.\ Diakonov and V.Yu.\ Petrov, Nucl.\ Phys.\ {\bf B272} (1986) 457-489.
\bibitem{SNRV71}      
 J.J. de Swart, M.M. Nagels, T.A. Rijken, and P.A. Verhoeven,
 Springer Tracts in Modern Physics, Vol.\ {\bf 60} (1971) 137.
\bibitem{NRS77} M.M.\ Nagels, Th.A.\ Rijken, and J.J.\ de\ Swart, 
Phys.\ Rev.\ {\bf D 15} (1977) 2547.
\bibitem{Gloz96a}
L.Ya.\ Glozman and D.O.\ Riska, Physics Reports, {\bf 268} (1996) 263-303.
\bibitem{Greenberg64}
O.W.\ Greenberg, Phys.\ Rev.\ Lett.\ {\bf 13} (1964) 598; 
J.J.J.\ Kokkedee, {\it The quark model}, W.A. Benjamin, New York, 1969.
\bibitem{Rij.nnqq14} Th.A.\ Rijken, {\it Compositeness and NN-potentoals II}, 
 IMAPP preprint, June 2014, 2014 (unpublished).
\bibitem{Weinberg75}
S.\ Weinberg, Phys.\ Rev.\ {\bf D11} (1975) 3583.
\bibitem{GtH76} 
 G.\ 't\ Hooft, Phys.\ Rev.\ {\bf D14} (1976) 3432.
\bibitem{DY-PE84a}
D.I.\ Diakonov and V.Yu.\ Petrov, Phys.\ Lett.\ {\bf 147B} (1984) 351.
\bibitem{DY-PE84b}
D.I.\ Diakonov and V.Yu.\ Petrov, Nucl.\ Phys.\ {\bf B245} (1984) 259-292.
\bibitem{SHUR84}
 E.V.\ Shuryak, Phys.\ Rep.\ {\bf C115} (1984) 152.     
\bibitem{SHUR82a}
 E.V.\ Shuryak, Nucl.\ Phys.\ {\bf B203} (1982) 93-115.
\bibitem{SVZ79}
M.A.\ Shiffman, A.I.\ Vainshtein, and V.I.\ Zakharov, Nucl.\ Phys.\ 
{\bf B147} (1979) 385, 448.
\bibitem{Lead96}
 E.\ Leader and E.\ Predazzi, {\it An introduction to gauge theories and 
 modern particle physics}, Cambridge University Press 1996, 
 see Vol. 1, section 17.1.2.
\bibitem{remark-spin}
This seems at first sight not to be true. Below the treatment of the 
axial-vector meson coupling will reveal that the ¨spin-crisis¨ 
suggests that most of the spin is not carried by the quarks, but by
the gluons. However, in the constituent-quark model there is no 
gluon-sea but only quarks. Therefore, we suppose that the 
coupling of the mesons at the quark-level has different vertices
than at the nucleon-level, and maintain that the nucleon spin is
the result of the total spin of the constituent quarks.

\bibitem{CC1}
The hermiticity of $\Delta{\cal L} \equiv \Delta J^\mu_a\cdot A_\mu$ 
follows from 
$[(\partial_\alpha\bar{\psi})\gamma_\nu(\partial_\beta\psi]^\dagger =
[(\partial_\beta\bar{\psi})\gamma_\nu(\partial_\alpha\psi]$. The (-)-sign
from $ i \rightarrow -i$ is compensated by the antisymmetry of 
$\epsilon^{\mu\nu\alpha\beta}=-\epsilon^{\mu\nu\beta\alpha}$.\\
The charge-conjugation properties follow from 
${\cal C}[(\partial_\alpha\bar{\psi})\gamma_\nu(\partial_\beta\psi]{\cal C}^{-1}=
-[(\partial_\beta\bar{\psi})\gamma_\nu(\partial_\alpha\psi]$, and again this
(-)-sign is compensated by the antisymmetry of the $\epsilon$-symbol. So the
hadronic current $\Delta J^\mu_a$ has $J^{PC} = 1^{++}$, {\it i.e.} the same as
the axial-vector meson field.

The extra Lagrangian can be written as
\begin{eqnarray*}
\Delta{\cal L} &\sim& 
-\epsilon^{\mu\nu\alpha\beta}
[\bar{\psi} \gamma_\nu \partial_\beta \psi]\ (\partial_\alpha A_\mu) 
 = -\frac{1}{2} [\bar{\psi} \gamma_\nu \partial_\beta \psi]\ 
 \frac{\partial_\alpha X_\mu}{\partial A^\nu} \\
&=& 
 -\frac{1}{4}\epsilon^{\mu\nu\alpha\beta}
 [\bar{\psi}\left(\gamma_\mu \partial_\nu -\gamma_\nu \partial_\mu\right)
 \psi]\ (\partial_\alpha A_\beta-\partial_\beta A_\alpha).    
\end{eqnarray*}
Here, $X_\mu \equiv 2\epsilon_{\mu\nu\alpha\beta} 
A^\nu(\partial^\alpha A^\beta)$ the conserved anomalous 
axial current \cite{anomaly}.
 
\bibitem{anomaly}
{\bf Note}: The axial-vector vertex is also connected to the famous 
Adler-Bell-Jackiw {\it axial anomaly} phenomenon \cite{Peskin95}.
For the axial isospin currents,
\begin{eqnarray*}
 \partial_\mu J^{5,a}_\mu &=& -\frac{g^2}{16\pi^2} \epsilon^{\alpha\beta\mu\nu}
 G^c_{\alpha\beta} G^d_{\mu\nu}\cdot tr[\tau^at^ct^d],
\end{eqnarray*}
where $G^c_{\mu\nu}$ is a gluon field strength, $\tau^a$ is an isospin matrix,
$t^c$ is a color matrix. the trace is 
$tr[\tau^at^ct^d] = tr[\tau^a]tr[t^ct^d]=0$. So the isospin currents are 
unaffected by the axial anomaly. However, for the isospin singlet this is not
the case:
\begin{eqnarray*}
 \partial_\mu J^{5}_\mu &=& -\frac{g^2n_f}{32\pi^2} \epsilon^{\alpha\beta\mu\nu}
 G^c_{\alpha\beta} G^d_{\mu\nu} \neq 0.                   
\end{eqnarray*}
This explains why the strong interactions contain no pseudoscalar meson as light
as the pion.
We conclude that the anomaly is not connected to a possible modification
of the strong axial-vector coupling to the quarks.
\bibitem{Peskin95}
M.E.\ Peskin and D.V.\ Schroeder, {\it An Introduction to Quantum Field
Theory}, Addison-Wesley publ. Company, 1995.

\bibitem{CC2}
The hermiticity of $\Delta{ L}'$ 
follows from the combination of $ i \rightarrow -i$ and partial integration.
For the latter the $\epsilon$-factor is important.\\
The charge-conjugation properties follow from 
${\cal C}[\bar{\psi}\gamma_\nu\psi]{\cal C}^{-1}=
-[\bar{\psi}\gamma_\nu\psi]$, and this
(-)-sign is compensated by partial integration to bring the differentiations
back in the original form.\\
The parity properties are checked easily: (i) $\mu=0$: all indices of the 
angular momentum operator or space-like, and hence 
${\cal P} A_0 {\cal P}^{-1}= -A_0$ is compensated by 
$\gamma_0 \gamma_n \gamma_0 = -\gamma_n$.
(ii) $\mu = m$: one easily checks that 
$[\bar{\psi}{\cal M}_{\nu\alpha\beta}\psi]$ is parity invariant due to the 
restrictions on the indices, again because of the $\epsilon$-factor.
So, writing $\Delta{\cal L}' = J'_{A,\mu}\ A^\mu$ the current has
$(-)^{PC}= 1^{++}$.


\bibitem{NRS78}
M.M.\ Nagels, Th.A.\ Rijken, and J.J.\ de Swart, 
Phys.\ Rev.\ {\bf D17} (1978) 768.

\bibitem{PTP185.a} Th.A.\ Rijken, M.M.\ Nagels, and Y.\ Yamamoto, 
Progres of Theoretical Physics Suppl. No. {\bf 185} (2010) 14.

\bibitem{YFYR13}
Y. Yamamoto, T. Furumoto, N. Yasutake, and Th.A. Rijken,
Phys.\ Rev.\ {\bf C 88} (2013) 022801(R).
\bibitem{Mil89}  G.A.\ Miller, Phys.\ Rev.\ {\bf C 39} (1989) 1563.
\bibitem{Wil94}
K.G.\ Wilson, et al., Phys.\ Rev.\ {\bf D49}, 6720 (1994); R.J.\ Perry,
arXiv:hep-th/9710175v1, 1997.

 \bibitem{YYR22}
 Y.\ Yamamoto, N.\ Yasutake, and Th.A.\ Rijken, Phys.\ Rev.\ C {\bf 105} 015804 (2022).
 \bibitem{YYR23}
 Y.\ Yamamoto, N.\ Yasutake, and Th.A.\ Rijken, Phys.\ Rev.\ C {\bf 108} 035811 (2023).
 \bibitem{YYR24}
 Y.\ Yamamoto, N.\ Yasutake, and Th.A.\ Rijken, Phys.\ Rev.\ C {\bf 110} 025805 (2024).

\bibitem{Close79}
 F.E. Close, {\it An Introduction to Quarks and Partons}, Academic
Press, London New York San Francisco, 1979.

\bibitem{Rij91} Th.A. Rijken, Ann.\ Phys.\ (N.Y.) {\bf 208} (1991) 253.
\bibitem{notation1}     
At this point it is suitable to change the notation of the initial and
final momenta. We use from now on the notations ${\bf p}_i\equiv {\bf p}$, 
${\bf p}_f\equiv {\bf p}'$ for both on-shell and off-shell momenta.
\bibitem{MRS89} P.M.M.\ Maessen, Th.A.\ Rijken, and J.J.\ de Swart,
         Phys.\ Rev.\ C {\bf 40} (1989) 2226.
\bibitem{IZ80}
 C.\ Itzykson and J-B\ Zuber, 'Quantum Field Theory', McGraw-Hill Inc. 1980.
\bibitem{Nak72}           
 N.\ Nakanishi, Suppl. Progr. Theor. Phys. {\bf 51} (1972) 1 
\bibitem{NO90}            
 N.\ Nakanishi and I.\ Ojima, 'Covariant Operator Formalism
of Gauge Theories and quantum Gravity', section 2.4.2, World Scientific Lecture Notes
 in Physics, Vol. 27, World Scientific Pub. Co 1990.

\bibitem{BD65}       
 J.D. Bjorken and S.D. Drell, {\it I. Relativistic Quantum Mechanics} and 
 {\it II. Relativistic Quantum Fields}, McGraw-Hill Publishing Company 1965.

\bibitem{dercopax}      
Note that in this paper we suppose that $f_A$ does not contain the 
one-pion-pole etc. In momentum space $\widetilde{f}_A({\bf k}^2)$ is a 
smooth function of ${\bf k}^2$.
\bibitem{Rij04a} Th.A.\ Rijken, Phys.\ Rev.\ {\bf C73} (2006) 44007.            
\bibitem{Rij04b} Th.A.\ Rijken and Y.\ Yamamoto, 
 Phys.\ Rev.\ {\bf C73} (2006) 44008.            
  \end{thebibliography}
\end{document}